{}%disable hyper at arxiv
{}%use pdflatex at arxiv

\documentclass[12pt]{article}
\usepackage{amssymb}
\usepackage{amsmath}
\usepackage[space]{cite}
\usepackage{graphicx}
\usepackage[normalem]{ulem}
\usepackage{tikz}
\usepackage{multirow}

\numberwithin{equation}{section}

\newcommand{\be}{\begin{equation}}
\newcommand{\ee}{\end{equation}}
\newcommand{\non}{\nonumber}

\newcommand{\tr}{\mathop{\rm tr}\nolimits}
\newcommand{\diag}{\mathop{\rm diag}\nolimits}

\newcommand{\RN}[1]{%
  \textup{\uppercase\expandafter{\romannumeral#1}}%
}

\begin{document}

\begin{titlepage}
	\strut\hfill UMTG--315
	\vspace{.5in}
	\begin{center}
		
		{\LARGE Flag Integrable Models and Generalized Graded Algebras
		}\\
		\vspace{1in}
		\large 
		Marius de Leeuw\footnote{School of Mathematics \& Hamilton 
			Mathematics Institute, Trinity College Dublin, Dublin, Ireland, m.deleeuw1@gmail.com}, Rafael I. Nepomechie\footnote{Physics Department,
			P.O. Box 248046, University of Miami, Coral Gables, FL 33124 USA, nepomechie@miami.edu}
		and Ana L. Retore\footnote{School of Mathematics \& Hamilton 
			Mathematics Institute, Trinity College Dublin, Dublin, Ireland}$ ^, $\footnote{Department of Mathematical Sciences, Durham University, Durham DH1 3LE, UK, ana.retore@durham.ac.uk, (Current)} 
		
	\end{center}
	
	\vspace{.5in}
	
	\begin{abstract}
		We introduce new classes of integrable models that exhibit a 
		structure similar to that of flag vector spaces.	
		We present their Hamiltonians, $R$-matrices and Bethe-ansatz solutions. 
		These models have a new type of generalized graded algebra 
		symmetry. 
	\end{abstract}
	
\end{titlepage}

\setcounter{footnote}{0}

\newpage

%%%%%%%%%%%%%%%%%%%%%%%%%%%%%%%%%%%%%%%%%%%%%%%%%%%%%%%%%%%%%%%%%%%%%%%%%%%%%%%%
%%%%%%%%%%%%%%%%%%%%%%%%%%%%%%%%%%%%%%%%%%%%%%%%%%%%%%%%%%%%%%%%%%%%%%%%%%%%%%%%

\tableofcontents
\section{Introduction}

The study of integrable spin chains is by now a mature subject — many
infinite families of such models have already been identified and
solved.  Many of these models were derived from quantum (super) algebras
\cite{Kulish:1981gi,Faddeev:1987ih,Reshetikhin:1990sq,jimbo1990yang}.
The best known examples are of course Yangians
\cite{Drinfeld:1985rx,Drinfeld:1986in} and quantum affine algebras
\cite{jimbo1983solitons,jimbo1988integrable}.  In fact, there is even
a close relation between the functional form of the $R$-matrix and the
symmetry algebra \cite{belavin1982solutions}.  Rational $R$-matrices
typically have a symmetry of Yangian type, while trigonometric 
$R$-matrices typically have a symmetry of a quantum affine type.  
Hence, it may come as a surprise that new {\em rational}
solutions of the Yang-Baxter equation, and corresponding integrable
spin chains, can still be found.

Recently, a more direct approach to classifying solutions of the 
Yang-Baxter equation has been put forward which employs the so-called 
boost operator \cite{deLeeuw:2019zsi, deLeeuw:2019vdb, 
deLeeuw:2020xrw,deLeeuw:2020ahe}. One of the advantages of this 
approach is that it does not rely on symmetry arguments and gives a 
complete classification. Several new solutions of the Yang-Baxter 
equation have been found that are rational, trigonometric and 
elliptic. The natural follow-up question is then whether there are quantum algebras that underlie these models. 
For some of the new models, the algebras seem closely related to 
centrally extended algebras \cite{deLeeuw:2021ufg}. However, in 
\cite{deLeeuw:2019vdb} very simple rational 
solutions (Models 4 and 
6) were found for which the symmetry algebra was still unclear.  More 
precisely, Models 4 and 6 from \cite{deLeeuw:2019vdb} have a 
4-dimensional Hilbert space at each site, and have  
$16\times 16$
$R$-matrices that take the form
\begin{align}
R \sim u \,\mathbb{I}^{(4, 4)}-\mathbb{P}^{(4, 4)}  + u\, 
\left(\mathbb{I}^{(2, 4)} - \mathbb{P}^{(2, 4)}\right) \,,
\end{align}
where $\mathbb{I}^{(4, 4)}$ and $\mathbb{P}^{(4, 4)}$ are the usual identity and permutation matrix, 
but $\mathbb{I}^{(2, 4)}$ and $\mathbb{P}^{(2, 4)}$ are the identity and permutation operator 
restricted to a two-dimensional subspace, see \eqref{eq:P}, \eqref{eq:I}. These models look like combinations of simple XXX type models. 
Similar models were found in work on so-called multiplicity
$A$-models \cite{Maassarani:1998} (building on earlier work 
in \cite{Maassarani:1997kon, Maassarani:1997}), which
were further studied and generalized in \cite{Drummond:2007gt} and in \cite{Kagan:2008} .  

\begin{figure}[h!]
\begin{center}
\includegraphics[scale=.45,angle=0]{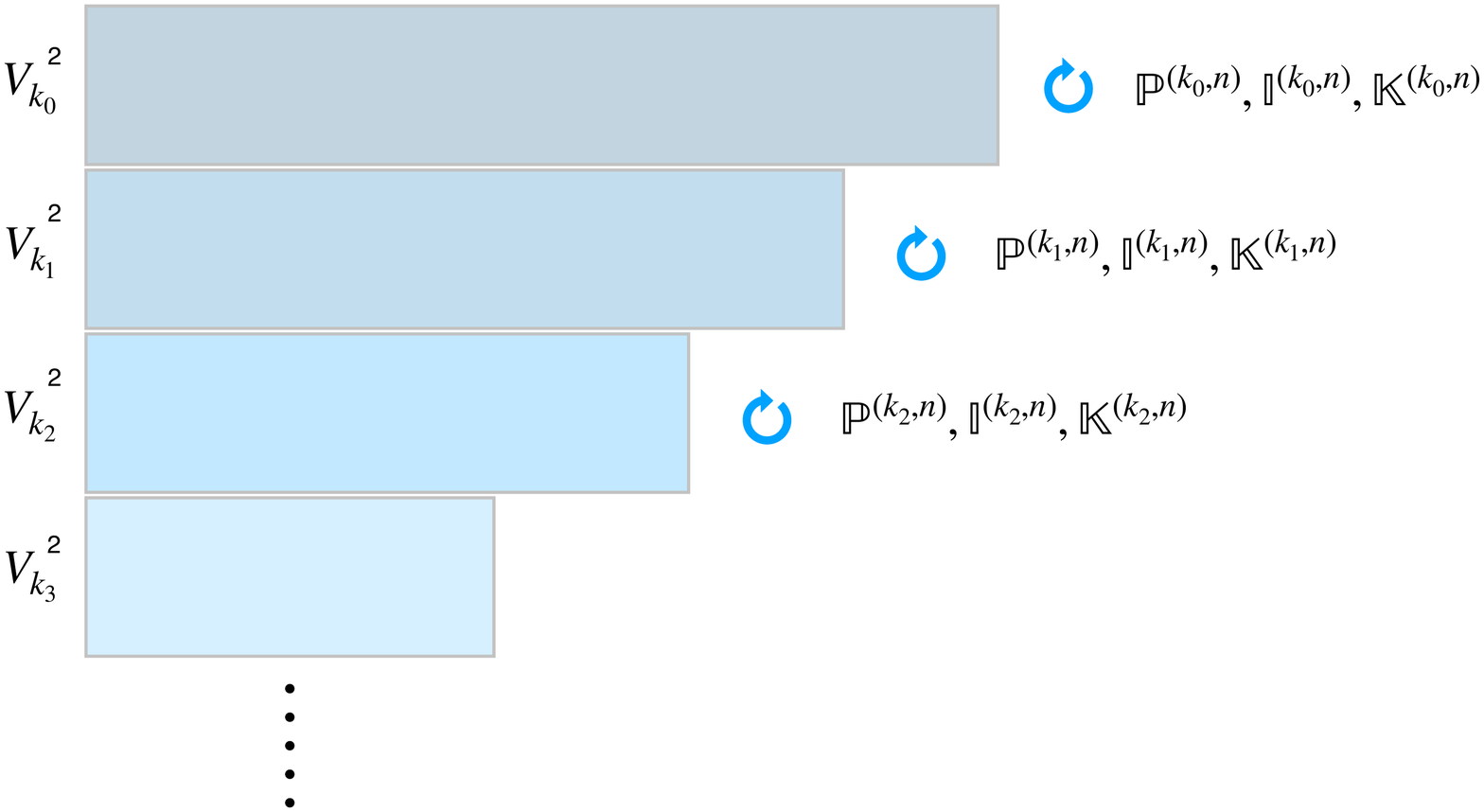}
\caption{We consider a flag vector space with operators $\mathbb{P},\mathbb{I},\mathbb{K}$ acting on the tensor products of various subspaces. In particular, $\mathbb{I}^{(k_i,n)}$ is the characteristic function of that subspace, \textit{i.e.} it is the identity for vectors in the subspace and 0 for the complement. The other operators are similarly defined in the case of the permutation and the trace operator. }\label{fig:Flag}
\end{center}
\end{figure} 

Inspired by this, we consider here a generalization of these types of
models where we take the $R$-matrix to be a linear combination of the
identity, permutation and trace operators, see \eqref{eq:P}-\eqref{eq:K}, that are
restricted to subspaces $V_{k_{d-1}} \subset \ldots \subset
V_{k_1}\subset V_{k_0}$ see Figure \ref{fig:Flag}.  We recall that, 
in linear algebra, a \textit{flag}
refers to such an increasing sequence of subspaces of a vector space,
and hence we name these solutions \textit{flag integrable models}.

By using the boost operator method, we find three non-trivial infinite families of integrable spin chains
that have such a flag structure. We refer to these as models 
\RN{1}, \RN{2} and  \RN{3}. These models are 
characterized by a set  $\vec{k}$ of $d$ decreasing positive integers
\begin{align}
&\vec{k} = \{ k_0, k_1,\, \ldots, \,k_{d-1}\} \,,
&& n = k_0 > k_1 > \ldots > k_{d-1}\ge 1 \,,
\label{eq:veckrules}
\end{align}
where $n$ is the dimension of the Hilbert space at each site. A subset of model \RN{2}
can be related to a subset of the model in \cite{Maassarani:1998}.
Despite the simplicity of their Hamiltonians and $R$-matrices, these models have nontrivial spectra,
symmetries and degeneracies.  We find a fourth model, model \RN{4}, whose spectrum is purely combinatorial.
For given values of $n$ and $d$, the number of possible models are 
${n-1 \choose d-1}$ for model \RN{1} and \RN{2}, and ${n-3 \choose d-2}$ for 
models \RN{3} and \RN{4}, respectively, as we will see below.

We will show that our models exhibit a type of generalized graded Lie
algebra symmetry, which we will denote by
$\mathfrak{gl}(k_0-k_1|\ldots|k_{d-2}-k_{d-1}|k_{d-1})$.  When the
flag has only two stripes \textit{i.e.} $d=2$, then we return to the
usual Lie superalgebra $\mathfrak{gl}(n-k|k)$.  We furthermore show
that Model \RN{1} admits a Yangian extension of this algebra and is
uniquely fixed by it.

We will also work out the nested algebraic Bethe ansatz for
models \RN{1}, \RN{2} and \RN{3}. Surprisingly, many of the transfer-matrix
eigenvalues are described by infinite, singular and/or continuous
Bethe roots.

\section{Derivation of the models}

In this section we derive the form of the flag models.  Motivated by
our work on Hubbard-type models and the Maassarani-Matthieu models, we
will consider Hamiltonians built out of restrictions of the identity,
permutation and trace operators.

\subsection{The Hamiltonians}

We begin by studying the direct generalization of Models 4 and 6 from
\cite{deLeeuw:2019vdb}. We will see that these models have
$R$-matrices that are rational and of difference form, and are similar to
XXX-type models.

\paragraph{Notation}
Let us first define the restricted operators that we will use to construct our integrable models. We denote
\begin{align}
& \mathbb{P}^{(m,n)}=\sum_{i,j=1}^{m}e_{i,j}\otimes e_{j,i} \,,\label{eq:P}\\
& \mathbb{I}^{(m,n)}=\sum_{i,j=1}^{m}e_{i,i}\otimes e_{j,j} \,,\label{eq:I}\\
& \mathbb{K}^{(m,n)}=\sum_{i,j=1}^{m}e_{i,j}\otimes e_{i,j} \,, 
\label{eq:K}
\end{align}
where $e_{i,j}$ is an $n\times n$ matrix such that $(e_{i,j})_{\alpha,\beta}=\delta_{i,\alpha}\delta_{j,\beta}$, and $1\le m \le n$. For $m=n$, the operator $ \mathbb{P}^{(m,n)} $ becomes the usual permutation operator for a Hilbert space of dimension $ n $, and similarly, $ \mathbb{I}^{(m,n)} $ reduces to the identity matrix.

\paragraph{Hamiltonian} Inspired by the simple form of Models 4 and 6
from \cite{deLeeuw:2019vdb}, we consider a similar nested
structure where we combine general Hamiltonians that are built out of
the building blocks of $SO(n)$ spin chains.  Consider a set
$\vec{k}$ of decreasing positive integers
\begin{align}
&\vec{k} = \{ k_0, k_1,\, \ldots, \,k_{d-1}\} \,,
&& n = k_0 > k_1 > \ldots > k_{d-1}\ge 1 \,,
\label{eq:veckrules}
\end{align}
where $n=k_{0}$ is the dimension of the Hilbert space at each site. We take our Hamiltonian to be of the form
\begin{align}
\mathcal{H}^{\vec{k}} = \sum_{i=0}^{d-1} \left( a_i \, 
\mathbb{I}^{(k_i,n)} + b_i\,  \mathbb{P}^{(k_i,n)} + c_i\, 
\mathbb{K}^{(k_i,n)} \right) \,.
\label{HAnsatz}
\end{align}
At this point we do not assume the $R$-matrix is of difference form
and hence the coefficients $a_i, b_i, c_i$ can depend on the
inhomogeneities $\theta$ of the spin chain.  We will suppress the
explicit $\theta$-dependence in our notation.  Nevertheless, when
solving the integrability conditions, we shall see that these
coefficients are in fact constants and the corresponding $R$-matrix is
of difference form.

\paragraph{Boost operator formalism} 
We now  proceed to insert the Ansatz \eqref{HAnsatz} in the general boost
operator formalism of \cite{deLeeuw:2020ahe} and classify all possible
integrable Hamiltonians of this form. 
\iffalse
Let us briefly recall the key ingredients from \cite{deLeeuw:2020ahe}.
The most important operator in the approach of \cite{deLeeuw:2020ahe}
is the boost operator $\mathcal{B}[H]$
%
\begin{align}
\mathcal{B}[H] =\partial_\theta +  \sum_{n} n \mathcal{H}_{n,n+1}(\theta).
\end{align}
%
 This operator depends on the Hamiltonian and generates the higher conserved charges. In particular 
 %
\begin{align}
Q_3 \sim [\mathcal{B}[H] ,H].
 \end{align}
 %
 \fi
In order for this system to be integrable, a criterion is derived in
\cite{deLeeuw:2020ahe} that gives a set of first-order differential
equations for the coefficients of the Hamiltonian.
%It is also shown that this method is complete, \textit{i.e.} in this way we find all integrable models with a Hamiltonian of the form \eqref{HAnsatz}.

\paragraph{Recursion relations}
We can obtain recursion relations for the coefficients in the 
Hamiltonian by acting on subspaces of our total vector space 
$V_{k_0}$. For instance, if we take a tensor product of vectors from 
the complement of $V_{k_1}$, then the only operators that act 
non-trivially on it will be the operators 
$\mathbb{P}^{(0,n)},\mathbb{I}^{(0,n)},\mathbb{K}^{(0,n)}$. In this 
paper, we are looking for solutions that are compatible with the
general flag structures.  There exist special solutions when $\vec{k}$
takes specific values; while these solutions are potentially
interesting, we do not consider them in this paper.

When imposing the integrability condition on the complement of $V_{k_1}$, we see that only the terms with
$a_0,b_0,c_0$ will contribute to the integrability condition
%\cite{deLeeuw:2020ahe} 
and, consequently, they have to give an
integrable Hamiltonian by themselves.  We find the following equations
\begin{align}
&b_0 \dot{c}_0 = c_0 \dot{b}_0 \,,
&&b_0 c_0 \left(b_0+ \frac{k_0-2}{2} c_0\right) = 0 \,,
\end{align}
where the dot denotes differentiation with respect to $\theta$.
There are three possible solutions to these integrability 
conditions, all of which are constant, namely
\begin{align}
&b_0=0,
&&c_0=0,
&&b_0 = - \frac{k_0-2}{2} c_0.
\end{align}
We easily recognize the usual $SU(n)$ when $c_0=0$, and the $SO(n)$
spin chain when $b_0 = \frac{2-k_0}{2}c_0$.  The last case $b_0=0$ is
a generalization of a spin chain with $SO(n)$ symmetry that was found
for the case $n=4$ in \cite{deLeeuw:2019vdb} (see formula (4.4) in
that reference).

Next we take vectors from the complement of $V_{k_{2}}$ and then
$\mathbb{P}^{(1,n)},\mathbb{I}^{(1,n)},\mathbb{K}^{(1,n)}$ will
contribute as well.  We find equations that relate the coefficients
$a_0,b_0,c_0$ and $a_1,b_1,c_1$.  We generate the corresponding set of
equations in \texttt{Mathematica}.  There are on the order of 50
(dependent) equations.
 
Nevertheless, it can be quickly seen that the case where $c_0\neq 0$
implies that $a_1=b_1=c_1=0$.  By induction this implies that there is
only the contribution to our Hamiltonian from the leading part, and we
keep the spin chains that we identified in the first step.  We find
that we need to take $c_0=0$ to get a new and interesting solution.
When $c_0=0$, we can normalize our Hamiltonian such that we find two
possible cases $b_0=0,1$.  Note that $a_0$ can be arbitrary, since it 
multiplies the identity operator, and a shift of the Hamiltonian that 
is proportional to the identity operator is harmless.

Let us first consider the case $b_0=1$. The equations for $a_1,b_1,c_1$ coupled to $a_0,b_0,c_0$ can then be solved to give three different non-trivial solutions 
\begin{itemize}
\item $a_1 = 1, ~ b_1 = -1,~ c_1=0$
\item $a_1 = -1,~ b_1 = -1,~ c_1=0$
\item $a_1 = 0, ~b_1 = -2, ~c_1=0$
\end{itemize}
At the next level, we consider vectors from the complement of $V_{k_2}$ and we see that the first two solutions impose that $a_i,b_i,c_i$ all vanish for $i>1$. Hence, for these solutions our recursion terminates. The third solution, however, offers a continuation at the next level and again gives rise to three cases
\begin{itemize}
\item $a_2 = 1, ~ b_2 = 1,~ c_2=0$
\item $a_2 = -1,~ b_2 = 1,~ c_2=0$
\item $a_2 = 0, ~b_2 = 2, ~c_2=0$
\end{itemize}
Also in this instance, the first two solutions terminate the recursion again. Repeating this process, we see that we are left with two types of models. First, there is the model with the third-type solution repeated to the end:
\begin{align}
H^{\RN{1},\vec{k}}=a_0\,\mathbb{I}^{(n,n)} + 
b_{0}\,\Big[\mathbb{P}^{(n,n)}+2\sum_{j=1}^{d-1}(-1)^j\mathbb{P}^{(k_j,n)} \Big] \,.
\label{eq:HmodelA}
\end{align}
It is natural to introduce $m\in[1,n]$, and to define $\bar{m}$ by
\begin{align}
\bar{m} = \left\{
 \begin{array}{cc}
 1 & k_1< m \leq k_0 \\
 2 & k_2< m \leq k_1 \\
 \vdots & \vdots 	\\
 d & 0< m < k_{d-1}
 \end{array}\right.  \,.
 \label{bar}
\end{align}
The barred index indicates in which subspace our vector takes values. We can then rewrite 
\begin{align}
H^{\RN{1},\vec{k}}=a_0\,\mathbb{I}^{(n,n)} + 
b_{0}\ \mathcal{P}^{\vec{k}},
%\label{eq:HmodelA}
\end{align}
where 
\begin{align}
\mathcal{P}^{\vec{k}} (e_i \otimes e_j) = (-1)^{\min (\bar{i}, \bar{j})} e_j \otimes e_i.
\end{align}
is a generalization of the usual graded permutation
operator\footnote{We thank the referee for pointing out this elegant
form.}.  For a flag with two stripes this is just proportional to the
usual graded permutation operator.  To the best of our knowledge, this
simple rational model has not been found in the literature before.

Second is the case where in the last step one of the other solutions is used
\begin{align}
H^{\RN{2}_{\pm},\vec{k}}=a_0\,\mathbb{I}^{(n,n)} + 
b_{0}\,\Bigg[\mathbb{P}^{(n,n)}+2\Big(\sum_{j=1}^{d-2}(-1)^j\mathbb{P}^{(k_j,n)}\Big)  - (-1)^{d}\mathbb{P}^{(k_{d-1},n)} \pm \mathbb{I}^{(k_{d-1},n)}\Bigg] \,.
\label{eq:HmodelB}
\end{align}
Third, there is a special case when $k_{d-1} =2$. In this case, we find that $\mathbb{K}^{(2)}$ can appear. Hence, we arrive at a 
third model given by
\begin{align}
H^{\RN{3}_\pm,\vec{k}} = a_0\, \mathbb{I}^{(n,n)} + 
b_0\Bigg[\mathbb{P}^{(n,n)}+2\Big(\sum_{j=1}^{d-2}(-1)^j\mathbb{P}^{(k_j,n)}\Big)-(-1)^{d}\mathbb{P}^{(2,n)}\pm \mathbb{K}^{(2,n)}\mp\mathbb{I}^{(2,n)}\Bigg] \,.
\label{eq:HmodelC}
\end{align}
Notice that the only possible $SO(N)$ type integrable Hamiltonian that is compatible with the imposed flag structure is the usual $SO(N)$ spin chain. The only other instance in which the trace operator appears is in the case $k_{d-1} = 2$ as in Model III.

Let us finally consider the case with $b_0=c_0=0$. Since we can set $a_0=0$ without loss of generality, 
we find at the next step that $b_1=c_1=0$, and that $a_1$ is constant. By induction, this structure goes
through to the other levels as well, and generically one arrives at a
diagonal Hamiltonian, which is trivially integrable.  However, also
here there is a special case when $k_{d-1}=2$.  When this is the case,
we find a non-trivial Hamiltonian.  This is our fourth model, which we
denote by
\begin{align}
H^{\RN{4},\vec{k}}=b_{d-1}\,(\mathbb{P}^{(2,n)} - 
\mathbb{K}^{(2,n)})+\sum_{j=0}^{d-1}a_{j}\mathbb{I}^{(k_j,n)} \,.
\label{eq:HmodelD}
\end{align}
This model, however, is different from the previous ones since its
spectrum is purely combinatorial: all the eigenvalues are simply
integer multiples of the coefficients $a_i$ and $b_{d-1}$.  Hence we
will not consider this model much further.

\subsection{$R$-matrices}\label{sec:models}

In order to prove that these models are integrable, we compute the
$R$-matrices that generate the Hamiltonians. We emphasize that we restrict throughout 
this paper to non-graded R-matrices, which satisfy the non-graded
(ordinary) Yang-Baxter equation.
Unsurprisingly, the $R$-matrices can be expressed in terms of the same
operators as the Hamiltonians, and are easily found from the Sutherland
equation \cite{deLeeuw:2019zsi, deLeeuw:2019vdb, 
deLeeuw:2020xrw,deLeeuw:2020ahe}
\begin{align}
\label{eqn:Sutherland}
\left[R_{13} R_{23}, H_{12}(u)\right] = \dot{R}_{13} R_{23} - R_{13} \dot{R}_{23}\, ,
\end{align}
where the dot indicates the derivative with respect to the first spectral parameter $\dot{R}(u,v) = \partial_u R(u,v) $.
The Sutherland equations can be derived from the Yang-Baxter equation
and give a set of non-linear first-order differential equations for
the $R$-matrix in terms of the Hamiltonian.  Given that the $R$-matrix
needs to satisfy the boundary conditions $R(u,u)=P$ and $\dot{R}(u,u)
=PH $, we find that a given Hamiltonian leads to a \textit{unique}
$R$-matrix which is a solution of the Yang-Baxter equation.

\subsubsection{Model \RN{1}}\label{subsec:model0}

%The Hamiltonian for the first model we found is given by
%%
%\begin{align}
%H^{A,\vec{k}}=a_0\,\mathbb{I}^{(n)} + b_{0}\,\Big[\mathbb{P}^{(n)}+2\sum_{j=1}^{d-1}(-1)^j\mathbb{P}^{(k_j)} \Big],
%\label{eq:Hmodel0}
%\end{align}

The $R$-matrix corresponding to the Hamiltonian \eqref{eq:HmodelA} is given by
\begin{align}
R^{\RN{1},\vec{k}}(u) 
=(u+1)\left(\eta\,\mathbb{P}^{(n,n)}+ 
u\,\mathbb{I}^{(n,n)} + 2 
\,u\,\sum_{j=1}^{d-1}(-1)^j\mathbb{I}^{(k_j,n)} \right) \,,
\label{eq:Rmodel0}
\end{align}
where we hereby set
\begin{equation}
a_{0} = \eta\,, \qquad b_{0} = 1 \,,
\end{equation}
 where $\eta$ has the interpretation of a quantum parameter (Planck's 
constant) rather than an anisotropy parameter. We can do this since we are free to choose a normalization of the $R$-matrix and also redefine our spectral parameter.
The form of this $R$-matrix is evidently very simple. 

We can now decompose the $R$-matrix into the 
sum of the permutation matrix and a simple diagonal matrix, namely
\begin{align}
&R^{\RN{1},\vec{k}}(u) =(u+1)\Big[\eta\,\mathbb{P}^{(n,n)} + 2 \,u\, \mathcal{I}^{\vec{k}} \Big],
%&& I^{\vec{k}} = 
%\sum_m (-1)^m 
%\diag( 1,\ldots,1 ,-1_{k_1},\ldots,-1,1_{k_2},\ldots ) \,,
\label{eq:RmodelA}
\end{align}
where $\mathcal{I}^{\vec{k}}$ is a diagonal matrix with the following $\pm1$ entries
\begin{align}
\mathcal{I}^{\vec{k}} (e_i \otimes e_j) = (-1)^{\min (\bar{i}, \bar{j})} e_i \otimes e_j.
\end{align}
To the best of our knowledge this is a new $R$-matrix.

\subsubsection{Model \RN{2}}\label{subsec:model1}

The $R$-matrix corresponding to the Hamiltonian \eqref{eq:HmodelB} is 
\begin{align}
R^{\RN{2} \pm,\vec{k}}(u) = 
(u+1)\left(u\mathbb{I}^{(n,n)}+\eta\,\mathbb{P}^{(n,n)}+2u\sum_{j=1}^{d-2}(-1)^j\mathbb{I}^{(k_j,n)}-u(-1)^{d}\,\mathbb{I}^{(k_{d-1},n)}\pm u\mathbb{P}^{(k_{d-1},n)}\right) \,.
\label{eq:Rmodel1}
\end{align}
The first three terms coincide with the $R$-matrix for Model \RN{1}, but 
with vector $\{k_0,\ldots, k_{d-2}\}$. Hence, we can write it as
\begin{align}
R^{\RN{2} \pm,\vec{k}}(u) = R^{\RN{1},\vec{k}-1}(u)  - u 
(u+1)\Big[(-1)^{d}\,\mathbb{I}^{(k_{d-1,n)}}\pm 
\mathbb{P}^{(k_{d-1},n)}\Big] \,,
\end{align}
where by $R^{\RN{1},\vec{k}-1}(u)$ we denote the $R$-matrix of Model 
\RN{1}  corresponding to $\vec{k}$ with the last element dropped. 

At this point, let us spell out more clearly that this R-matrix
actually describes a family of models indexed by $\vec{k}$ and the
$\pm$ sign.  For fixed values of $n$ and $d$,  there are ${n-1\choose 
d-1}$ possible sets of $\vec{k}$'s.  For $ n=k_{0}=5$, for example, we have:
\begin{itemize}
	\item $d=1:$ only one model, corresponds to XXX;
	\item $d=2:$ $\{k_1\}$ can be equal to 
	$k_1=\{\{1\},\{2\},\{3\},\{4\}\}$, so there are four sets of $\vec{k}$'s;
	\item $d=3:$ $\{k_1,k_2\}$ can be equal to $\{k_1,k_2\}=\{\{2,1\},\{3,1\},\{4,1\},\{3,2\},\{4,2\},\{4,3\}\}$, resulting in six different sets of $\vec{k}$'s;
	\item $d=4:$ $\{k_1,k_2,k_3\}$ can be equal to 
	$\{\{3,2,1\},\{4,2,1\},\{4,3,1\},\{4,3,2\}\}$, which corresponds to four sets of $\vec{k}$'s;
	\item $d=5:$ $\{k_1,k_2,k_3,k_4\}$ can be only equal to $\{4,3,2,1\}$.
\end{itemize}
%
%In principle these models have different spectra and symmetries, but it turns out that there is an unusual relation between them which we will discuss in the next section.

We note that a subset of Model \RN{2} can be related to the models 
found in \cite{Maassarani:1998}. Setting in the latter all $x_{\alpha, 
\alpha'}=1$ and $\gamma=0$, we find the following dictionary

\begin{center}
\begin{tabular}[h]{c|c}
Model $\RN{2}^{+}$ with $d=2$	 & Maassarani's model \cite{Maassarani:1998}\\
\hline
$n$ & $n$ \\
$\vec k = \{k_{0}, k_{1}\} = \{n, n-m+1\} $  &  $\vec n = \{n_{1}, n_{2},\ldots, n_{m}\} = \{1,1, 
\ldots, 1, n-m+1\}$ 
\end{tabular}	
\end{center}
The mapping between the R-matrices is as follows: removing from the 
$R$-matrix \eqref{eq:Rmodel1} the overall factor $(1+u)$, setting 
$d=2$ and $\eta=i$, we have
\begin{align}
R^{\RN{2}+,\{k_{0}, k_{1}\}}(u) = 
u(\mathbb{I}^{(n,n)}-\,\mathbb{I}^{(k_{1},n)})+i\,\mathbb{P}^{(n,n)}+
u\mathbb{P}^{(k_{1})} \,.
\label{eq:Rmodel1mod}
\end{align}
Then
\begin{equation}
R^{\{1,1, \ldots, 1, k_{1}\}}_{\rm Maassarani}(u) = (V \otimes V) 
R^{\RN{2}+,\{k_{0}, k_{1}\}}(u) (V \otimes V) \,,
\end{equation}
where $V$ is the $ n\times n $ anti-diagonal unit matrix
\be
V = \begin{pmatrix}
    0 & 0 & \ldots & 0  & 1\\
	0 & 0 & \ldots & 1  & 0\\
	\vdots & \vdots &   & \vdots & \vdots \\
	0 & 1 & \ldots & 0  & 0\\
	1 & 0 & \ldots & 0  & 0
	\end{pmatrix} =\sum_{i=1}^{n}e_{i,n-i+1} \,.
\label{Vmat}
\ee

Inspired by the presentation of \cite{Maassarani:1998}, we find that 
we can rewrite our $R$-matrix \eqref{eq:Rmodel1} as
\begin{equation}
R^{\RN{2}\pm,\vec{k}}(u)=(u+1)\left(\eta \mathbb{P}^{(n,n)}+\,u\, 
\mathbb{F}^{(\pm,\vec{k})}\right) \,,
\label{eq:ModelBv2}
\end{equation}
where $\mathbb{F}^{(\pm,\vec{k})}$ is defined by
\begin{equation}
\mathbb{F}^{(\pm,\vec{k})}=\mathbb{I}^{(n,n)}-(-1)^{d}\,\mathbb{I}^{(k_{d-1},n)}\pm \mathbb{P}^{(k_{d-1},n)}+2\sum_{j=1}^{d-2}(-1)^j\mathbb{I}^{(k_{j},n)} \,,
\label{eq:FmodelBdef}
\end{equation}
which satisfies
\begin{align}
\mathbb{F}^{(\pm,\vec{k})}\, \mathbb{F}^{(\pm,\vec{k})} 
&=\mathbb{I}^{(n,n)} \,,
\label{eq:eq1forFModelB} \\
\mathbb{F}_{12}^{(\pm,\vec{k})}\mathbb{F}_{13}^{(\pm,\vec{k})}\mathbb{F}_{23}^{(\pm,\vec{k})}&=\mathbb{F}_{23}^{(\pm,\vec{k})}\mathbb{F}_{13}^{(\pm,\vec{k})}\mathbb{F}_{12}^{(\pm,\vec{k})} \,.
\label{eq:eq3forFModelB}
\end{align}
Moreover, $\check{\mathbb{F}}^{(\pm,\vec{k})}:=\mathbb{P}^{(n,n)}\, 
\mathbb{F}^{(\pm,\vec{k})}$ satisfies
\begin{equation}
\check{\mathbb{F}}^{(\pm,\vec{k})}_{12}\check{\mathbb{F}}^{(\pm,\vec{k})}_{23}\check{\mathbb{F}}^{(\pm,\vec{k})}_{12}=\check{\mathbb{F}}^{(\pm,\vec{k})}_{23}\check{\mathbb{F}}^{(\pm,\vec{k})}_{12}\check{\mathbb{F}}^{(\pm,\vec{k})}_{23} \,.
\label{eq:eq2forFModelB}\\
\end{equation}
In other words, $\mathbb{F}$ is a constant solution of the 
Yang-Baxter equation, and we can view the total $R$-matrix of Model \RN{2} as a Baxterization of $\mathbb{F}$ with the constant solution.
The proof for \eqref{eq:eq1forFModelB} for any $ n $, $ d $ and $ k_i $ for both Models \RN{2}$ ^+ $ and \RN{2}$ ^- $ is straightforward.

\subsubsection{Model \RN{3}}\label{subsec:model2}

Model \RN{3} is very similar to Model \RN{2}, and only differs in 
a new two-dimensional term. The R-matrix for Model \RN{3} \eqref{eq:HmodelC} is given by
\begin{align}
R^{\RN{3} \pm,\vec{k}}(u)&=(u+1)\Bigg(u\,\mathbb{I}^{(n,n)}+\eta\,\mathbb{P}^{(n,n)}+2\,u\,\sum_{j=1}^{d-2}(-1)^j\,\mathbb{I}^{(k_j,n)}\nonumber\\
&\hspace{1cm}+(-1)^{d+1}\, u\,\mathbb{I}^{(2,n)}\pm 
u\,\mathbb{K}^{(2,n)}\mp u\,\mathbb{P}^{(2,n)}\Bigg)\,,
\label{eq:Rmodel2}
\end{align}
where $k_{d-1}=2$.

\subsubsection{Properties of R-matrices for models \RN{1}, \RN{2} and 
\RN{3}}

The $R$-matrices for models \RN{1}, \RN{2}, \RN{3} satisfy some additional relations. First, we note that they all are symmetric 
\begin{align}
R^t = R \,.
\end{align}
Second, they are also trivially parity invariant 
\begin{align}
R_{21} = R_{12} \,.
\end{align}
Third, the $R$-matrices satisfy braiding unitarity 
\begin{align}
R_{12}(u)\, R_{21}(-u) \sim 1 \,.
\label{braidingunit}
\end{align}
We found also some more general relations
\begin{equation}
R(u)\, \mathbb{P}^{(m,n)}\, R(-u)\, \mathbb{P}^{(m,n)} \sim \mathbb{I}^{(m,n)} 
\,,
\label{eq:braidingmodel2}
\end{equation}
for $m \in [1,n]$ for models \RN{1} and \RN{2}, and for $m\ne n-1$ 
for model \RN{3}.
For $m=n$, this corresponds to braiding unitarity \eqref{braidingunit}.
Additionally, these $R$-matrices satisfy
\begin{equation}
R(u)\, \mathbb{I}^{(m,n)}\, R(-u)\, \mathbb{I}^{(m,n)} \sim \mathbb{I}^{(m,n)} 
\,,
\label{eq:identitymodel2}
\end{equation}
where again $m \in [1,n]$ for models \RN{1} and \RN{2}, and $m\ne 
n-1$ for model \RN{3}.

In general, the $R$-matrices do not satisfy crossing symmetry, except for a few specific values of $k_i$.

\subsubsection{Model \RN{4}}\label{subsec:model3}

The R-matrix for Model \RN{4} \eqref{eq:HmodelD} is given by
\begin{align}
R^{\RN{4},\vec{k}}(u)=&(\eta\,u+1)\left[\mathbb{P}^{(n,n)}-\left(\prod_{j=1}^{d-2}e^{a_j\,u}\right)\left(1-e^{a_{d-1}\,u}\,\cosh u\right)\mathbb{P}^{(2,n)}+\right.\nonumber\\
&\left.+\left(\prod_{j=1}^{d-1}e^{a_j\,u}\right)
\sinh u\left(\mathbb{K}^{(2,n)}-\mathbb{I}^{(2,n)}\right)-\sum_{j=1}^{d-2}\left(1-e^{a_j\,u}\right)\left(\prod_{i=1}^{j-1}e^{a_i\,u}\right)\mathbb{P}^{(k_j,n)}\right] \,.
\label{eq:Rmodel3}
\end{align} 
We see that, despite the simple form of the Hamiltonian 
\eqref{eq:HmodelD}, the corresponding R-matrix
is more involved and is in fact of trigonometric type.

\section{Generalized graded algebra}\label{sec:symmetries}

Usually, understanding the symmetries of the underlying models helps
with explaining the degeneracies of the spectrum and further properties of
the model.  Given the closeness of the models to usual XXX-type
models, we expect some sort of Yangian symmetry to be present.  In
this section we will demonstrate that models \RN{1}, \RN{2} and \RN{3} exhibit a new
type symmetry.  We can fully fix model \RN{1} by symmetry considerations,
but for models \RN{2} and \RN{3} a symmetry derivation seems to be out of reach.
The new symmetry is particularly interesting because it seems to
describe a generalized type of Fermi statistics.
For this reason we call them \textit{generalized graded algebras}. 

\subsection{Definition}

Let us first look at Model \RN{1}, since this will be the model with the
most symmetry.  Let us define the \textit{stripes} of the flag as the complements $V_{k_i} \backslash V_{k_{i+1}}$.
Then we see that Model \RN{1} obviously has
$\mathfrak{gl}(k_0-k_1)\oplus\ldots\oplus
\mathfrak{gl}(k_{d-2}-k_{d-1})\oplus \mathfrak{gl}(k_{d-1})$ symmetry.
In particular, on each stripe of the flag, we can transform the basis
vectors into each other by the appropriate $\mathfrak{gl}$
transformation.  For example the first stripe  $V_{k_0} \backslash V_{k_{1}}$ is a $k_0-k_1$ dimensional subspace
and has the corresponding factor of $\mathfrak{gl}(k_0-k_1)$ in the symmetry algebra.

However, the symmetry generators that map between the
different stripes of the flag take on a different form.  This can be
seen by considering the large $u$ limit on $R^\RN{1}$ \eqref{eq:RmodelA}, where it becomes
diagonal but not proportional to the identity operator.  This is 
reminiscent of the appearance of a braiding charge from the AdS/CFT correspondence
\cite{Beisert:2014hya}. So, let us try to emulate the discussion in 
that paper and consider the RTT representation of a Yangian algebra from the $R$-matrix \eqref{eq:RmodelA}.

\iffalse
%
\begin{figure}[h!]
\begin{center}
\includegraphics[scale=.4]{FiguresFlag.pdf}
\caption{The operators $E_{ij}$ map between the different stripes in the flag. Their commutation relations and coproduct depend on the stripes between which they map. For two stripes we recover the usual graded algebra.}\label{GenGradedAlgebra}
\end{center}
\end{figure} 
\fi

In \cite{Beisert:2014hya} the braiding charge appears at the lowest order in the expansion of the RTT algebra. If we do a similar expansion here, however, we find that in contradistinction to a braiding charge, the corresponding element here is not central. Hence, we are led to the introduction of a set 
of elements $\Gamma_{\bar{k}\bar{l}}$ that generalize the notion of a braiding charge but can have non-trivial commutation relations. 
Now, expanding our $R$-matrix further at large $u$, we find the next order to be the standard matrix unities $E_{ij}$. 

Combining these observations, we introduce a new type of Hopf
algebra $\mathcal{A}_\gamma$ which is a general braided version of
$\mathfrak{gl}(n)$ and depends on some constants $\gamma = \pm1$.
This new algebra will contain the symmetry for model \RN{1}, graded
models as well as braided coproducts.

\paragraph{Algebra} Let us now define this new algebra. Consider generators $\Gamma_{\bar{k}\bar{l}}$ and $E_{ij}$ that satisfy the following (anti-)commutation relations
\begin{align}
&E_{ij}E_{kl} - \gamma_{ijkl} E_{kl} E_{ij} = \delta_{jk} E_{il}  - 
\gamma_{ijkl} \delta_{il} E_{kj} \,, \label{alg:EE}\\
&E_{ij}\Gamma_{\bar{k}\bar{l}} - \gamma_{ijkl} 
\Gamma_{\bar{k}\bar{l}} E_{ij} =  0 \,,\label{alg:SE} \\
%&\Gamma_{\bar{i}\bar{j}} E_{kl} - \gamma_{ijkl} E_{kl} 
%\Gamma_{\bar{i}\bar{j}} = 0 \,,\label{alg:SE2} \\
& [\Gamma_{\bar{a}\bar{b}},\Gamma_{\bar{c}\bar{d}}] = 0 \,.\label{alg:SS}
\end{align}
Notice that from \eqref{alg:EE} it follows that $\gamma_{ij,kl} = \gamma_{kl,ij}$.

\paragraph{Coalgebra} We then introduce the following coproduct structure
\begin{align}
\Delta \Gamma_{\bar{a}\bar{b}} &=  \Gamma_{\bar{a}\bar{b}}  \otimes  
\Gamma_{\bar{a}\bar{b}} \,, \\
\Delta E_{ij}  & =  E_{ij} \otimes  1 + \Gamma_{\bar{i}\bar{j}} 
\otimes E_{ij} \,. \label{DeltaE}
\end{align}
This coproduct is easily seen to be coassociative but it only
constitutes an algebra homomorphism for certain cases.  It is
straightforward to check that the coproduct is compatible with
\eqref{alg:SE} and \eqref{alg:SS}.  However, let us now apply the
coproduct to \eqref{alg:EE}.  We find
\begin{align}
\Delta (E_{ij}E_{kl} - \gamma_{ijkl} E_{kl} E_{ij} ) &= \,
(E_{ij} E_{kl}- \gamma_{ijkl} E_{kl}E_{ij}) \otimes 1 + \Gamma_{\bar{i}\bar{j}}\Gamma_{\bar{k}\bar{l}} \otimes (E_{ij} E_{kl}- \gamma_{ijkl} E_{kl}E_{ij})\nonumber \\
&\quad + \big[ E_{ij} \Gamma_{\bar{k}\bar{l}}-\gamma_{ijkl} \Gamma_{\bar{k}\bar{l}} E_{ij} \big] \otimes E_{kl} + 
\big[ \Gamma_{\bar{i}\bar{j}} E_{kl} - \gamma_{ijkl} E_{kl} \Gamma_{\bar{i}\bar{j}} \big] \otimes E_{ij}, \nonumber\\
&=( \delta_{jk} E_{il}  - \gamma_{ijkl} \delta_{il} E_{kj}) \otimes 1 + \Gamma_{\bar{i}\bar{j}}\Gamma_{\bar{k}\bar{l}} \otimes ( \delta_{jk} E_{il}  - \gamma_{ijkl} \delta_{il} E_{kj}) ,\nonumber \\
&= \delta_{jk} \Big[E_{il} \otimes 1 + \Gamma_{\bar{i}\bar{j}}\Gamma_{\bar{k}\bar{l}} \otimes  E_{il}  \Big] 
- \gamma_{ijkl} \delta_{il}\Big[ E_{kj} \otimes 1 + 
\Gamma_{\bar{i}\bar{j}}\Gamma_{\bar{k}\bar{l}} \otimes  E_{kj}  \Big] \,,
\end{align}
where we used that the second line vanishes because of \eqref{alg:SE}.
On the other hand, 
\begin{align}
\Delta( \delta_{jk} E_{il}  - \gamma_{ijkl} \delta_{il} E_{kj} ) &= 
\delta_{jk} ( E_{il} \otimes  1 + \Gamma_{\bar{i}\bar{l}} \otimes 
E_{il} ) - \gamma_{ijkl} \delta_{il} ( E_{kj} \otimes  1 + 
\Gamma_{\bar{k}\bar{j}} \otimes E_{kj} ) \,.
\end{align}
Hence we see that the coproduct defines an algebra homomorphism if and only if
\begin{align}
%& \delta_{il}  \Gamma_{\bar{i}\bar{j}}\Gamma_{\bar{k}\bar{l}}  = 
%\delta_{il}  \Gamma_{\bar{k}\bar{j}} \,,
& \delta_{kj}  \Gamma_{\bar{i}\bar{j}}\Gamma_{\bar{k}\bar{l}}  = 
\delta_{kj}  \Gamma_{\bar{i}\bar{l}} \,. \label{eq:extraS}
\end{align}
This puts additional relations on our braiding functions $\Gamma$ that need to be satisfied for this to define a bialgebra.

\paragraph{Antipode} The antipode $\Sigma$ would satisfy
\begin{align}
& \Sigma(\Gamma_{\bar{i}\bar{j}}) \Gamma_{\bar{i}\bar{j}} = 1 \,,
&& \Sigma(E_{ij}) \Gamma_{\bar{i}\bar{j}} = -E_{ij} \,. \label{antipode}
\end{align}
This means that for any coefficients $i,j$ there should be some $i',j'$ such that $\Gamma_{\bar{i}\bar{j}} \Gamma_{\bar{i}'\bar{j}'}  = 1$. This imposes some further constraints on our generators in order to give a Hopf algebra.

\subsection{Examples}

Let us now give some examples of explicit realizations of our algebra.
\paragraph{Standard Lie algebra}

Setting $\Gamma = 1 = \gamma_{ijkl}$ simply gives us the usual $\mathfrak{gl}(n)$ Lie algebra.

\paragraph{Grading}
Let us consider a flag with two stripes and let us choose $\Gamma$ to
be such that it does not commute with all algebra elements, but is
idempotent $\Gamma^2 = 1$.  Consider a representation of the algebra
elements $E_{ij}$ and introduce a matrix $J$ that acts on the same
space.  We then define
\begin{align}
\Gamma_{\bar a \bar b} = \left\{
 \begin{array}{cc}
 \bar a<\bar b & J \\
 \bar a=\bar b & 1 \\
 \bar a>\bar b & J
 \end{array}\right. \,,
\end{align}
where the matrix $J$ satisfies $J^{2}=1$.
Then the antipode maps $\Gamma$ to itself and \eqref{eq:extraS} is
satisfied as well.  Let us now have a look on how to interpret this
model.  For conciseness, let us restrict to two dimensions 
$(n=2)$.  The
coproduct takes the form
\begin{align}
&\Delta E_{11} = E_{11} \otimes 1 + 1 \otimes E_{11} \,, && \Delta 
E_{22} = E_{22} \otimes 1 + 1 \otimes E_{22} \,, \\
&\Delta E_{12} = E_{12} \otimes 1 + J \otimes E_{12} \,, && \Delta 
E_{21} = E_{21} \otimes 1 + J \otimes E_{21} \,.
\end{align}
This exactly yields the well-known way to implement the graded tensor product using the standard tensor product by interpreting $J$ as the graded identity matrix. So, let us set 
\begin{align}
\gamma_{1212} = \gamma_{1221} = \gamma_{2112} = \gamma_{2121} = -1,
\end{align}
and the other $\gamma$'s equal to 1. Then we precisely recover 
$\mathfrak{gl}(1|1)$ where the coproduct is realized by using the grading matrix $J = \mathrm{diag}(1,-1)$, and we see that all the Hopf algebra relations are indeed satisfied. This straightforwardly generalizes to $\mathfrak{gl}(m|n)$.

\paragraph{AdS/CFT type braiding}
We can make $\Gamma$ central and set $\gamma_{ijkl}=1$.  This
automatically satisfies all the algebra relations
\eqref{alg:EE}-\eqref{alg:SS}.  However, the additional constraints
\eqref{eq:extraS} and \eqref{antipode} put restrictions on our choice
of a braiding factor.  Inspired by the braiding in AdS/CFT, let us
consider the flag with two stripes, so $n=k_0 > k_1$.  Hence the
indices on $\Gamma$ only take the values $1,2$.  Now, let us define
\begin{align}
\Gamma_{\bar a \bar b} = \left\{
 \begin{array}{cc}
 \bar a<\bar b & e^{ip} \\
\bar a= \bar b & 1 \\
 \bar a>\bar b & e^{-ip}
 \end{array}\right. \,.
\end{align}
Then it is easy to check that \eqref{eq:extraS} and \eqref{antipode}
are satisfied assuming that the antipode maps $p\mapsto -p$,
\textit{i.e.} we find that $\Sigma(\Gamma_{ij}) = \Gamma_{ji}$.  We
see that the algebra is undeformed, but that the coproduct is deformed
by a central element usually referred to as a braiding factor.  This
algebra is simply $\mathfrak{gl}(n)$ with a braided coproduct similar
to the one found in the AdS/CFT correspondence \cite{Plefka:2006ze}.

\paragraph{Flag models}
Our flag models \RN{1}, \RN{2} and \RN{3} satisfy a generalization of the graded algebra given above. The braiding elements $\Gamma$ are again not commutative and idempotent $\Gamma^2=1$. However, they take different values between different stripes of the flag. We will work out this algebra in detail in Section \ref{sec:SymModelA} and discuss its properties.

\subsection{Algebra for flag model \RN{1}}\label{sec:SymModelA}

Let us focus here on flag model \RN{1}, whose $R$-matrix has an extended symmetry 
that we denote by $\mathfrak{gl}(k_0-k_1|\ldots|k_{d-2}-k_{d-1}|k_{d-1})$, which we will interpret as a generalized graded algebra. 

We find that for Model \RN{1}, we need to make the choice that if $\bar{i}=\bar{j}$, then $\Gamma_{\bar{i}\bar{j}}=1$. 
If $\bar{i}\ne\bar{j}$, then $\Gamma_{\bar{i}\bar{j}}$ is given by
\begin{equation}
\Gamma_{\bar{i}\bar{j}} = 
\Gamma_{\bar{j}\bar{i}} = \prod_{l={\rm min}(\bar{i},\bar{j})}^{{\rm max}(\bar{i},\bar{j})-1}\Gamma_{k_{l}}\,,
\label{Sigma}
\end{equation}
where 
$\sigma_l$ is the $n \times n$ diagonal matrix defined by
\begin{equation}
\sigma_l = \diag(-1, \ldots, -1_{l}, 1, \ldots, 1_{n}) \,.	
\label{sig}
\end{equation}
Hence we see that just like for the graded algebra, $\Gamma$ takes the form of a diagonal matrix with $\pm1$.

It is easy to check that the $R$-matrix for model \RN{1} \eqref{eq:Rmodel0} has 
$\mathfrak{gl}(k_0-k_1|\ldots|k_{d-2}-k_{d-1}|k_{d-1})$ symmetry
\begin{equation}
\Delta^{op} E_{ij}\, R(u) =  R(u)\, 
\Delta E_{ij}  \,, \qquad i, j \in [1, n] \,,
\label{RAgl}
\end{equation}
where $\Delta E_{ij}$ is given by \eqref{DeltaE} and 
$\Delta^{op} E_{ij}$ is similarly given by
\begin{equation}
\Delta^{op} E_{ij}  =  E_{ij} \otimes  \Gamma_{\bar{i}\bar{j}}  + 
1 \otimes E_{ij} \,. 
\label{DeltaoE}
\end{equation}

We can determine the constants $\gamma_{ijkl}$ from the $\Gamma$ 
matrices: 
multiplying \eqref{alg:SE} on the right by $E_{ji}$, we obtain
\begin{equation}
E_{ij}\Gamma_{\bar{k}\bar{l}} E_{ji} = \gamma_{ijkl} 
\Gamma_{\bar{k}\bar{l}} E_{ii} \,,
\end{equation}
where there is no summation over repeated indices.
Since the $\Gamma$ matrices are diagonal, we see that
\begin{equation}
\left(\Gamma_{\bar{k}\bar{l}}\right)_{jj} E_{ii} = \gamma_{ijkl} 
\left(\Gamma_{\bar{k}\bar{l}}\right)_{ii} E_{ii} \,,
\end{equation}
which implies
\begin{equation}\label{eq:gamma}
\gamma_{ijkl} = \left(\Gamma_{\bar{k}\bar{l}}\right)_{jj}/\left(\Gamma_{\bar{k}\bar{l}}\right)_{ii} \,. 
\end{equation}
Hence we also find that in this case $\gamma=\pm1$, meaning that we are dealing with a mixture of commutation relations and anti-commutation relations.

The easiest way to see that this model is not just a usual graded
algebra in disguise is the fact that $\Gamma$ appearing in the
coproducts will be different depending on the operator.  For usual
superalgebras, all even and odd generators share the same braiding
factor.  As an example, let us work out the case $\vec{k} =
\{3,2,1\}$.  This is the first non-trivial example since it
corresponds to a flag with 3 stripes.  The diagonal operators $E_{ii}$
have the standard coproduct
\begin{align}
\Delta E_{ii} = E_{ii} \otimes 1 + 1 \otimes E_{ii}.
\end{align}
Then there are three other possibilities $E_{12},E_{13},E_{23}$, 
which are the operators that relate basis vectors belonging to the
different stripes in the flag.  This corresponds to the algebra
$\mathfrak{gl}(1|1|1)$.  The elements $E_{21},E_{31},E_{32}$ are
simply related by transposition, which also shows that $\Gamma_{ij} =
\Gamma_{ji}$.

The easiest way to represent this algebra is by taking $E_{ij}$ to be
the standard matrix unities; from \eqref{eq:extraS} it is then easy
to see that $\Gamma_{12}\Gamma_{13} = \Gamma_{23}$, and we find
\begin{align}
&\Gamma_{12} = \begin{pmatrix}
-1 & 0 & 0 \\
0 & 1 & 0 \\
0 & 0 & 1 \\
\end{pmatrix}, 
&&\Gamma_{13} = \begin{pmatrix}
1 & 0 & 0 \\
0 & -1 & 0 \\
0 & 0 & 1 \\
\end{pmatrix}.
\end{align}
From this we can compute $\gamma$ from \eqref{eq:gamma}, and we can nicely package $\gamma$ in a table
\begin{align}
\begin{array}{c|ccccccccc}
 \gamma & E_{11} & E_{21} & E_{31} & E_{12} & E_{22} & E_{32} & E_{13} & E_{23} & E_{33} \\\hline
 E_{11} & 1 & 1 & 1 & 1 & 1 & 1 & 1 & 1 & 1 \\
 E_{21} &1 & -1 & -1 & -1 & 1 & 1 & -1 & 1 & 1 \\
 E_{31} & 1 & -1 & 1 & -1 & 1 & -1 & 1 & -1 & 1 \\
 E_{12} & 1 & -1 & -1 & -1 & 1 & 1 & -1 & 1 & 1 \\
 E_{22} & 1 & 1 & 1 & 1 & 1 & 1 & 1 & 1 & 1 \\
 E_{32} & 1 & 1 & -1 & 1 & 1 & -1 & -1 & -1 & 1 \\
 E_{13} & 1 & -1 & 1 & -1 & 1 & -1 & 1 & -1 & 1 \\
 E_{23} & 1 & 1 & -1 & 1 & 1 & -1 & -1 & -1 & 1 \\
 E_{33} & 1 & 1 & 1 & 1 & 1 & 1 & 1 & 1 & 1 \\
\end{array}
\end{align}
Let us now have a look at the commutation relations.  We see that
$E_{12}$ and $E_{23}$ satisfy anti-commutations relation with itself
since $\gamma_{1212} = \gamma_{2323} =-1$.  Hence, these behave as 
odd generators.  However, between each other they satisfy a usual
commutation relation since $\gamma_{1223}=1$.  On the other hand,
$E_{13}$ and $E_{31}$ seem to be even generators ($\gamma_{1331}=1$),
but satisfy \textit{anti}-commutation relations with $E_{12}$ and
$E_{23}$.

We conclude that we are left with a generalized graded algebra which
is characterized by the number of stripes in the flag.  Given the fact
that model \RN{1} is unique, we see that this is the unique extension
of a graded-type algebra that includes multiple types of generators.  A
generator will satisfy either commutation or anti-commutation relations
depending on which stripes it relates.

\subsection{Generalized graded Yangians}\label{sec:GGY}

There is a natural way to extend our algebra to a generalized graded Yangian. Consider the level-1 Yangian generators $\hat{E}_{ij}$ such that the following commutation relations hold
\begin{align}
&E_{ij}\hat{E}_{kl} - \gamma_{ijkl} \hat{E}_{kl} E_{ij} = \delta_{jk} \hat{E}_{il}  - \gamma_{ijkl} \delta_{il} \hat{E}_{kj}, \label{alg:EEb}\\
&\hat E_{ij}{E}_{kl} - \gamma_{ijkl} {E}_{kl} \hat E_{ij} = \delta_{jk} \hat{E}_{il}  - \gamma_{ijkl} \delta_{il} \hat{E}_{kj}, \label{alg:EEb}\\
&\hat{E}_{ij}\Gamma_{\bar{k}\bar{l}} - \gamma_{ijkl} \Gamma_{\bar{k}\bar{l}} \hat{E}_{ij} =  0,\label{alg:SEb} \\
&\Gamma_{\bar{i}\bar{j}} \hat{E}_{kl} - \gamma_{ijkl} \hat{E}_{kl} \Gamma_{\bar{i}\bar{j}} = 0,\label{alg:SE2b}.
\end{align}
We then introduce the standard Yangian-type coproduct 
\begin{align}
\Delta \hat{E}_{ij} = \hat{E}_{ij} \otimes 1 + \Gamma_{\bar i\bar j}
\otimes \hat{E}_{ij} - \frac{\eta}{2} \sum_{k} \Big[ E_{kj}\otimes \Gamma_{\bar k\bar j} E_{ik}  - \gamma_{ikjk}E_{ik}\otimes \Gamma_{\bar i\bar k} E_{kj}
\Big].
\label{yangian}
\end{align}
For this to define a proper Hopf algebra, we must in principle impose additional 
restrictions on $\Gamma,\gamma$. However, we can check that for our 
generalized graded algebra  $\mathfrak{gl}(k_0-k_1|\ldots|k_{d-2}-k_{d-1}|k_{d-1})$ for Model \RN{1} everything is compatible. Hence, if we consider the evaluation representation $\hat{E} = u E$ and the corresponding coproduct
\begin{align}
\Delta\hat{E}_{ij} = u_1 E_{ij} \otimes 1 + \Gamma_{\bar i\bar j} \otimes  u_2 E _{ij}  - \frac{\eta}{2} \sum_{k} \Big[ E_{kj}\otimes \Gamma_{\bar k\bar j} E_{ik}  - \gamma_{ikjk}E_{ik}\otimes \Gamma_{\bar i\bar k} E_{kj}  \Big],
\end{align}
we find that it is a symmetry of the $R$-matrix of Model \RN{1}. In 
fact, we find that the $R$-matrix of Model \RN{1} is \textit{completely fixed} by its generalized Yangian symmetry.

\subsection{Symmetries for model \RN{2}}

Let us discuss the symmetries of the $R$-matrix for model \RN{2}. As 
is clear from the form of the Hamiltonian and $R$-matrix, there is a 
large overlap with model \RN{1}. Because of this, there is also a large overlap in symmetries. Let $e_i$ be the basis vectors of $V$, then the $R$-matrices of models \RN{2} and \RN{1} have the same action on $e_i\otimes e_j$ where ${i,j}>k_1$. Hence, we find that the model exhibits a $\mathfrak{gl}(k_1-k_2|\ldots|k_{d-2}-k_{d-1}|k_{d-1})$ symmetry as well as the manifest $\mathfrak{gl}(k_0-k_1)$ that acts on the first indices. Moreover, also the Yangian generators $\Delta \hat{E}_{ij}$ are a symmetry for ${i,j}>k_1$. However, this is clearly not enough to fully fix the $R$-matrix.

Model \RN{2} exhibits some additional discrete symmetries. First, models 
\RN{1} and \RN{2} are both invariant under parity. Second we have that
\begin{align}
&[R^{\RN{2}-} ,  E_{\bar{1}\bar{2}} \otimes E_{\bar{1}\bar{2}} ]=0,
&&[R^{\RN{2}+} ,  E_{\bar{1}\bar{3}}\otimes E_{\bar{1}\bar{3}} ]=0.
\end{align}
Unfortunately, this is still not enough symmetry to fix the $R$-matrix. We have not been able to identify a remaining (discrete) symmetry that fully fixes the model.

\section{Bethe ansatz for model \RN{2}}\label{sec:BAmodelB}

We now analyze model \RN{2} using nested algebraic Bethe ansatz (see 
e.g. \cite{Maassarani:1998, Maassarani:1997kon, Kagan:2008,
Kulish:1979cr, Kulish:1983rd, Babelon:1981un, Kulish:1985, Belliard:2008di, 
Wang2015, Levkovich-Maslyuk:2016kfv} and references therein), restricting to $k_{d-1}>1$.

\subsection{First level of nesting}\label{sec:Bfirstlevel}

If we try to perform the nested Bethe ansatz procedure for model \RN{2}
with the R-matrix as written in Eq. \eqref{eq:Rmodel1}, we obtain
exchange relations that are not useful.  A very simple local basis
transformation solves this problem.  We therefore use instead
\begin{equation}
\tilde{R}^{\RN{2}\pm,\vec{k}}(u)=\left(V\otimes 
V\right) R^{\RN{2}\pm,\vec{k}}(u) \left(V^{-1}\otimes V^{-1}\right) \,,
\label{eq:newRmodelB}
\end{equation}
where the $ n\times n $ matrix $V$ is defined in \eqref{Vmat}.  
This is exactly the same model as before because local basis
transformations do not change the spectrum.

We can write the monodromy matrix for a chain of length $L$ as
\begin{align}
T_{0}(u;\{\theta_j\})&=\tilde{R}^{\RN{2}\pm, \vec 
k}_{01}(u-\theta_1)\, \tilde{R}^{\RN{2}\pm, \vec 
k}_{02}(u-\theta_2)\,\ldots\, \tilde{R}^{\RN{2}\pm, \vec k}_{0L}(u-\theta_L)\\
&=\begin{pmatrix}
\mathcal{T}_{0,0}(u;\{\theta_j\}) & \mathcal{B}_1(u;\{\theta_j\})   & \mathcal{B}_2(u;\{\theta_j\})     & \cdots & \mathcal{B}_{n-1}(u;\{\theta_j\})\\
\mathcal{C}_1(u;\{\theta_j\})    & \mathcal{T}_{1,1}(u;\{\theta_j\}) & \mathcal{T}_{1,2}(u;\{\theta_j\})  & \cdots & \mathcal{T}_{1,n-1}(u;\{\theta_j\})\\
\mathcal{C}_2(u;\{\theta_j\})    & \mathcal{T}_{2,1}(u;\{\theta_j\}) & \mathcal{T}_{2,2}(u;\{\theta_j\})  & \cdots & \mathcal{T}_{2,n-1}(u;\{\theta_j\})\\
\vdots    & \vdots     & \vdots      & \ddots    & \vdots   \\
\mathcal{C}_{n-1}(u;\{\theta_j\})    & \mathcal{T}_{n-1,1}(u;\{\theta_j\}) & \mathcal{T}_{n-1,2}(u;\{\theta_j\}) & \cdots & \mathcal{T}_{n-1,n-1}(u;\{\theta_j\})
\end{pmatrix} \,,
\label{eq:monodromy}
\end{align}
where $ \{\theta_j \} $ are the inhomogeneities, and we suppress the 
superscripts $\RN{2}\pm, \vec k$ on the monodromy matrix to lighten the notation.
The transfer matrix is therefore given by 
\begin{equation}
t(u;\{\theta_j \})= \tr_{0} T_{0}(u;\{\theta_j\}) = 
\mathcal{T}_{0,0}(u;\{\theta_j \})+\sum_{\alpha=1}^{n-1}\mathcal{T}_{\alpha,\alpha}(u;\{\theta_j \}).
\label{eq:transfer}
\end{equation}
For a reference state such as 
\begin{equation}
|0\rangle = \begin{pmatrix}
1\\
0\\
\vdots\\
0\\
\end{pmatrix}^{\otimes L} \,,
\label{allup}
\end{equation}
we can see that 
\begin{align}
&\mathcal{C}_{\alpha}(u;\{\theta_j\})|0\rangle=0 \quad \forall \, 
\alpha=1,\ldots,n-1\,, \label{eq:Cinvac}\\
&\mathcal{T}_{00}(u;\{\theta_j\})|0\rangle 
=\prod_{j=1}^{L}\left(\eta + u-\theta_j\right)\left(1+u-\theta_j\right)|0\rangle \,,\label{eq:T00invac}\\
&\mathcal{T}_{\alpha\beta}(u;\{\theta_j\})|0\rangle=\delta_{\alpha\beta}\, \prod_{j=1}^{L}(u-\theta_j)(1+u-\theta_j)|0\rangle  \quad \forall \,
\alpha\,, \beta =1,\ldots,n-1 \,.\label{eq:Tabinvac}
\end{align}
The operators $\mathcal{B}_{\alpha}(u;\{\theta_j \})$ act as creation operators. So, we can use them to define excited states
\begin{equation}
|\psi\rangle = \sum_{\{a_{1}, \ldots, 
a_{m}\}}\prod_{i=1}^{m}\mathcal{B}_{a_i}(u_i;\{\theta_j\})F^{a_1,\cdots, a_m}|0\rangle
\label{eq:psi}
\end{equation} 
where $\{a_{i}\}$ can assume values from $1$ to $n-1$, and $\{u_i\}$ are
the Bethe roots.  By continuing the Bethe ansatz procedure we will
obtain the conditions that the Bethe roots must satisfy in order for 
$|\psi\rangle$ to be an eigenvector of the transfer matrix $t(u;\{\theta_j \})$.

We have seen that the transfer matrix is given by Eq. \eqref{eq:transfer},
and we know how $\mathcal{T}_{i,j}(u,\{\theta_j\})$ acts on the
reference state.  When acting with $t(u;\{\theta_j \})$ on 
$|\psi\rangle$, we need a way to
pass through all the $\mathcal{B}_{a_i}(u_1;\{\theta_j\})$ operators.
The exchange relations which allow us to do that are obtained from the RTT relation
\begin{equation}
\tilde{R}_{ab}^{\RN{2}\pm,\vec{k}}(u-v)\, T_a(u;\{\theta_j\})\, 
T_b(v;\{\theta_j\})=T_b(v;\{\theta_j\})\, T_a(u;\{\theta_j\})\, \tilde{R}_{ab}^{\RN{2}\pm,\vec{k}}(u-v) \,.
\label{eq:RTT}
\end{equation}
By substituting $\tilde{R}_{ab}^{\RN{2}\pm,\vec{k}}(u)$ from 
\eqref{eq:newRmodelB} and $T(u;\{\theta_j\})$ as in \eqref{eq:monodromy}, we obtain several exchange relations. The useful ones are
\begin{align}
\mathcal{T}_{0,0}(v;\{\theta_j\})\, 
\mathcal{B}_{\alpha}(u;\{\theta_j\})&=\frac{\eta+u-v}{u-v}\mathcal{B}_{\alpha}(u;\{\theta_j\})\,\mathcal{T}_{0,0}(v;\{\theta_j\})\nonumber\\
&\hspace{0.3cm}-\frac{\eta}{u-v}\mathcal{B}_{\alpha}(v;\{\theta_j\})\, \mathcal{T}_{0,0}(u;\{\theta_j\}) \,,
\label{eq:commutT00}
\end{align}
where $ \alpha=1,\ldots,n-1 $; and
\begin{align}
\mathcal{T}_{\alpha,\beta}(v;\{\theta_j\})\, \mathcal{B}_{\gamma}(u;\{\theta_j\})
&=\sum_{\tau, \eta}f(u-v)\mathbb{R}_{\beta,\gamma}^{\tau,\eta}(u-v)
\mathcal{B}_\eta(u;\{\theta_j\})\, \mathcal{T}_{\alpha,\tau}(v;\{\theta_j\})\nonumber\\
&\hspace{0.3cm}+g(u-v)\mathcal{B}_\beta(v;\{\theta_j\})\, \mathcal{T}_{\alpha,\gamma}(u;\{\theta_j\}) \,,
\label{eq:commutTab}
\end{align}
where $f(u)$, $g(u)$ and $\mathbb{R}_{\beta,\gamma}^{\tau,\eta}(u)$ 
depend on the model, see \eqref{eq:mapforB+BA}-\eqref{gforB-BA} below. 

Let us see how $\mathcal{T}_{0,0}(u;\{\theta_j \})$ acts on 
$|\psi\rangle$:
\begin{align}
\mathcal{T}_{0,0}(u;\{\theta_j \})|\psi\rangle 
&=\sum_{\{a\}}\mathcal{T}_{0,0}(u;\{\theta_j \}) \prod_{i=1}^{m_1}\mathcal{B}_{a_i}(u_i;\{\theta_j\})F^{a_1,\cdots, a_{m_1}}|0\rangle\label{eq:T00eq1}\\
&=\sum_{\{a\}}\mathcal{T}_{0,0}(u;\{\theta_j 
\})\mathcal{B}_{a_1}(u_1;\{\theta_j\})\prod_{i=2}^{m_1}\mathcal{B}_{a_i}(u_i;\{\theta_j\})F^{a_1, \cdots, a_{m_1}}|0\rangle,\label{eq:T00eq2}\\
&=\sum_{\{a\}}\frac{\eta+u_1-u}{u_1-u}\mathcal{B}_{a_1}(u_1;\{\theta_j\})\mathcal{T}_{0,0}(u;\{\theta_j \})\prod_{i=2}^{m_1}\mathcal{B}_{a_i}(u_i;\{\theta_j\})F^{a_1,\cdots, a_{m_1}}|0\rangle\nonumber\\
&\hspace{0.5cm}-\frac{\eta}{u_1-u}\sum_{\{a\}}\mathcal{B}_{a_1}(u;\{\theta_j\})\mathcal{T}_{0,0}(u_1;\{\theta_j \})\prod_{i=2}^{m_1}\mathcal{B}_{a_i}(u_i;\{\theta_j\})F^{a_1,\cdots, a_{m_1}}|0\rangle\label{eq:T00eq3}\\
&=\prod_{j=1}^{L}\left(\eta+u-\theta_j\right)\left(1+u-\theta_j\right)\prod_{i=1}^{m_1}\frac{\eta+u_i-u}{u_i-u}|\psi\rangle \nonumber\\
&\hspace{1cm}+\text{unwanted terms}.\label{eq:T00eq4}
\end{align}
In passing from \eqref{eq:T00eq2} to \eqref{eq:T00eq3}, we use once
\eqref{eq:commutT00}.  We see that the second term depends on 
$\mathcal{B}_{a_1}(u;\{\theta_j\})$, so it cannot be written in terms 
of $|\psi\rangle$.  As we continue to use the exchange relations to pass
$\mathcal{T}_{0,0}$ through all the $\mathcal{B}$'s, we will
get more and more such terms, called ``unwanted 
terms,'' which we ignore for now.  In passing from 
\eqref{eq:T00eq3} to \eqref{eq:T00eq4}, we just continue to use the
exchange relations; and when $\mathcal{T}_{0,0}$ hits $|0\rangle$,
we use \eqref{eq:T00invac}. 

Let us now see how $\mathcal{T}_{\alpha,\alpha}((u;\{\theta_j\})$ acts on 
$|\psi\rangle$:
\begin{align}
\mathcal{T}_{\alpha,\alpha}&(u;\{\theta_j\})|\psi\rangle 
=\sum_{\{a\}}\mathcal{T}_{\alpha,\alpha}(u;\{\theta_j 
\})\prod_{i=1}^{m_1}\mathcal{B}_{a_i}(u_i;\{\theta_j\})F^{a_1,\cdots, 
a_{m_1}}|0\rangle \,, \label{eq:Taaeq1}\\
&=\sum_{\{a\}}\mathcal{T}_{\alpha,\alpha}(u;\{\theta_j 
\})\mathcal{B}_{a_1}(u_1;\{\theta_j\})\prod_{i=2}^{m_1}\mathcal{B}_{a_i}(u_i;\{\theta_j\})F^{a_1,\cdots, a_{m_1}}|0\rangle \,, \label{eq:Taaeq2}\\
&=\sum_{\{a\}}\sum_{\tau_{1}, 
b_{1}}f(u_1-u)\mathbb{R}_{\alpha,a_1}^{\tau_1,b_1}(u_1-u)\mathcal{B}_{b_1}(u_1;\{\theta_j\})\mathcal{T}_{\alpha,\tau_1}(u;\{\theta_j \})\prod_{i=2}^{m_1}\mathcal{B}_{a_i}(u_i;\{\theta_j\})F^{a_1,\cdots, a_{m_1}}|0\rangle \nonumber\\
&\hspace{0.3cm}+\sum_{\{a\}}g(u_1-u)\mathcal{B}_{b_1}(u;\{\theta_j\})\mathcal{T}_{\alpha,a_1}(u_1;\{\theta_j \})\prod_{i=2}^{m_1}\mathcal{B}_{a_i}(u_i;\{\theta_j\})F^{a_1,\cdots, a_{m_1}}|0\rangle\,, \label{eq:Taaeq3}\\
&=f(u_1-u)f(u_2-u)\sum_{\{a\}}\sum_{\tau_{1}, \tau_{2}}\sum_{b_{1}, b_{2}}\mathbb{R}_{\alpha,a_1}^{\tau_1,b_1}(u_1-u)\mathbb{R}_{\tau_1,a_2}^{\tau_2,b_2}(u_2-u)\mathcal{B}_{b_1}(u_1;\{\theta_j\})\mathcal{B}_{b_2}(u_2;\{\theta_j\}) \nonumber\\
&\hspace{0.8cm}\times \mathcal{T}_{\alpha,\tau_2}(u;\{\theta_j 
\})\prod_{i=3}^{m_1}\mathcal{B}_{a_i}(u_i;\{\theta_j\})F^{a_1,\cdots, 
a_{m_1}}|0\rangle+\text{unwanted terms}\,, \label{eq:Taaeq4}
\end{align}
\begin{align}
&=\prod_{i=1}^{m_1}f(u_i-u)\sum_{\{b\}}\prod_{l=1}^{m_1}\mathcal{B}_{b_l}(u_l;\{\theta_j\})\sum_{\{a\}, \{\tau\}}\mathbb{R}_{\alpha,a_1}^{\tau_1,b_1}(u_1-u)\mathbb{R}_{\tau_1,a_2}^{\tau_2,b_2}(u_2-u)\cdots \mathbb{R}_{\tau_{m_1-1},a_{m_1}}^{\tau_{m_1},b_{m_1}}(u_{m_1}-u) \nonumber\\
&\hspace{0.8cm}\times F^{a_1,\cdots, 
a_{m_1}}\mathcal{T}_{\alpha,\tau_{m_1}}(u;\{\theta_j 
\})|0\rangle+\text{unwanted terms}\,, \label{eq:Taaeq5}\\
&=\prod_{i=1}^{m_1}f(u_i-u)\sum_{\{b\}}\prod_{l=1}^{m_1}\mathcal{B}_{b_l}(u_l;\{\theta_j\})\sum_{\{a\}, \{\tau\}}\mathbb{R}_{\alpha,a_1}^{\tau_1,b_1}(u_1-u)\mathbb{R}_{\tau_1,a_2}^{\tau_2,b_2}(u_2-u)\cdots \mathbb{R}_{\tau_{m_1-1},a_{m_1}}^{\alpha,b_{m_1}}(u_{m_1}-u) \nonumber\\
&\hspace{0.8cm}\times F^{a_1,\cdots, 
a_{m_1}}\prod_{j=1}^{L} (u-\theta_j)(1+u-\theta_j)|0\rangle+\text{unwanted terms}\,. \label{eq:Taaeq6}
\end{align}

We conclude that the action of the transfer matrix 
\eqref{eq:transfer} on $|\psi\rangle$ \eqref{eq:psi} is given by 
\begin{align}
&t(u,\{\theta_j\})|\psi\rangle=\mathcal{T}_{00}(u,\{\theta_j\})|\psi\rangle+\sum_{\alpha=1}^{n-1}\mathcal{T}_{\alpha\alpha}(u,\{\theta_j\})|\psi\rangle \nonumber\\
&=\prod_{j=1}^{L}\left(\eta+u-\theta_j\right)\left(1+u-\theta_j\right)\prod_{i=1}^{m_1}\frac{\eta+u_i-u}{u_i-u}|\psi\rangle \nonumber\\
&+\sum_{\alpha=1}^{n-1}\left[\prod_{i=1}^{m_1}f(u_i-u)\sum_{\{b\}}\prod_{l=1}^{m_1}\mathcal{B}_{b_l}(u_l;\{\theta_j\})\sum_{\{a\}, \{\tau\}}\mathbb{R}_{\alpha,a_1}^{\tau_1,b_1}(u_1-u)\mathbb{R}_{\tau_1,a_2}^{\tau_2,b_2}(u_2-u)\cdots \mathbb{R}_{\tau_{m_1-1},a_{m_1}}^{\alpha,b_{m_1}}(u_{m_1}-u)\right.\nonumber\\
&\left. \quad\times\, 
F^{a_1\,\cdots\,a_{m_1}}\prod_{j=1}^{L} (u-\theta_j)(1+u-\theta_j)\right]|0\rangle +\text{unwanted terms}\,. 
\end{align}
If $|\psi\rangle$ is an eigenvector of $t(u,\{\theta_j\})$ so that the 
unwanted terms vanish, then the corresponding eigenvalue is given by
\begin{align}
\Lambda(u,\{\theta_j\})
&=\prod_{j=1}^{L}\left(\eta+u-\theta_j\right)\left(1+u-\theta_j\right)\prod_{i=1}^{m_1}\frac{\eta+u_i-u}{u_i-u} \nonumber\\
&+\begin{cases}
(n-1) \prod_{j=1}^{L} (u-\theta_j)(1+u-\theta_j) & m_1=0 \\
\Lambda_{\text{aux}}(u)\prod_{j=1}^{L} 
(u-\theta_j)(1+u-\theta_j)\prod_{i=1}^{m_1}f(u_i-u) & m_1\ge 1 
\end{cases} \,,
\label{nestedresult}
\end{align} 
where $\Lambda_{\text{aux}}(u)$ is an eigenvalue of the auxiliary 
transfer matrix defined by
\begin{align}
t_{\text{aux}}(u) &= \sum_{\alpha=1}^{n-1} \sum_{\{\tau\}} 
\mathbb{R}_{\alpha,a_1}^{\tau_1,b_1}(u_1-u)\mathbb{R}_{\tau_1,a_2}^{\tau_2,b_2}(u_2-u)\cdots \mathbb{R}_{\tau_{m_1-1},a_{m_1}}^{\alpha,b_{m_1}}(u_{m_1}-u) \nonumber\\
&= \Big[ \tr_{0} \mathbb{R}_{01}(u_1-u)\, \mathbb{R}_{02}(u_2-u)\, 
\ldots \mathbb{R}_{0 m_1}(u_{m_1}-u) \Big]^{b_1 b_2 \ldots b_{m_1}}_{a_1 a_2 \ldots a_{m_1}} \,.
\label{taux}
\end{align}

Starting with model $\RN{2}^+$ with a local Hilbert space of dimension $n$ 
(the R-matrix is $n^2\times n^2$), the corresponding 
$\mathbb{R}(u)$ in \eqref{taux} is $(n-1)^2\times (n-1)^2$ and is given by
\begin{equation}
\mathbb{R}(u)=\begin{cases}
\mathbb{P}^{(n-1,n-1)} & \text{for initial model with } k_1=k_0-1\text{ and } d=2\\
\frac{1}{1-u}\frac{1}{\eta-u}\tilde{R}^{(\RN{2}+,k_0-1,k_1,..,k_{d-1})}(-u)& \text{for initial model with } k_1<k_0-1\text{ and } d\ge2\\
\frac{1}{1+u}\frac{1}{\eta+u}\tilde{R}^{(\RN{2}-,k_1,..,k_{d-1})}(u)& \text{for initial model with  } k_1=k_0-1 \text{ and } d>2\\
\end{cases}.
\label{eq:mapforB+BA}
\end{equation}
Also, 
\begin{equation}
f(u)=\begin{cases}
-\frac{\eta-u}{u} & \text{for all the cases with auxiliary problem given by }\tilde{R}^+(-u) \text{ or } \mathbb{P}^{(n-1,n-1)}\\
-\frac{\eta+u}{u} & \text{for all the cases with auxiliary problem given by }\tilde{R}^-(u)
\end{cases}
\end{equation}
while 
\begin{equation}
g(u)=\frac{\eta}{u},
\end{equation}
for all cases.
In particular, starting with the R-matrix for model $\RN{2}^+$, we are led
to an auxiliary problem that can be either related to $\RN{2}^+$ or to 
$\RN{2}^-$ depending on the values of $d$ and $\vec k$ according to
Eq. \eqref{eq:mapforB+BA}. 

Similarly, starting with model $\RN{2}^-$, the $\mathbb{R}(u)$ in 
\eqref{taux} is given by
\begin{equation}
\mathbb{R}(u)=\begin{cases}
\mathbb{P}^{(n-1,n-1)} & \text{for initial model with } k_1=k_0-1\text{ and } d=2\\
\frac{1}{1-u}\frac{1}{\eta-u}\tilde{R}^{(\RN{2}-,k_0-1,k_1,..,k_{d-1})}(-u)& \text{for initial model with } k_1<k_0-1 \text{ and } d\ge 2\\
\frac{1}{1+u}\frac{1}{\eta+u}\tilde{R}^{(\RN{2}+,k_1,..,k_{d-1})}(u)& \text{for initial model with  } k_1=k_0-1 \text{ and } d>2.\\
\end{cases}.
\label{eq:mapforB-BA}
\end{equation}
Also, 
\begin{equation}
f(u)=\begin{cases}
-\frac{\eta-u}{u} & \text{for all the cases with auxiliary problem given by }\tilde{R}^-(-u)\\
-\frac{\eta+u}{u} & \text{for all the cases with auxiliary problem given by }\tilde{R}^+(u) \text{ or } \mathbb{P}^{(n-1,n-1)}
\end{cases}
\end{equation}
while 
\begin{equation}
g(u)=\frac{\eta}{u},
\label{gforB-BA}
\end{equation}
for all cases.

\subsection{Transfer-matrix eigenvalues}\label{sec:BEandTQ}

We now proceed to determine the transfer-matrix eigenvalues and 
Bethe equations of Model $\RN{2}$. To this end, it is useful to 
introduce some further notations. Starting from the R-matrix
$\tilde{R}^{\RN{2}\pm,\vec{k}}(u)$ \eqref{eq:newRmodelB}, where $\vec k$ 
is the vector $\vec k = \{k_{0}, k_{1}, \ldots, k_{d-1}\}$ with 
dimension $|\vec k| := d$,
we define a sequence of R-matrices
\begin{equation}
	\tilde{R}^{(l)}(u) \equiv \tilde{R}^{\mu_{l}, \vec{k}^{(l)}}(u)\,, \qquad 
	l = 0, 1, \ldots\,,
\end{equation}
where $\tilde{R}^{(0)}(u) = \tilde{R}^{\RN{2}\pm,\vec{k}}(u)$, with 
$\mu_{0} = \pm 1$ for $\RN{2}\pm$, respectively; and $\vec{k}^{(0)} = 
\vec{k}$. Moreover, the vectors $\vec{k}^{(l)}$, as well 
as the parameters $\mu_{l}$, $\gamma_{l}$ and $\delta_{l}$, are 
defined for $l\ge 1$ recursively as follows:
\begin{align}
&\text{If   } k_{1}^{(l-1)} < k_{0}^{(l-1)}-1\,, \quad \text{then  }	
\vec{k}^{(l)} = \vec{k}^{(l-1)} - \vec \epsilon\,, \quad \mu_{l} = 
\mu_{l-1}\,, \quad \gamma_{l}=\delta_{l}=1 \,; \non \\
&\text{if   } k_{1}^{(l-1)} = k_{0}^{(l-1)}-1 \text{ and } |\vec 
k^{(l-1)}|>2 \,, \quad \text{then  }
\vec{k}^{(l)} = \hat{\vec{k}}^{(l-1)}\,, \quad \mu_{l} = 
-\mu_{l-1}\,, \quad \gamma_{l}=\delta_{l}=-1 \,; \non \\
&\text{if   } k_{1}^{(l-1)} = k_{0}^{(l-1)}-1 \text{ and } |\vec 
k^{(l-1)}|=2 \,, \quad \text{then  } \mu_{l} = 
-\mu_{l-1}\,, \quad \gamma_{l}=-1\,, \quad \delta_{l}=\mu_{l-1} \,,
\label{iteration}
\end{align}
where $l=1, 2, \ldots$, and $\gamma_{0} = \delta_{0} =1$.
In the first line of \eqref{iteration}, $\vec \epsilon$ is the vector $\vec 
\epsilon=\{1,0,\ldots,0\}$ that has the same dimension as 
$\vec{k}^{(l-1)}$, i.e. $|\vec \epsilon\,| = 
|\vec{k}^{(l-1)}|$. In the second line, the hat denotes dropping the 
first (left-most) component; hence, since  $\vec{k}^{(l-1)} = \{k^{(l-1)}_{0}\,, 
k^{(l-1)}_{1}\,, \ldots \}$, then $\hat{\vec{k}}^{(l-1)} = \{ 
k^{(l-1)}_{1}\,, \ldots \}$.

The sequence $\vec{k}^{(0)}, \vec{k}^{(1)}, \ldots$ terminates with
\begin{equation}
\vec{k}^{(l)} = \{ k_{d-1}+1\,, k_{d-1} \} \qquad \mbox{where   } 
l=k_{0}-k_{d-1}-1 \,.
\label{terminus}
\end{equation}
Indeed, it follows from \eqref{iteration} that 
$\vec{k}^{(k_{0}-k_{j})}= \{k_{j}\,, k_{j+1}\,, \ldots \,, 
k_{d-1}\}$ with $j=0, 1, \ldots, d-2$. Hence, $\vec{k}^{(k_{0}-k_{d-2})}= \{k_{d-2}\,, 
k_{d-1}\}$, and therefore 
\begin{align}
\vec{k}^{(k_{0}-k_{d-1}-1)}	
&= \vec{k}^{(k_{0}-k_{d-2}+(k_{d-2}-k_{d-1}-1))}
= \{k_{d-2}-(k_{d-2}-k_{d-1}-1)\,, 
k_{d-1}\} \non\\
&= \{ k_{d-1}+1\,, k_{d-1} \} \,.
\end{align}
Examples of such sequences of $\vec{k}^{(l)}$ and $\mu_{l}$ are shown in 
Table \ref{modelB+n=5}.

\begin{table}[h!]
	\setlength{\tabcolsep}{0.5em} % for the horizontal padding
	{\renewcommand{\arraystretch}{0.3}
		{\renewcommand{\arraystretch}{0.5}
			\begin{center}
				\begin{tabular}{c|ccc|ccc|c}
					\multicolumn{8}{c}{\textbf{$ \hspace{0.5cm} $}}\\
					\multicolumn{8}{c}{\textbf{Model \RN{2}$ \mathbf{^+} $  $ (\mathbf{n=5}) $}}\\
					\multicolumn{8}{c}{\textbf{$ \hspace{0.5cm} $}}\\
					\hline
					$ \hspace{0.5cm} $  &$ \hspace{0.5cm} $  & $ \hspace{0.5cm} $ & $ \hspace{0.5cm} $ & $ \hspace{0.5cm} $& $ \hspace{0.5cm} $& $ \hspace{0.5cm} $ & $ \hspace{0.5cm} $\\
					$ \hspace{0.5cm}$ & $ \hspace{0.5cm}$ & $ \mathbf{d=2}$ & $ \hspace{0.5cm}$ & $ \hspace{0.5cm}$ & $ \mathbf{d=3}$ & $ \hspace{0.5cm}$ & $ \mathbf{d=4}$\\
					$ \hspace{0.5cm} $  &$ \hspace{0.5cm} $  & $ \hspace{0.5cm} $ & $ \hspace{0.5cm} $ & $ \hspace{0.5cm} $& $ \hspace{0.5cm} $& $ \hspace{0.5cm} $ & $ \hspace{0.5cm} $\\
					\hline
					$ \hspace{0.5cm} $  &$ \hspace{0.5cm} $  & $ \hspace{0.5cm} $ & $ \hspace{0.5cm} $ & $ \hspace{0.5cm} $& $ \hspace{0.5cm} $& $ \hspace{0.5cm} $ & $ \hspace{0.5cm} $\\
					$ \mathbf{l=0} $ & $ \{5,4\}^+ $ & $ \{5,3\}^+ $ & $ \{5,2\}^+ $ & $ \{5,4,3\}^+ $ & $ \{5,4,2\}^+ $ & $ \{5,3,2\}^+ $ & $ \{5,4,3,2\}^+ $\\
					$ \hspace{0.5cm} $  &$ \hspace{0.5cm} $  & $ \hspace{0.5cm} $ & $ \hspace{0.5cm} $ & $ \hspace{0.5cm} $& $ \hspace{0.5cm} $& $ \hspace{0.5cm} $ & $ \hspace{0.5cm} $\\
					$ \hspace{0.5cm} $  &$ \big\downarrow $  & $ \big\downarrow $ & $ \big\downarrow $ & $ \big\downarrow $ & $ \big\downarrow $ & $ \big\downarrow $  & $ \big\downarrow $\\
					$ \hspace{0.5cm} $  &$ \hspace{0.5cm} $  & $ \hspace{0.5cm} $ & $ \hspace{0.5cm} $ & $ \hspace{0.5cm} $& $ \hspace{0.5cm} $& $ \hspace{0.5cm} $ & $ \hspace{0.5cm} $\\
					$ \mathbf{l=1} $  &$ \mathbb{P}^{(4)} $ & $ \{4,3\}^+ $& $ \{4,2\}^+ $& $ \{4,3\}^- $& $ \{4,2\}^- $ & $ \{4,3,2\}^+ $& $ \{4,3,2\}^- $\\
					$ \hspace{0.5cm} $  &$ \hspace{0.5cm} $  & $ \hspace{0.5cm} $ & $ \hspace{0.5cm} $ & $ \hspace{0.5cm} $& $ \hspace{0.5cm} $& $ \hspace{0.5cm} $ & $ \hspace{0.5cm} $\\
					$ \hspace{0.5cm} $  &$ \hspace{0.5cm} $  & $ \big\downarrow $ & $ \big\downarrow $ & $ \big\downarrow $ & $ \big\downarrow $ & $ \big\downarrow $  & $ \big\downarrow $\\
					$ \hspace{0.5cm} $  &$ \hspace{0.5cm} $  & $ \hspace{0.5cm} $ & $ \hspace{0.5cm} $ & $ \hspace{0.5cm} $& $ \hspace{0.5cm} $& $ \hspace{0.5cm} $ & $ \hspace{0.5cm} $\\
					$ \mathbf{l=2} $  &$\hspace{0.5cm} $ & $  \mathbb{P}^{(3)} $& $  \{3,2\}^+ $  & $ \mathbb{P}^{(3)}  $ & $ \{3,2\}^- $& $ \{3,2\}^- $& $ \{3,2\}^+ $\\
					$ \hspace{0.5cm} $  &$ \hspace{0.5cm} $  & $ \hspace{0.5cm} $ & $ \hspace{0.5cm} $ & $ \hspace{0.5cm} $& $ \hspace{0.5cm} $& $ \hspace{0.5cm} $ & $ \hspace{0.5cm} $\\
					$ \hspace{0.5cm} $  &$ \hspace{0.5cm} $  & $ \hspace{0.5cm} $ & $ \big\downarrow $ & $ \hspace{0.5cm} $ & $ \big\downarrow $ & $ \big\downarrow $  & $ \big\downarrow $\\
					$ \hspace{0.5cm} $  &$ \hspace{0.5cm} $  & $ \hspace{0.5cm} $ & $ \hspace{0.5cm} $ & $ \hspace{0.5cm} $& $ \hspace{0.5cm} $& $ \hspace{0.5cm} $ & $ \hspace{0.5cm} $\\
					$ \mathbf{l=3} $  & $ \hspace{0.5cm} $  & $  \hspace{0.5cm} $   &$ \mathbb{P}^{(2)} $ & $ \hspace{0.5cm}  $   & $ \mathbb{P}^{(2)} $& $  \mathbb{P}^{(2)}  $& $ \mathbb{P}^{(2)}$
				\end{tabular}
	\end{center}}}
	\caption{$\vec k^{(l)}$ and $\mu_{l}$ for $l=0, 1, \ldots, $
	$k_{0}-k_{d-1}$ for 
		model \RN{2} with 
		$\mu_{0}=+1$ and $n=5$.}
	\label{modelB+n=5}
\end{table}

Note that the $\gamma_{l}$'s satisfy
\begin{equation}
	\gamma_{l} = \begin{cases}
	-1 & \text{ if }  l\in \{ n-k_{1}\,,  n-k_{2}\,, 
	\ldots, n-k_{d-1} \} \\
	\ \ 1 & \text{ otherwise } 
	 \end{cases} \,.
\label{gammas}
\end{equation}	
Moreover, the  $\delta_{l}$'s satisfy
\begin{equation}
	\delta_{l} = \begin{cases}
	\gamma_{l} & \text{ if   }  l = 0, 1, \ldots 
	\ldots, n-k_{d-1}-1  \\
	\mu_{l-1} & \text{ if   } l = n-k_{d-1}
	 \end{cases} \,.
\label{deltas}
\end{equation}	

We further define
\begin{align}
	\mathbb{R}^{(l)}(u) &= \begin{cases}
	\frac{1}{(1-\gamma_{l} u)(\eta-\gamma_{l}  
	u)}\tilde{R}^{(l)}(-\gamma_{l} u) & l= 1, 2, \ldots, 
	k_{0}-k_{d-1}-1 \\[0.1in]
	\mathbb{P}^{(k_{d-1}\,, k_{d-1})} & l= k_{0}-k_{d-1}
	\end{cases} \,, \non \\
	f^{(l)}(u) &=\frac{-\eta + \delta_{l} u}{u} \,,
\label{RRfl}
\end{align}
see \eqref{eq:mapforB+BA},\eqref{eq:mapforB-BA}.

Let us define the sequence of transfer matrices $t^{(l)}(u; \{u_{j}^{(l)}\})$ by
\begin{equation}
t^{(l)}(u; \{u_{j}^{(l)}\}) = 
\tr_{0}\tilde{R}^{(l)}_{01}(u-u_1^{(l)})\,\ldots\,\tilde{R}^{(l)}_{0 
m_{l}}(u-u_{m_{l}}^{(l)}) \,, \qquad l = 0, 1, \ldots, 
k_{0}-k_{d-1}\,,
\label{transftild}
\end{equation}
and let us denote the corresponding eigenvalues by $\Lambda^{(l)}(u; 
\{u_{j}^{(l)}\})$. 
Note that the original transfer matrix
$t(u;\{\theta_j \})$ in \eqref{eq:transfer} 
is equal (up to a similarity transformation, see \eqref{eq:newRmodelB}) 
to $t^{(0)}(u; \{\theta_j\})$.
We wish to determine 
$\Lambda(u; \{\theta_{j}\}) := \Lambda^{(0)}(u; \{u_{j}^{(0)}\})$,
where
\begin{equation}
m_{0} := L \,, \qquad  u_{j}^{(0)} := \theta_{j} \,.
\label{initial}
\end{equation}
It follows from the result \eqref{nestedresult} that 
\begin{align}
&\Lambda^{(l)}(u; \{u_{j}^{(l)}\}) = \prod_{j=1}^{m_{l}}(\eta+u-u_j^{(l)}) (1+u-u_j^{(l)})
\prod_{i=1}^{m_{l+1}}\frac{\eta+u_i^{(l+1)}-u}{u_i^{(l+1)}-u}\non\\
&+ \Lambda_{\text{aux}}^{(l+1)}(u; \{u_{j}^{(l+1)}\})
\prod_{j=1}^{m_{l}}(u-u_j^{(l)})(1+u-u_j^{(l)})
\prod_{i=1}^{m_{l+1}}f^{(l+1)}(u_i^{(l+1)}-u) \,, \qquad l = 0, 1, 
\ldots \,, 
\label{anagen}
\end{align}
where $\Lambda_{\text{aux}}^{(l)}(u; \{u_{j}^{(l)}\})$ is an 
eigenvalue of the auxiliary transfer matrix $t_{\text{aux}}^{(l)}(u; \{u_{j}^{(l)}\})$, 
which is given by\footnote{For $l=1$, $t_{\text{aux}}^{(l)}$ 
coincides with $t_{\text{aux}}$ \eqref{taux}.}
\begin{align}
t_{\text{aux}}^{(l)}(u; \{u_{j}^{(l)}\}) &= 
\tr_{0}\mathbb{R}^{(l)}_{01}(u_1^{(l)}-u)\,\ldots\,\mathbb{R}^{(l)}_{0 m_{l}}(u_{m_{l}}^{(l)}-u) \,, \non\\
&=\left(\prod_{j=1}^{m_{l}}\frac{1}{(1-\gamma_{l} (u_{j}^{(l)}-u))
(\eta-\gamma_{l} (u_{j}^{(l)}-u))}\right)\,
t^{(l)}(\gamma_{l} u; \{\gamma_{l} u_{j}^{(l)}\}) \,,
\label{transfRR}
\end{align}
where we have passed to the second equality using \eqref{RRfl} and 
\eqref{transftild}. Hence,
\begin{equation}
\Lambda_{\text{aux}}^{(l)}(u; \{u_{j}^{(l)}\})
=\left(\prod_{j=1}^{m_{l}}\frac{1}{(1-\gamma_{l} (u_{j}^{(l)}-u))
(\eta-\gamma_{l}(u_{j}^{(l)}-u))}\right)\,
\Lambda^{(l)}(\gamma_{l} u; \{ \gamma_{l} u_{j}^{(l)}\}) \,.
\label{lambaux}
\end{equation}
We see from \eqref{anagen} that
\begin{align}
&\Lambda^{(l)}(\gamma_{l} u; \{ \gamma_{l} u_{j}^{(l)}\})	
=\prod_{j=1}^{m_{l}}(\eta + \gamma_{l} (u-u_j^{(l)})) 
(1+\gamma_{l}(u-u_j^{(l)}))
\prod_{i=1}^{m_{l+1}}\frac{\eta+ u_i^{(l+1)}-\gamma_{l} u}{u_i^{(l+1)}- \gamma_{l} u}\non\\
&+ \Lambda_{\text{aux}}^{(l+1)}(\gamma_{l} u; \{u_{j}^{(l+1)}\})
\prod_{j=1}^{m_{l}}\gamma_{l} (u-u_j^{(l)})(1+\gamma_{l}(u-u_j^{(l)}))
\prod_{i=1}^{m_{l+1}}f^{(l+1)}(u_i^{(l+1)}-\gamma_{l} u) \,.
\label{anagen2}
\end{align}

We conclude that the
eigenvalue of the auxiliary transfer matrix $t_{\text{aux}}^{(l)}(u; \{u_{j}^{(l)})$ is given by
\begin{align}
&\Lambda_{\text{aux}}^{(l)}(u; \{u_{j}^{(l)}\})
= \prod_{i=1}^{m_{l+1}}\frac{\eta+ u_i^{(l+1)}-\gamma_{l} u}{u_i^{(l+1)}- \gamma_{l} u}\non \\
&+ \Lambda_{\text{aux}}^{(l+1)}(\gamma_{l} u; \{u_{j}^{(l+1)}\})
\prod_{j=1}^{m_{l}}\frac{\gamma_{l} (u-u_j^{(l)})}
{\eta + \gamma_{l} (u-u_j^{(l)})}
\prod_{i=1}^{m_{l+1}}\frac{-\eta+\delta_{l+1} (u_i^{(l+1)}-\gamma_{l} u)}
{u_i^{(l+1)}-\gamma_{l} u} \,, \non \\
&\qquad\qquad l=1, \ldots, k_{0}-k_{d-1}-1\,,
\label{Lamaux}
\end{align}
where we have used \eqref{RRfl}, \eqref{lambaux} and \eqref{anagen2}. 
For $l=k_{0}-k_{d-1}$, we see from \eqref{RRfl} that $\mathbb{R}^{(l)} = 
\mathbb{P}^{(k_{d-1}\,, k_{d-1})}$ is independent of the spectral parameter, 
and we find
\begin{equation}
\Lambda_{\text{aux}}^{(k_{0}-k_{d-1})}	
= \begin{cases}
\exp\left(\frac{2\pi i p}{m_{k_{0}-k_{d-1}}}\right) \,, \quad p = 0, 1, \ldots, 
m_{k_{0}-k_{d-1}}-1\,,    & \text{   if   } m_{k_{0}-k_{d-1}} \ne 0 \\
k_{d-1} & \text{   if   } m_{k_{0}-k_{d-1}} = 0\\
\end{cases} \,.
\label{lamblast}
\end{equation}
Let us define $\Lambda_{\text{aux}}^{(0)}(u;\{\theta_{j}\})$ by \eqref{Lamaux} with 
$l=0$, keeping in mind \eqref{initial}. That is,
\begin{align}
\Lambda_{\text{aux}}^{(0)}(u;\{\theta_{j}\}) &=
\prod_{i=1}^{m_{1}}\frac{\eta+u_i^{(1)}- u}{u_i^{(1)}- u} 
\non\\
&+ \Lambda_{\text{aux}}^{(1)}(u; \{u_{j}^{(1)}\})
\prod_{j=1}^{L}\frac{u-\theta_{j}}{\eta + u-\theta_{j}}
\prod_{i=1}^{m_{1}}\frac{-\eta +\delta_{1} (u_i^{(1)}-u)}{u_i^{(1)}- u} \,.
\label{lamb0}
\end{align}
It follows from \eqref{nestedresult} that
\begin{equation}
\Lambda(u; \{\theta_{j}\}) = \prod_{j=1}^{L} (\eta + 
u-\theta_{j})(1 + u - \theta_{j}) \Lambda_{\text{aux}}^{(0)}(u; \{\theta_{j}\}) \,,
\label{TQ0}
\end{equation}
where $\Lambda_{\text{aux}}^{(0)}(u; \{\theta_{j}\})$ can be determined recursively 
using \eqref{lamb0}, \eqref{Lamaux} and \eqref{lamblast}. 

\subsection{Bethe equations}\label{sec:BEmodeI}

The conditions that the expressions \eqref{Lamaux} for $\Lambda_{\text{aux}}^{(l)}(u; 
\{u_{j}^{(l)}\})$ have vanishing residues at the poles 
$u= \gamma_{l} u_i^{(l+1)}$ lead (after the shift $l \mapsto l-1$) to the following Bethe equations
for $\{ u_{i}^{(l)} \}$   
\begin{align}
\prod_{j=1}^{m_{l-1}}\frac{\eta - \gamma_{l-1}(u_{j}^{(l-1)} - 
\gamma_{l-1} u_{i}^{(l)})}{\gamma_{l-1}(\gamma_{l-1} u_{i}^{(l)} - u_{j}^{(l-1)})}
&= \Lambda^{(l)}_{\text{aux}}(u_{i}^{(l)}; \{ u_{j}^{(l)} \})
\prod_{j \ne i; j=1}^{m_{l}} \frac{\delta_{l} (u_{j}^{(l)} - 
u_{i}^{(l)})-\eta}{u_{j}^{(l)} - u_{i}^{(l)} + \eta} 
 \,,  \non \\
&\qquad i=1, 2, \ldots, m_{l}\,, 
\qquad l= 1, 2, \ldots, k_{0} - k_{d-1} \,, \label{BE1}\\
\Lambda^{(l)}_{\text{aux}}(u_{i}^{(l)}; \{ u_{j}^{(l)} \})&= 
\prod_{j=1}^{m_{l+1}}\frac{\eta + u_{j}^{(l+1)} - \gamma_{l} 
u_{i}^{(l)}}{u_{j}^{(l+1)} - \gamma_{l}  u_{i}^{(l)}}\,, 
\quad l = 1, 2, \ldots, 
k_{0} - k_{d-1}-1  \,, \label{BE2}
\end{align}
and $\Lambda^{(k_{0}-k_{d-1})}_{\text{aux}}$ is given in \eqref{lamblast}.

In summary, the eigenvalues $\Lambda(u; \{\theta_{j}\})$ of the 
transfer matrix $t(u;\{\theta_j \})$
\eqref{eq:transfer} are given by \eqref{Lamaux}-\eqref{TQ0}, where $\{ u_{i}^{(l)} \}$  
are solutions of the Bethe equations \eqref{BE1}, \eqref{BE2}.

Remarkably, these Bethe equations can be brought to a form similar 
to those of usual $\mathfrak{gl}(m|n-m)$ spin chains.\footnote{We thank the 
referee for bringing this fact to our attention.} Indeed, let us define 
the rescaled Bethe roots
\begin{equation}
	\tilde{u}^{(l)}_{j} := \chi_{l}\, u^{(l)}_{j} \,, \qquad
	\chi_{l} := \prod_{l' < l} \gamma_{l'} \,,
	\label{chi}
\end{equation}
in terms of which the Bethe equations \eqref{BE1}, 
\eqref{BE2} can be rewritten as
\begin{align}
	(\delta_{l})^{m_{l}-1} &= \prod_{j=1}^{m_{l-1}} 
	\frac{\tilde{u}^{(l)}_{i} - \tilde{u}^{(l-1)}_{j}}
	{\tilde{u}^{(l)}_{i} - \tilde{u}^{(l-1)}_{j} + \chi_{l} \eta}
	\prod_{j\ne i}^{m_{l}} 
	\frac{\tilde{u}^{(l)}_{i} - \tilde{u}^{(l)}_{j} + \delta_{l} \chi_{l} \eta}
	{\tilde{u}^{(l)}_{i} - \tilde{u}^{(l)}_{j} - \chi_{l} \eta} \non\\
	& \times \prod_{j=1}^{m_{l+1}} 
	\frac{\tilde{u}^{(l)}_{i} - \tilde{u}^{(l+1)}_{j} - \chi_{l+1} \eta}
	{\tilde{u}^{(l)}_{i} - \tilde{u}^{(l+1)}_{j}} 
	\,, \qquad i = 1, \ldots, m_{l}\,, \qquad l = 1, \ldots, k_{0}-k_{d-1}-1\,, \non\\
	\frac{(\delta_{l})^{m_{l}-1}}{\Lambda^{(k_{0}-k_{d-1})}_{\text{aux}}} &=
	\prod_{j=1}^{m_{l-1}} 
	\frac{\tilde{u}^{(l)}_{i} - \tilde{u}^{(l-1)}_{j}}
	{\tilde{u}^{(l)}_{i} - \tilde{u}^{(l-1)}_{j} + \chi_{l} \eta}
	\prod_{j\ne i}^{m_{l}} 
	\frac{\tilde{u}^{(l)}_{i} - \tilde{u}^{(l)}_{j} + \delta_{l} \chi_{l} \eta}
	{\tilde{u}^{(l)}_{i} - \tilde{u}^{(l)}_{j} - \chi_{l} \eta}\,, \quad 
	i = 1, \ldots, m_{l}\,, \quad l = k_{0}-k_{d-1} \,.
	\label{BEtilde}
\end{align}
Finally, in terms of the shifted Bethe roots
\begin{equation}
	\tilde{\tilde{u}}^{(l)}_{j} := \tilde{u}^{(l)}_{j} + 
	\frac{\eta}{2}\sum_{i=1}^{l} \chi_{i} \,,
	\label{doubletildebetheroots}
\end{equation}
the Bethe equations \eqref{BEtilde} take the more symmetric form
\begin{align}
	(\delta_{l})^{m_{l}-1} &= \prod_{j=1}^{m_{l-1}} 
	\frac{\tilde{\tilde{u}}^{(l)}_{i} - \tilde{\tilde{u}}^{(l-1)}_{j} - \frac{\eta}{2}\chi_{l}}
	{\tilde{\tilde{u}}^{(l)}_{i} - \tilde{\tilde{u}}^{(l-1)}_{j} + \frac{\eta}{2}\chi_{l}}
	\prod_{j\ne i}^{m_{l}} 
	\frac{\tilde{\tilde{u}}^{(l)}_{i} - \tilde{\tilde{u}}^{(l)}_{j} + \delta_{l} \chi_{l} \eta}
	{\tilde{\tilde{u}}^{(l)}_{i} - \tilde{\tilde{u}}^{(l)}_{j} - \chi_{l} \eta} 
	\prod_{j=1}^{m_{l+1}} 
	\frac{\tilde{\tilde{u}}^{(l)}_{i} - \tilde{\tilde{u}}^{(l+1)}_{j} - \frac{\eta}{2}\chi_{l+1}}
	{\tilde{\tilde{u}}^{(l)}_{i} - \tilde{\tilde{u}}^{(l+1)}_{j} + \frac{\eta}{2}\chi_{l+1}} 
	\,, \non\\
	&\qquad\qquad\qquad i = 1, \ldots, m_{l}\,, 
	\qquad l = 1, \ldots, k_{0}-k_{d-1}-1\,, \non\\
	\frac{(\delta_{l})^{m_{l}-1}}{\Lambda^{(k_{0}-k_{d-1})}_{\text{aux}}} &=
	\prod_{j=1}^{m_{l-1}} 
	\frac{\tilde{\tilde{u}}^{(l)}_{i} - \tilde{\tilde{u}}^{(l-1)}_{j} - \frac{\eta}{2}\chi_{l}}
	{\tilde{\tilde{u}}^{(l)}_{i} - \tilde{\tilde{u}}^{(l-1)}_{j} + \frac{\eta}{2}\chi_{l}}
	\prod_{j\ne i}^{m_{l}} 
	\frac{\tilde{\tilde{u}}^{(l)}_{i} - \tilde{\tilde{u}}^{(l)}_{j} + \delta_{l} \chi_{l} \eta}
	{\tilde{\tilde{u}}^{(l)}_{i} - \tilde{\tilde{u}}^{(l)}_{j} - \chi_{l} \eta}\,, \quad 
	i = 1, \ldots, m_{l}\,, \quad l = k_{0}-k_{d-1} \,,
	\label{BEtildetilde} 
\end{align}
where $\Lambda^{(k_{0}-k_{d-1})}_{\text{aux}}$ is defined in equation \eqref{lamblast}. 

\begin{figure}[t]
\begin{center}
\unitlength=2.4pt
\begin{picture}(40,18)
\thicklines 
\put(22,10){\line(1,0){6}}\put(32,10){\line(1,0){6}} 
\put(20,10){\circle{4}} \put(30,10){\circle{4}}\put(40,10){\circle{4}} 
\put(17.69,8.53){\large$\times$} \put(27.69,8.53){\large$\times$}
\end{picture}
\caption{Dynkin diagram for the model II case $\{5,4,3,2\}^{+}$
\label{fig:dynkinII}}
\end{center}
\end{figure}
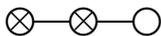

The Bethe equations \eqref{BEtildetilde} are therefore 
simply given by
\begin{equation}
 	(\delta_{l})^{m_{l}-1} z^{-1}_l(p) =	\prod_{l'=1}^{k_{0}-k_{d-1}} \prod_{j=1}^{m_{l'}}{}^{'}
\frac{\tilde{\tilde{u}}^{(l)}_{i} - \tilde{\tilde{u}}^{(l')}_{j} + 
\frac{\eta}{2}c_{l, l'}}
{\tilde{\tilde{u}}^{(l)}_{i} - \tilde{\tilde{u}}^{(l')}_{j} - 
\frac{\eta}{2}c_{l, l'}} \,, \qquad i = 1, \ldots, m_{l}\,,
 \qquad l = 1, \ldots, k_{0}-k_{d-1}\,, 
 \label{BAIIcompact}
\end{equation}
where the primed product omits the $j=i$ term if $l'=l$, and $c_{l, l'}$ is given by
\begin{align}
	\text{diagonal:} & &c_{l,l} &= \begin{cases}
	2 \chi_{l} & \text{ if } \quad  \delta_{l} = +1 \\
	0          & \text{ if } \quad  \delta_{l} = -1
	\end{cases} \,, \non \\
	\text{off-diagonal:} &  &c_{l,l'} &= \begin{cases}
	-\chi_{l} & \text{ if } \quad  l' = l - 1 \\
	-\chi_{l'} & \text{ if } \quad  l' = l + 1 \\
	\ \ 0        & \text{ otherwise }
	\end{cases} \,,
\end{align}
where $\chi_{l}$ \eqref{chi} is $\pm 1$. The function $z_l(p)$ is given by
\begin{equation}
	 z_l(p)= \begin{cases}
		1  & \text{ if } \qquad l = 1, \ldots, k_{0}-k_{d-1}-1 \\
		\Lambda^{(k_{0}-k_{d-1})}_{\text{aux}} & \text{ if } \qquad  l=k_{0}-k_{d-1}
	\end{cases} \,.
\label{simplifiedBEmodel2}
\end{equation}
Note that $c_{l,l+1} = 
c_{l+1,l}$, and therefore $c_{l,l'}$ is 
symmetric. Moreover, in view of \eqref{deltas},
\begin{equation}
\sum_{l'} c_{l,l'} = -\chi_{l} (1 + \gamma_{l}) + c_{l,l} = 0 \,, 
\qquad
l = 2, \ldots, k_{0}-k_{d-1}-1 \,.\label{CsmodelII}
\end{equation}
Hence, 
$c_{l,l'}$ can be identified as the Cartan matrix for a (potentially 
non-distinguished) $\mathfrak{gl}(m|n-m)$ Kac-Dynkin diagram. For example, 
for the case $\{5,4,3,2\}^{+}$ (for which $\delta_{1} =  \delta_{2} = 
-1\,, \delta_{3} = 1$), the corresponding diagram is shown in Fig. 
\ref{fig:dynkinII}; fermionic nodes (for which $c_{l,l}=0$) 
are denoted by a cross. However, compared 
with usual $\mathfrak{gl}(m|n-m)$ spin chains, the LHS of 
the Bethe equations \eqref{BEtildetilde} has additional phases; 
moreover, the transfer-matrix eigenvalues \eqref{TQ0}, which can be 
re-expressed in terms of the redefined Bethe roots 
$\tilde{\tilde{u}}^{(l)}_{i}$, are not the 
standard ones.

Since model II has rank $k_0-1$, one would expect it to have an equal 
number of Bethe equations; however, there are in fact only 
$k_0-k_{d-1}$ such equations \eqref{BAIIcompact}. The ``missing'' Bethe equations are hidden in the
condition \eqref{lamblast}. For example, the Kac-Dynkin diagram in Fig. \ref{fig:dynkinII} for a model of rank four 
has one less node than expected. Therefore, despite the similarities with
$\mathfrak{gl}(m|n-m)$, this model is significantly different. 

We have checked the completeness of this Bethe ansatz solution numerically
for small values of $L, d, \vec k$ by using \eqref{Lamaux}-\eqref{BE2}
to solve for the eigenvalues of the homogeneous transfer matrix (all 
$\theta_{j} =0$), and comparing with the corresponding results obtained 
by exact diagonalization, see e.g. Tables 
\ref{table:B+d2k42},\ref{table:B+d3k432},\ref{table:B-d3k432}. 
We observe the presence of infinite Bethe roots, as well as 
singular (exceptional) solutions of the Bethe equations.\footnote{Singular solutions for 
the XXX chain are discussed in e.g. \cite{Avdeev:1985cx, 
Nepomechie:2013mua, Hao:2013jqa, Nepomechie:2014hma}. Infinite Bethe 
roots have been noted in various models, see e.g. \cite{Frahm:2010ei, Frahm:2012eb, Frahm:2019czs,
Nepomechie:2019zrx}.}
While we can account for all distinct eigenvalues (although not their 
degeneracies), there is one caveat: 
we find instances with repeated singular Bethe roots
(such as the last line of Table \ref{table:B+d3k432}), where
the roots indeed give the eigenvalue through the TQ equation
\eqref{TQ0}, but the Bethe equations are not all satisfied (at 
least naively), which we leave as a problem for future investigation.
Based on these studies, we conjecture that the values of $\{m_{l} 
\}$ can be restricted as follows
\begin{equation}
m_{0} \ge m_{1} \ge m_{2} \ge \ldots \ge m_{k_{0}-k_{d-1}} \,,
\end{equation}
where $m_{0} :=L$.

\section{Bethe ansatz for model \RN{1}}\label{sec:BAmodelA}

We now analyze model \RN{1} using nested algebraic Bethe ansatz, again restricting to $k_{d-1}>1$. 

\subsection{First level of nesting}\label{sec:Afirstlevel}

Similarly to model \RN{2}, for model \RN{1} we perform the Bethe ansatz 
for the gauge-transformed R-matrix
\begin{equation}
\tilde{R}^{\RN{1},\vec{k}}(u)=\left(V\otimes 
V\right) R^{\RN{1},\vec{k}}(u) \left(V^{-1}\otimes V^{-1}\right) \,,
\label{eq:newRmodelA}
\end{equation}
where $V$ is defined in \eqref{Vmat}, and $R^{\RN{1},\vec{k}}(u)$ is given 
in \eqref{eq:Rmodel0}.

This model has much in common with model \RN{2}, so for the parts of the
analysis that coincide, we refer to the previous section in order to 
avoid repeating formulas.  The equations from \eqref{eq:monodromy} to
\eqref{eq:psi} remain the same.  In particular, the action of the
operators $\mathcal{C}_{\alpha}, \, \mathcal{T}_{00}$ and 
$\mathcal{T}_{\alpha\beta}$ on the reference state $|0\rangle$ do
not change. The exchange relations are again given by \eqref{eq:commutT00} and 
\eqref{eq:commutTab}, except $\mathbb{R}(u)$ is now given by 
\begin{equation}
\mathbb{R}(u)=\begin{cases}
\frac{1}{\eta+u}R^{(gl_{n-1})}(u) & \text{for initial model with } k_1=k_0-1 \text{ and } d=2\\
\frac{1}{1+u}\frac{1}{\eta+u}\tilde{R}^{(\RN{1},k_0-1,k_1,..,k_{d-1})}(u)& \text{for initial model with  } k_1=k_0-1 \text{ and } d>2\\
\frac{1}{1-u}\frac{1}{\eta-u}\tilde{R}^{(\RN{1},k_1,..,k_{d-1})}(-u)& \text{for initial model with } k_1<k_0-1 \text{ and } d\ge 2\\
\end{cases}
\label{eq:mapforA}
\end{equation}
where $R^{gl_{m}}(u)$ is the $m^2\times m^2$ R-matrix given by
\begin{equation}
R^{(gl_{m})}(u)=\eta \mathbb{P}^{(m,n)}+ u \mathbb{I}^{(m,n)}.
\label{eq:Rglm}
\end{equation}
The functions $f(u)$ and $g(u)$ are defined as
\begin{align}
& f(u)=\begin{cases}
\frac{\eta- u}{-u} & \text{for initial model with } k_1<k_0-1 \text{ and } d\ge 2\\
\frac{\eta+ u}{-u} & \text{for initial model with } k_1=k_0-1 \text{ and } d\ge 2\\
\end{cases}\\
& g(u)=\frac{\eta}{u}.
\end{align}
Hence, except for the explicit forms of $\mathbb{R}(u)$ and $f(u)$, 
Eqs. \eqref{eq:Taaeq1}-\eqref{eq:Taaeq6} remain the same. We conclude 
that the eigenvalues of the transfer matrix are given by
\begin{align}
\Lambda(u,\{\theta_j\})&=\prod_{j=1}^{L}\left(\eta+u-\theta_j\right)\left(1+u-\theta_j\right)
\prod_{j=1}^{m_1}\frac{\eta+u_j-u}{u_j-u} \nonumber\\
&+\Lambda_{\text{aux}}(u)\prod_{j=1}^{L} (u-\theta_j)(1+u-\theta_j)\prod_{j=1}^{m_1}f(u_j-u)\,,
\label{eq:nestedresultmodelA}
\end{align} 
where $\Lambda_{\text{aux}}(u)$ is an eigenvalue of the auxiliary transfer matrix \eqref{taux}.

\subsection{Transfer-matrix eigenvalues}\label{sec:BEandTQmodelA}

In the following we will recursively construct the TQ and Bethe
equations for this model, in a similar way as for model \RN{2}. As
we will show, the main difference is that in the
``last'' step of the nesting procedure, for $d=2$ and $k_1=k_0-1$, we
have $\mathbb{R}(u)\sim R^{gl_{k_{d-1}}}(u)$ for model \RN{1}, instead of
$\mathbb{R}(u)\sim \mathbb{P}^{(k_{d-1},k_{d-1})}$ for
model \RN{2}. Consequently, an extra recursion procedure will be 
needed for model \RN{1}.

We define a sequence of R-matrices
\begin{equation}
\tilde{R}^{(l)}(u) \equiv \tilde{R}^{ \vec{k}^{(l)}}(u)\,, \qquad 
l = 0, 1, \ldots\,,k_{d-1}
\end{equation}
\noindent
where $\tilde{R}^{(0)}(u) = \tilde{R}^{\vec{k}}(u)$ and $\vec{k}^{(0)} = 
\vec{k}$. Moreover, the vectors $\vec{k}^{(l)}$, as well 
as the parameter  $\gamma_{l}$, similarly to section \ref{sec:BAmodelB} are 
defined for $l\ge 1$ recursively as follows:
\begin{align}
&\text{If   } k_{1}^{(l-1)} < k_{0}^{(l-1)}-1\,, \quad \text{then  }	
\vec{k}^{(l)} = \vec{k}^{(l-1)} - \vec \epsilon\,,  \quad \gamma_{l}=1 \,; \non \\
&\text{if   } k_{1}^{(l-1)} = k_{0}^{(l-1)}-1 \text{ and } |\vec 
k^{(l-1)}|>2 \,, \quad \text{then  }
\vec{k}^{(l)} = \hat{\vec{k}}^{(l-1)}\,,  \quad \gamma_{l}=-1 \,; \non \\
&\text{if   } k_{1}^{(l-1)} = k_{0}^{(l-1)}-1 \text{ and } |\vec 
k^{(l-1)}|=2 \,, \quad \text{then  } \gamma_{l}=-1 \,.
\label{iterationmodelA}
\end{align}
As before, $\vec \epsilon$ is the vector $\vec 
\epsilon=\{1,0,\ldots,0\}$ that has the same dimension as 
$\vec{k}^{(l-1)}$, i.e. $|\vec \epsilon\,| = 
|\vec{k}^{(l-1)}|$. Furthermore, the hat denotes dropping the 
first (left-most) component; hence, since  $\vec{k}^{(l-1)} = \{k^{(l-1)}_{0}\,, 
k^{(l-1)}_{1}\,, \ldots \}$, then $\hat{\vec{k}}^{(l-1)} = \{ 
k^{(l-1)}_{1}\,, \ldots \}$. Examples of such $\vec{k}^{(l)}$ sequences are shown in Table \ref{modelAn=5}.

The $\gamma_{l}$'s again satisfy \eqref{gammas}, for $ l=1,...,k_0-k_{d-1} $. 

We also define
\begin{align}
\mathbb{R}^{(l)}(u) &= \begin{cases}
\frac{1}{(1-\gamma_{l} u)(\eta-\gamma_{l} u)}\tilde{R}^{(l)}(-\gamma_{l} u) & l= 1, 2, \ldots, 
k_{0}-k_{d-1}-1 \\[0.1in]
\frac{1}{\eta+ u}R^{gl_{k_{d-1}}}(u) & l= k_{0}-k_{d-1}
\end{cases} \,, \non \\
f^{(l)}(u) &=\frac{-\eta + \gamma_{l} u}{u} \,,
\label{RRflModelA}
\end{align}
where $R^{gl_{k_{d-1}}}(u)$ is given by \eqref{eq:Rglm}.

\begin{table}[h!]
	\setlength{\tabcolsep}{0.5em} % for the horizontal padding
	{\renewcommand{\arraystretch}{0.2}
		{\renewcommand{\arraystretch}{0.5}
			\begin{center}
				\begin{tabular}{c|ccc|ccc|c}
					\multicolumn{8}{c}{\textbf{$ \hspace{0.5cm} $}}\\
					\multicolumn{8}{c}{\textbf{Model \RN{1} $ (\mathbf{n=5} $})}\\
					\multicolumn{8}{c}{\textbf{$ \hspace{0.5cm} $}}\\
					\hline
					$ \hspace{0.5cm} $  &$ \hspace{0.5cm} $  & $ \hspace{0.5cm} $ & $ \hspace{0.5cm} $ & $ \hspace{0.5cm} $& $ \hspace{0.5cm} $& $ \hspace{0.5cm} $ & $ \hspace{0.5cm} $\\
					$ \hspace{0.5cm}$ & $ \hspace{0.5cm}$ & $ \mathbf{d=2}$ & $ \hspace{0.5cm}$ & $ \hspace{0.5cm}$ & $ \mathbf{d=3}$ & $ \hspace{0.5cm}$ & $ \mathbf{d=4}$\\
					$ \hspace{0.5cm} $  &$ \hspace{0.5cm} $  & $ \hspace{0.5cm} $ & $ \hspace{0.5cm} $ & $ \hspace{0.5cm} $& $ \hspace{0.5cm} $& $ \hspace{0.5cm} $ & $ \hspace{0.5cm} $\\
					\hline
					$ \hspace{0.5cm} $  &$ \hspace{0.5cm} $  & $ \hspace{0.5cm} $ & $ \hspace{0.5cm} $ & $ \hspace{0.5cm} $& $ \hspace{0.5cm} $& $ \hspace{0.5cm} $ & $ \hspace{0.5cm} $\\
					$ \mathbf{l=0} $ & $ \{5,4\} $ & $ \{5,3\} $ & $ \{5,2\} $ & $ \{5,4,3\} $ & $ \{5,4,2\} $ & $ \{5,3,2\} $ & $ \{5,4,3,2\} $\\
					$ \hspace{0.5cm} $  &$ \hspace{0.5cm} $  & $ \hspace{0.5cm} $ & $ \hspace{0.5cm} $ & $ \hspace{0.5cm} $& $ \hspace{0.5cm} $& $ \hspace{0.5cm} $ & $ \hspace{0.5cm} $\\
					$ \hspace{0.5cm} $  &$ \big\downarrow $  & $ \big\downarrow $ & $ \big\downarrow $ & $ \big\downarrow $ & $ \big\downarrow $ & $ \big\downarrow $  & $ \big\downarrow $\\
					$ \hspace{0.5cm} $  &$ \hspace{0.5cm} $  & $ \hspace{0.5cm} $ & $ \hspace{0.5cm} $ & $ \hspace{0.5cm} $& $ \hspace{0.5cm} $& $ \hspace{0.5cm} $ & $ \hspace{0.5cm} $\\
					$ \mathbf{l=1} $  &$ \text{gl}_{4} $ & $ \{4,3\} $& $ \{4,2\} $& $ \{4,3\} $& $ \{4,2\} $ & $ \{4,3,2\} $& $ \{4,3,2\} $\\
					$ \hspace{0.5cm} $  &$ \hspace{0.5cm} $  & $ \hspace{0.5cm} $ & $ \hspace{0.5cm} $ & $ \hspace{0.5cm} $& $ \hspace{0.5cm} $& $ \hspace{0.5cm} $ & $ \hspace{0.5cm} $\\
					$ \hspace{0.5cm} $  &$ \hspace{0.5cm} $  & $ \big\downarrow $ & $ \big\downarrow $ & $ \big\downarrow $ & $ \big\downarrow $ & $ \big\downarrow $  & $ \big\downarrow $\\
					$ \hspace{0.5cm} $  &$ \hspace{0.5cm} $  & $ \hspace{0.5cm} $ & $ \hspace{0.5cm} $ & $ \hspace{0.5cm} $& $ \hspace{0.5cm} $& $ \hspace{0.5cm} $ & $ \hspace{0.5cm} $\\
					$ \mathbf{l=2} $  &$\hspace{0.5cm} $ & $ \text{gl}_{3} $& $ (3,2) $  & $ \text{gl}_{3}  $ & $ \{3,2\} $& $ \{3,2\} $& $ \{3,2\} $\\
					$ \hspace{0.5cm} $  &$ \hspace{0.5cm} $  & $ \hspace{0.5cm} $ & $ \hspace{0.5cm} $ & $ \hspace{0.5cm} $& $ \hspace{0.5cm} $& $ \hspace{0.5cm} $ & $ \hspace{0.5cm} $\\
					$ \hspace{0.5cm} $  &$ \hspace{0.5cm} $  & $ \hspace{0.5cm} $ & $ \big\downarrow $ & $ \hspace{0.5cm} $ & $ \big\downarrow $ & $ \big\downarrow $  & $ \big\downarrow $\\
					$ \hspace{0.5cm} $  &$ \hspace{0.5cm} $  & $ \hspace{0.5cm} $ & $ \hspace{0.5cm} $ & $ \hspace{0.5cm} $& $ \hspace{0.5cm} $& $ \hspace{0.5cm} $ & $ \hspace{0.5cm} $\\
					$ \mathbf{l=3} $  & $ \hspace{0.5cm} $  & $  \hspace{0.5cm} $   &$ \text{gl}_{2} $ & $ \hspace{0.5cm}  $   & $  \text{gl}_{2}  $& $ \text{gl}_{2}  $& $ \text{gl}_{2} $\\
					$ \hspace{0.5cm} $  &$ \hspace{0.5cm} $  & $ \hspace{0.5cm} $ & $ \hspace{0.5cm} $ & $ \hspace{0.5cm} $& $ \hspace{0.5cm} $& $ \hspace{0.5cm} $ & $ \hspace{0.5cm} $\\
				\end{tabular}
	\end{center}}}
	\caption{$\vec k^{(l)}$ for $l=0, 1, \ldots, k_{0}-k_{d-1}$ for 
	model \RN{1} with $n=5$.}
	\label{modelAn=5}
\end{table}

The first part of this calculation is very similar to the one for
model \RN{2}, with only a few sign modifications.  
We again start by
defining a sequence of transfer matrices $t^{(l)}(u; \{u_{j}^{(l)}\})$
as in \eqref{transftild}
\begin{equation}
t^{(l)}(u; \{u_{j}^{(l)}\}) = 
\tr_{0}\tilde{R}^{(l)}_{01}(u-u_1^{(l)})\,\ldots\,\tilde{R}^{(l)}_{0 
	m_{l}}(u-u_{m_{l}}^{(l)}) \,, \qquad l = 0, 1, \ldots, 
k_{0}-k_{d-1}\,,
\label{transftildmodelA}
\end{equation}
and denoting the corresponding eigenvalues by $\Lambda^{(l)}(u; 
\{u_{j}^{(l)}\})$. 
% We wish to determine $ \Lambda^{(0)}(u; \{u_{j}^{(0)}\})$, where
% \begin{equation}
% m_{0} := L \,, \qquad  u_{j}^{(0)} :=\theta_j \,.
% \label{initialmodelA}
% \end{equation} 
%From \eqref{eq:nestedresultmodelA} 
We obtain as in \eqref{anagen}
\begin{align}
&\Lambda^{(l)}(u; \{u_{j}^{(l)}\}) = \prod_{j=1}^{m_{l}}(\eta+u-u_j^{(l)}) (1+u-u_j^{(l)})
\prod_{j=1}^{m_{l+1}}\frac{\eta+u_j^{(l+1)}-u}{u_j^{(l+1)}-u} \non  \\
&+ \Lambda_{\text{aux}}^{(l+1)}(u; \{u_{j}^{(l+1)}\})
\prod_{j=1}^{m_{l}}(u-u_j^{(l)})(1+u-u_j^{(l)})
\prod_{j=1}^{m_{l+1}}f^{(l+1)}(u_j^{(l+1)}-u) \,, \non \\
&\qquad\qquad\qquad\qquad\qquad l = 0, 1, \ldots \,, k_0-k_{d-1} \label{anagenModelA}
\end{align}
Eqs. \eqref{transfRR}-\eqref{anagen2} do not change, and from them and \eqref{anagenModelA} 
we obtain (cf. \eqref{Lamaux})
\begin{align}
&\Lambda_{\text{aux}}^{(l)}(u; \{u_{j}^{(l)}\})
= \prod_{j=1}^{m_{l+1}}\frac{\eta+u_j^{(l+1)}-\gamma_{l} u}{u_j^{(l+1)}- \gamma_{l} u}\non \\
&+ \Lambda_{\text{aux}}^{(l+1)}(\gamma_{l} u; \{u_{j}^{(l+1)}\})
\prod_{j=1}^{m_{l}}\frac{\gamma_{l} (u-u_j^{(l)})}{\eta + \gamma_{l}(u-u_j^{(l)})}
\prod_{j=1}^{m_{l+1}}\frac{-\eta+\gamma_{l+1} (u_j^{(l+1)}-\gamma_{l} u)}
{u_j^{(l+1)}-\gamma_{l} u} \,, \non \\
&\qquad\qquad l=1, \ldots, k_{0}-k_{d-1}-1\,.
\label{LamauxmodelA}
\end{align}

The second part of the calculation is dedicated to computing 
$\Lambda_{\text{aux}}^{(k_0-k_{d-1})}$, which appears in 
\eqref{LamauxmodelA} for the final value $l=k_0-k_{d-1}-1$. We shall see that this 
requires the diagonalization of $gl_{m}$-type transfer matrices.
We therefore first define a sequence of $gl_{m}$-type R-matrices
\begin{equation}
r^{(l)}(u) = R^{gl_{k_0-l}}(u)\,, \qquad 
l = k_0-k_{d-1},k_0-k_{d-1}+ 1, \ldots\,, k_0-2 \,,
\label{eq:rdef}
\end{equation}
where $R^{gl_{m}}$ is defined in \eqref{eq:Rglm}.
Consider then a sequence of transfer matrices $\tau^{(l)}(u;\{u_j^{(l)}\})$
\begin{equation}
\tau^{(l)}(u; \{u_{j}^{(l)}\}) = 
\tr_{0}r^{(l)}_{01}(u-u_1^{(l)})\,\ldots\,r^{(l)}_{0 m_{l}}(u-u_{m_{l}}^{(l)}) \,, \qquad 
l = k_0-k_{d-1}, \ldots, k_0-2 \,.
\label{transftildglk}
\end{equation}
The corresponding eigenvalue $\lambda^{(l)}(u;\{u_j^{(l)}\})$ is well known to be given by 
\begin{align}
\lambda^{(l)}(u; \{u_{j}^{(l)}\}) &= \prod_{j=1}^{m_{l}}(\eta+u-u_j^{(l)}) 
\prod_{j=1}^{m_{l+1}}\frac{\eta+u_j^{(l+1)}-u}{u_j^{(l+1)}-u}\non\\
&+ \lambda_{\text{aux}}^{(l+1)}(u; \{u_{j}^{(l+1)}\})
\prod_{j=1}^{m_{l}}(u-u_j^{(l)})
\prod_{j=1}^{m_{l+1}}\phi^{(l+1)}(u_j^{(l+1)}-u) \,, \non\\
&\qquad\qquad\qquad l = k_0-k_{d-1}, \ldots\,,k_0-2,
\label{genglk}
\end{align}
where $\phi^{(l)}(u) $ is given by
\begin{equation}
\phi^{(l)}(u)=\frac{-\eta+ u}{u} \,,
\label{eq:hatl}
\end{equation}
and $\lambda_{\text{aux}}^{(l)}(u; \{u_{j}^{(l)}\})$ is an eigenvalue 
of the auxiliary transfer matrix 
\begin{equation}
\tau_{\text{aux}}^{(l)}(u; \{u_{j}^{(l)}\}) = 
\tr_{0}\mathbb{R}^{(l)}_{01}(u_1^{(l)}-u)\,\ldots\,\mathbb{R}^{(l)}_{0 m_{l}}(u_{m_{l}}^{(l)}-u) \,, 
\label{tauaux}
\end{equation}
where $\mathbb{R}^{(l)}(u)$ is given by
\begin{equation}
\mathbb{R}^{(l)}(u)=\frac{1}{\eta- u }\,r^{(l)}(-u)\,, \qquad 
l = k_0-k_{d-1}+1, \ldots\,, k_0-2 \,,
\label{eq:auxRglk}
\end{equation}
and $ r^{(l)}(u) $ was defined in \eqref{eq:rdef}.
It follows from \eqref{eq:auxRglk}
that $\tau_{\text{aux}}^{(l)}$ \eqref{tauaux} can be related to 
$\tau^{(l)}$ \eqref{transftildglk}
\begin{equation}
\tau_{\text{aux}}^{(l)}(u; \{u_{j}^{(l)}\}) 
=\left(\prod_{j=1}^{m_{l}}
	\frac{1}{(\eta -  (u_{j}^{(l)}-u))}\right)\,
\tau^{(l)}( u; \{u_{j}^{(l)}\}) \,,
\label{transfRRglk}
\end{equation}
and similarly for the corresponding eigenvalues
\begin{align}
\lambda_{\text{aux}}^{(l)}(u; \{u_{j}^{(l)}\}) &=\left(\prod_{j=1}^{m_{l}}
\frac{1}{(\eta - (u_{j}^{(l)}-u))} \right)\,
\lambda^{(l)}( u; \{u_{j}^{(l)}\}) \label{genglkaux0} \\
&=\prod_{j=1}^{m_{l+1}}\frac{\eta + u_j^{(l+1)}-u}{u_j^{(l+1)}-u}\non \\
&+ \lambda_{\text{aux}}^{(l+1)}(u; \{u_{j}^{(l+1)}\})
\prod_{j=1}^{m_{l}}\frac{ u-u_j^{(l)}}{\eta + u-u_j^{(l)}}
\prod_{j=1}^{m_{l+1}}\frac{\eta + u-u_j^{(l+1)}}{u-u_j^{(l+1)}} \,, \non \\
&\qquad\qquad\qquad  l = k_0-k_{d-1}+1, \ldots\,, k_0-2 \,,
\label{genglkaux}
\end{align}
where we have passed to \eqref{genglkaux} using \eqref{genglk}, and
$\lambda_{\text{aux}}^{(k_0-1)}(u; \{u_{j}^{(k_0-1)}\})\equiv 1$.

We are finally ready to compute 
$\Lambda_{\text{aux}}^{(k_0-k_{d-1})}$. Recalling from 
\eqref{RRflModelA} that $\mathbb{R}^{(l)}$ for $l=k_0-k_{d-1}$ is 
proportional to $R^{gl_{k_{d-1}}}$, and recalling the definitions of 
the transfer matrices
$t_{\text{aux}}^{(l)}$ \eqref{transfRR} and $\tau^{(l)}$ \eqref{transftildglk}, we see that
\begin{align}
\Lambda_{\text{aux}}^{(k_0-k_{d-1})}(u;\{u_j^{(k_0-k_{d-1})}\}) 
&=\frac{\lambda^{(k_0-k_{d-1})}(-u;\{-u_j^{(k_0-k_{d-1})}\})}{\prod_{j=1}^{m_{k_0-k_{d-1}}}(\eta + u_j^{(k_0-k_{d-1})}-u)} \non\\
&=\lambda_{\text{aux}}^{(k_0-k_{d-1})}(-u;\{-u_j^{(k_0-k_{d-1})}\}) 
\,.
\label{eq:Lambdak0-kdm1v2}
\end{align}
To pass to the second line, we have used \eqref{genglkaux0} to define 
$\lambda_{\text{aux}}^{(l)}$ for $l=k_0-k_{d-1}$.

The full result for $\Lambda^{(0)}(u; \{u_{j}^{(0)}\})$ is therefore obtained by starting from $l=0$ in Eq.
\eqref{anagenModelA}, and then using Eqs.  \eqref{LamauxmodelA}, 
\eqref{eq:Lambdak0-kdm1v2} and \eqref{genglkaux}.

\subsection{Bethe equations}\label{sec:BEmodeII}

For this model, the Bethe equations for $\{u_j^{(l)}\}$
have two sources, depending on
whether $l$ is smaller or larger than $k_0-k_{d-1}$.  The first set of
Bethe equations, as in model \RN{2}, comes from shifting $l\mapsto l-1$
in Eq. \eqref{LamauxmodelA} and requiring the vanishing of residues
at the poles $u=\gamma_{l-1}u_i^{(l)}$.  This leads to
\begin{align}
\prod_{j=1}^{m_{l-1}}\frac{\eta - \gamma_{l-1}(u_{j}^{(l-1)} - 
	\gamma_{l-1} u_{i}^{(l)})}{\gamma_{l-1} (\gamma_{l-1} u_{i}^{(l)} - u_{j}^{(l-1)})}
&= 
\prod_{j \ne i; j=1}^{m_{l}} \frac{\gamma_{l} (u_{j}^{(l)} - 
	u_{i}^{(l)})-\eta}{u_{j}^{(l)} - u_{i}^{(l)} + \eta} \prod_{j=1}^{m_{l+1}}\frac{\eta + u_{j}^{(l+1)} - \gamma_{l} 
	u_{i}^{(l)}}{u_{j}^{(l+1)} - \gamma_{l}  u_{i}^{(l)}}
\,,  \non \\
&\qquad i=1, 2, \ldots, m_{l}\,,  
\qquad l= 1, 2, \ldots, k_{0} - k_{d-1} \,.
\label{BE1modelA}
\end{align}
Let us now turn to the second set of Bethe equations.
By requiring the vanishing of the residues at the poles $u=-u_i^{(k_0-k_{d-1}+1)}$ in 
$\lambda_{\text{aux}}^{(k_0-k_{d-1})}(-u,\{-u_{j}^{(k_0-k_{d-1})}\})$,  we obtain the Bethe equations for $l=k_0-k_{d-1}+1$. 
Similarly, by requiring vanishing residues at the poles 
$u=-u_i^{(l)}$ in $\lambda_{\text{aux}}^{(l-1)}(-u,\{u_{j}^{(l-1)}\})$, we obtain the 
Bethe equations for $l=k_0-k_{d-1}+2,\ldots, k_0-1$. Explicitly, 
the second set of Bethe equations is given by 
\begin{align}
1 &=\lambda_{\text{aux}}^{(l)}(u_i^{(l)};\{u_j^{(l)}\})\prod_{j=1}^{m_{l-1}}\frac{ u_i^{(l)}-\gamma_{l-1}u_j^{(l-1)}}{u_i^{(l)}-\gamma_{l-1}u_j^{(l-1)}+\eta}
\prod_{j \ne i; j=1}^{m_{l}} \frac{ u_{j}^{(l)} - 
	u_{i}^{(l)}-\eta}{u_{j}^{(l)} - u_{i}^{(l)} + \eta}\,, \non\\
&\qquad i=1,\ldots, m_l, \, \qquad l=k_0-k_{d-1}+1, \ldots, k_0-1 \,,
\label{BEset2modelA0} \\
&\lambda_{\text{aux}}^{(l)}(u_i^{(l)};\{u_j^{(l)}\})= \begin{cases}
\prod_{j=1}^{m_{l+1}}\frac{\eta+u_{j}^{(l+1)}-u_i^{(l)}}{u_{j}^{(l+1)}-u_i^{(l)}} & l=k_0-k_{d-1}+1, \ldots, k_0-2\\
1 & l=k_0-1
\end{cases} \,.
\label{BEset2modelA}
\end{align}

Notice that $\gamma_l$ had so far been defined only 
for $l=1,...,k_0-k_{d-1}$, with $\gamma_{k_0-k_{d-1}}=-1$ \eqref{gammas}. 
In \eqref{BEset2modelA0}, we introduced 
$k_{d-1} +1$  additional $\gamma_l$  defined by
\begin{equation}
	\gamma_l=1, \qquad l=k_0-k_{d-1}+1,...,k_0-1 \,.
\end{equation}

\begin{figure}[t]
\begin{center}
\unitlength=2.4pt
\begin{picture}(40,18)
\thicklines 
\put(22,10){\line(1,0){6}}
\put(32,10){\line(1,0){6}} 
\put(42,10){\line(1,0){6}} 
\put(20,10){\circle{4}} 
\put(30,10){\circle{4}}
\put(40,10){\circle{4}} 
\put(50,10){\circle{4}} 
\put(17.69,8.53){\large$\times$} 
\put(27.69,8.53){\large$\times$}
\put(37.69,8.53){\large$\times$}
\end{picture}
\caption{Dynkin diagram for the model I case $\{5,4,3,2\}$
\label{fig:dynkinI}}
\end{center}
\end{figure}
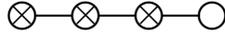

As we did for the Bethe equations of model II \eqref{BAIIcompact}, 
here we can also simplify the Bethe equations \eqref{BE1modelA}-\eqref{BEset2modelA}
using the transformations \eqref{chi} and \eqref{doubletildebetheroots}, 
resulting in simply
\begin{equation}
	(\gamma_{l})^{m_{l}-1} =	\prod_{l'=1}^{k_{0}-1} \prod_{j=1}^{m_{l'}}{}^{'}
	\frac{\tilde{\tilde{u}}^{(l)}_{i} - \tilde{\tilde{u}}^{(l')}_{j} + 
		\frac{\eta}{2}c_{l, l'}}
	{\tilde{\tilde{u}}^{(l)}_{i} - \tilde{\tilde{u}}^{(l')}_{j} - 
		\frac{\eta}{2}c_{l, l'}} \,, \qquad i = 1, \ldots, m_{l}\,,
	\qquad l = 1, \ldots, k_{0}-1\,, 
\label{BAIcompact}
\end{equation}
where the primed product omits the $j=i$ term  
if $l'=l$, and $c_{l, l'}$ 
is given by
\begin{align}
	\text{diagonal:} & &c_{l,l} &= \begin{cases}
		2 \chi_{l} & \text{ if } \quad  \gamma_{l} = +1 \\
		0          & \text{ if } \quad  \gamma_{l} = -1
	\end{cases} \,, \non \\
	\text{off-diagonal:} &  &c_{l,l'} &= \begin{cases}
		-\chi_{l} & \text{ if } \quad  l' = l - 1 \\
		-\chi_{l'} & \text{ if } \quad  l' = l + 1 \\
		\ \ 0        & \text{ otherwise }
	\end{cases} \,,\label{Csmodel1}
\end{align}
where $\chi_{l}$ \eqref{chi} is $\pm 1$.  As an example, the Dynkin diagram 
for the case $\{5,4,3,2\}$ (for which $\gamma_{1} =  \gamma_{2} =  
\gamma_{3} = -1\,, \gamma_{4} = 1$), is shown in Fig. 
\ref{fig:dynkinI}.

Contrary to model II,  the number of Bethe equations for model I is 
equal to the rank. However, the Bethe equations have extra factors of 
$(-1)^{m_l-1}$ in comparison with the usual $\mathfrak{gl}(m|n-m)$ model due to the LHS of 
\eqref{BAIcompact}; this point is discussed further
in Appendix \ref{app:modelAgl111} for the case $n=3$.

We have checked the completeness of this Bethe ansatz solution 
numerically for small values of $L, d, \vec k$  by using 
\eqref{anagenModelA}, \eqref{LamauxmodelA}, \eqref{genglkaux} - \eqref{BEset2modelA}
to solve for the eigenvalues, and comparing with 
corresponding results obtained by direct diagonalization of the transfer matrix, 
see Tables \ref{table:modelA42}, \ref{table:modelA43}, 
\ref{table:modelA432}.
As in the case of model \RN{2}, we observe the existence of infinite Bethe 
roots, as well as singular solutions of the Bethe equations. We also 
find some \emph{continuous} solutions (i.e., with \emph{arbitrary} 
Bethe roots) \footnote{Continuous solutions of Bethe equations have 
been noted previously in the context of the XXZ chain at roots of unity, see 
e.g. \cite{Baxter:1972wg, Fabricius:2000yx, Fabricius:2001yy, 
Baxter:2001sx, Tarasov:2003, Gainutdinov:2015vba, 
Gainutdinov:2016pxy}.}, which here is presumably related to the 
presence of infinite Bethe roots. The transfer-matrix eigenvalues do 
not depend on the values of the arbitrary Bethe roots.

\begin{figure}[h!]
	\hspace{3cm}	\includegraphics[width=10cm]{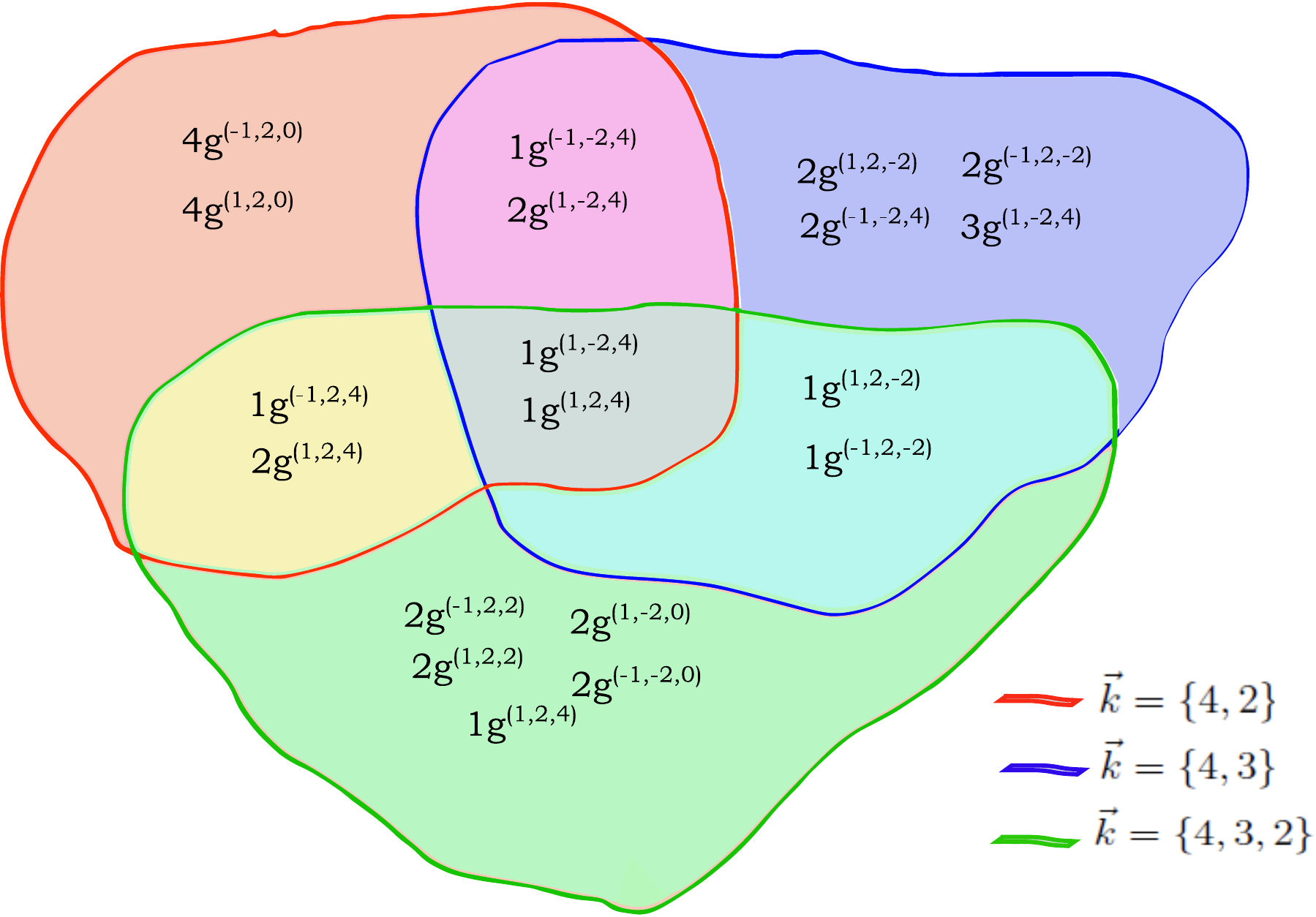}
	\caption{Transfer-matrix eigenvalues for model \RN{1} with $L=2$ 
	and different values of $d$ and $\vec k$, see Eq. 
	\eqref{eq:lambdadef}. Note that some 
	$\text{g}^{(\alpha_1,\alpha_2,\alpha_3)}$ (such as 
	$\text{g}^{(1,-2,4)}$) appear more than once.}
	\label{modelAn=4comparingds}
\end{figure}

Before closing this section, we remark that there is significant 
overlap in the spectra of transfer matrices for different values 
of $d$ and $\vec k$.  This is illustrated for model \RN{1} in Fig. 
\ref{modelAn=4comparingds}, where an eigenvalue $\Lambda(u)$ of the homogeneous transfer
matrix (i.e. all $\theta_j=0$) with $L=2$ is denoted by $q\, 
\text{g}^{(\alpha_1,\alpha_2,\alpha_3)}$, where 
\begin{equation}
\Lambda(u) = \text{g}^{(\alpha_1,\alpha_2,\alpha_3)} \equiv 
(1+u)^2\left(\alpha_1 \eta^2+\alpha_2 \eta u+\alpha_3 u^2\right) \,,
\label{eq:lambdadef}
\end{equation} 
and $q$ is its degeneracy (multiplicity).

\section{Bethe ansatz for model \RN{3}}\label{sec:BAmodelC}

\subsection{Transfer-matrix eigenvalues}\label{sec:BEandTQmodelC}

For model \RN{3}, we recall \eqref{eq:Rmodel2} that not only $k_0$ is fixed as $k_0=n$, but
also $k_{d-1}$ is fixed as $k_{d-1}=2$. The vector $\vec{k}$ is therefore given by
\begin{equation}
\vec{k}=(n,k_1,...,k_{d-2},2), \quad \text{with} \quad n=k_0>k_1>k_2>...>k_{d-2}>k_{d-1}=2.
\end{equation}
There are therefore ${n-3\choose d-2}$ models of \RN{3}${}^+$ type, and an equal number of \RN{3}${}^-$ type.

%\subsection{First level of nesting}\label{sec:Cfirstlevel}

The nested algebraic Bethe ansatz for model \RN{3} can be
performed in a similar way as for models \RN{1} and \RN{2}, the main difference 
appearing in the final step.
As before, we perform the Bethe ansatz for the gauge-transformed R-matrix
\begin{equation}
\tilde{R}^{\RN{3}\pm,\vec{k}}(u)=\left(V\otimes 
V\right) R^{\RN{3}\pm,\vec{k}}(u) \left(V^{-1}\otimes V^{-1}\right) \,,
\label{eq:newRmodelC}
\end{equation}
where $V$ is given in \eqref{Vmat}, and $R^{\RN{3}\pm,\vec{k}}(u)$ is 
given by \eqref{eq:Rmodel2}. The first level of nesting is 
basically the same as for model \RN{1}, also resulting in 
\eqref{eq:nestedresultmodelA};
what changes are the explicit forms of $\mathbb{R}$ and $f(u)$. For 
model \RN{3}${}^{\pm}$, the nesting procedure results in the following rule
\begin{equation}
\mathbb{R}(u)=\begin{cases}
\frac{1}{1-u}\frac{1}{\eta - u}\tilde{R}^{(\RN{3}\pm,k_0-1,k_1,...,k_{d-2},2)}(-u) & \text{if}\quad  k_1<k_0-1 \text{ and } k_0>3\\
\frac{1}{1+u}\frac{1}{\eta + u}\tilde{R}^{(\RN{3}\mp,k_1,...,k_{d-2},2)}(u) & \text{if}\quad  k_1=k_0-1  \text{ and } k_0>3\\
\frac{1}{\eta}r(u,\delta_0)& \text{if}\quad k_0=3  \text{ and }  k_1=2.
\end{cases}\,,
\label{ruleRmodelC}
\end{equation}
with 
\begin{equation}
f(u)=\begin{cases}\frac{-\eta+ u}{u} & \text{if}\quad  k_1<k_0-1 \text{ and } k_0>3\\
\frac{-\eta - u}{u} & \text{if}\quad  k_1=k_0-1 \text{ and } k_0>3\\
-\frac{\eta}{u}& \text{if}\quad k_0=3  \text{ and }  k_1=2
\end{cases}\,,
\label{fmodelC}
\end{equation}
while $g(u)$ is again defined as
\begin{equation}
g(u)=\frac{\eta}{u} \,.
\end{equation}

The nesting procedure ends with $\vec{k}=(3,2)$, for which (see Eq. 
\eqref{ruleRmodelC})
\begin{equation}
\mathbb{R}(u)=\frac{1}{\eta}r(u,\delta_0) \,,
\end{equation}
where 
\begin{equation}
r(u,\delta_0)=\begin{pmatrix}
\eta & 0 & 0 & \delta_0  u\\
0 & 0 & \eta-\delta_0  u & 0\\
0 & \eta-\delta_0  u & 0 & 0\\
\delta_0  u & 0 & 0 & \eta
\end{pmatrix}\,,
\label{eq:rmatrixunusual6v}
\end{equation}
and
\begin{equation}
\delta_0=\begin{cases}
-1 & \text{ if } \quad \vec{k}=\{3,2\}^+\\
+1 & \text{ if } \quad \vec{k}=\{3,2\}^-
\end{cases} \,.
\end{equation}
We bring this R-matrix to the usual six-vertex form by a basis transformation
\begin{align}
\tilde{r}(u,\delta_0)& =(U\otimes U)\, r(u,\delta_0)\, (U^{-1}\otimes U^{-1})\\
&=\begin{pmatrix}
\eta-\delta_0 u & 0 & 0 & 0\\
0 & \delta_0  u & \eta & 0\\
0 & \eta & \delta_0 u & 0\\
0 & 0 & 0 &\eta-\delta_0 u
\end{pmatrix}\,,
\label{eq:rtilde}
\end{align}
where 
\begin{equation}
U=\begin{pmatrix}
1 & i\\
i & 1
\end{pmatrix} \,.
\end{equation}
We henceforth use $\tilde{r}(u,\delta_0)$ instead of $r(u,\delta_0)$.

%\subsection{Transfer-matrix eigenvalues and Bethe equations}\label{sec:BEandTQmodelC}

We again define a sequence of R-matrices
\begin{equation}
\tilde{R}^{(l)}(u) \equiv \tilde{R}^{\mu_{l}, \vec{k}^{(l)}}(u)\,, \qquad 
l = 0, 1, \ldots\,k_0-2 \,,
\end{equation}
where $ \mu_l $ and $ \vec{k}^{(l)} $ are constructed via the iterative procedure
\begin{align}
&\text{if   } k_{1}^{(l-1)} < k_{0}^{(l-1)}-1\,, \quad \text{then  }	
\vec{k}^{(l)} = \vec{k}^{(l-1)} - \vec \epsilon\,, \quad \mu_{l} = 
\mu_{l-1}\,, \quad \gamma_{l}=+1 \,; \non \\
&\text{if   } k_{1}^{(l-1)} = k_{0}^{(l-1)}-1\,, \quad \text{then  }
\vec{k}^{(l)} = \hat{\vec{k}}^{(l-1)}\,, \quad \mu_{l} = 
-\mu_{l-1}\,, \quad \gamma_{l}=-1 \,; 
\label{iterationmodelC}
\end{align}
where $ \vec{\epsilon} $ and $ \hat{\vec{k}} $ are defined as in 
models \RN{1} and \RN{2}.
Examples of such sequences of $\vec{k}^{(l)}$ and $\mu_{l}$ are shown in 
Table \ref{modelC+n=5}.

\begin{table}[h!]
	\setlength{\tabcolsep}{0.5em} % for the horizontal padding
	{\renewcommand{\arraystretch}{0.3}
		{\renewcommand{\arraystretch}{0.5}
			\begin{center}
				\begin{tabular}{c|c|cc|c}
					\multicolumn{5}{c}{\textbf{$ \hspace{0.5cm} $}}\\
					\multicolumn{5}{c}{\textbf{Model \RN{3}$ \mathbf{^+} $  $ (\mathbf{n=5}) $}}\\
					\multicolumn{5}{c}{\textbf{$ \hspace{0.5cm} $}}\\
					\hline
					$ \hspace{0.5cm} $  &$ \hspace{0.5cm} $  & $ \hspace{0.5cm} $ & $ \hspace{0.5cm} $ & $ \hspace{0.5cm} $\\
					$ \hspace{0.5cm}$ &  $ \mathbf{d=2}$ & $ \hspace{0.5cm}$ &  $ \mathbf{d=3}$ & $ \mathbf{d=4}$\\
					$ \hspace{0.5cm} $  &$ \hspace{0.5cm} $  & $ \hspace{0.5cm} $ & $ \hspace{0.5cm} $ & $ \hspace{0.5cm} $\\
					\hline
					$ \hspace{0.5cm} $  &$ \hspace{0.5cm} $  & $ \hspace{0.5cm} $ & $ \hspace{0.5cm} $ & $ \hspace{0.5cm} $\\
					$ \mathbf{l=0} $ &  $ \{5,2\}^+ $ &  $ \{5,4,2\}^+ $ & $ \{5,3,2\}^+ $ & $ \{5,4,3,2\}^+ $\\
					$ \hspace{0.5cm} $  &$ \hspace{0.5cm} $  & $ \hspace{0.5cm} $ & $ \hspace{0.5cm} $ & $ \hspace{0.5cm} $\\
					$ \hspace{0.5cm} $  &$ \big\downarrow $  & $ \big\downarrow $ & $ \big\downarrow $ & $ \big\downarrow $ \\
					$ \hspace{0.5cm} $  &$ \hspace{0.5cm} $  & $ \hspace{0.5cm} $ & $ \hspace{0.5cm} $ & $ \hspace{0.5cm} $\\
					$ \mathbf{l=1} $  & $ \{4,2\}^+ $ & $ \{4,2\}^- $& $ \{4,3,2\}^+ $& $ \{4,3,2\}^-$\\
					$ \hspace{0.5cm} $  &$ \hspace{0.5cm} $  & $ \hspace{0.5cm} $ & $ \hspace{0.5cm} $ & $ \hspace{0.5cm} $\\
					$ \hspace{0.5cm} $  &$ \big\downarrow $  & $ \big\downarrow $ & $ \big\downarrow $ & $ \big\downarrow $ \\
					$ \hspace{0.5cm} $  &$ \hspace{0.5cm} $  & $ \hspace{0.5cm} $ & $ \hspace{0.5cm} $ & $ \hspace{0.5cm} $\\
					$ \mathbf{l=2} $  & $ \{3,2\}^+ $ & $ \{3,2\}^- $& $ \{3,2\}^- $& $ \{3,2\}^+$\\
					$ \hspace{0.5cm} $  &$ \hspace{0.5cm} $  & $ \hspace{0.5cm} $ & $ \hspace{0.5cm} $ & $ \hspace{0.5cm} $\\
					$ \hspace{0.5cm} $  &$ \big\downarrow $  & $ \big\downarrow $ & $ \big\downarrow $ & $ \big\downarrow $ \\
					$ \hspace{0.5cm} $  &$ \hspace{0.5cm} $  & $ \hspace{0.5cm} $ & $ \hspace{0.5cm} $ & $ \hspace{0.5cm} $\\
					$ \mathbf{l=3} $  & 6-vertex  & 6-vertex & 6-vertex & 6-vertex
				\end{tabular}
	\end{center}}}
	\caption{$\vec k^{(l)}$ and $\mu_{l}$ for $l=0, 1, \ldots, 
	k_{0}-2$ for model \RN{3} with $\mu_{0}=+1$ and $n=5$.}
	\label{modelC+n=5}
\end{table}

The $\gamma_{l}$'s again satisfy \eqref{gammas}, but only for $l=1, 
\ldots, k_0-3$. For the final two $l$-values, we define 
\begin{equation}
	\gamma_l=-1\,, \qquad l=k_0-2\,, k_0-1
\end{equation}
for later convenience, see \eqref{BE2modelC}, \eqref{BE3modelC}.

Also
\begin{align}
\mathbb{R}^{(l)}(u) &= \begin{cases}
\frac{1}{(1-\gamma_{l} u)(\eta-\gamma_{l} u)}\tilde{R}^{(l)}(-\gamma_{l} u) & l= 1, 2, \ldots, 
k_{0}-3 \\[0.1in]
\frac{1}{\eta}\tilde{r}(u,\delta_0) & l= k_{0}-2
\end{cases} \,,  \label{RRModelC}\\
f^{(l)}(u) &=\begin{cases}\frac{-\eta + \gamma_{l} u}{u} &  l=1,...,k_0-3\\
\frac{-\eta}{u} & l=k_0-2
\end{cases} \,,
\label{flModelC}
\end{align}
where $\tilde{r}(u,\delta_0)$ was defined in \eqref{eq:rtilde}.

Transfer matrices can be constructed as in \eqref{transftildmodelA}
\begin{equation}
t^{(l)}(u; \{u_{j}^{(l)}\}) = 
\tr_{0}\tilde{R}^{(l)}_{01}(u-u_1^{(l)})\,\ldots\,\tilde{R}^{(l)}_{0 
	m_{l}}(u-u_{m_{l}}^{(l)}) \,, \qquad l = 0, 1, \ldots, 
k_{0}-2\,,
\label{transftildmodelC}
\end{equation}
whose eigenvalues $\Lambda^{(l)}(u; \{u_{j}^{(l)}\})$ are given, as 
in \eqref{anagenModelA}, by
\begin{align}
&\Lambda^{(l)}(u; \{u_{j}^{(l)}\}) = \prod_{j=1}^{m_{l}}(\eta+u-u_j^{(l)}) (1+u-u_j^{(l)})
\prod_{j=1}^{m_{l+1}}\frac{\eta+u_j^{(l+1)}-u}{u_j^{(l+1)}-u} \non  \\
&+ \Lambda_{\text{aux}}^{(l+1)}(u; \{u_{j}^{(l+1)}\})
\prod_{j=1}^{m_{l}}(u-u_j^{(l)})(1+u-u_j^{(l)})
\prod_{j=1}^{m_{l+1}}f^{(l+1)}(u_j^{(l+1)}-u) \,, \non \\
&\qquad\qquad\qquad\qquad\qquad l = 0, 1, \ldots \,, k_0-2 \label{eigenmodelC}
\end{align}
%Our goal here is to compute $ \Lambda^{(0)}(u; \{u_{j}^{(0)}\})$ where $m_0$ and $u_{j}^{(0)}$ are like in \eqref{initial}. 
As before, $\Lambda_{\text{aux}}^{(l)}(u; \{u_{j}^{(l)}\})$ are eigenvalues of
$t_{\text{aux}}^{(l)}(u; \{u_{j}^{(l)}\})$, which is defined by
\begin{align}
t_{\text{aux}}^{(l)}(u; \{u_{j}^{(l)}\}) &= 
\tr_{0}\mathbb{R}^{(l)}_{01}(u_1^{(l)}-u)\,\ldots\,\mathbb{R}^{(l)}_{0 m_{l}}(u_{m_{l}}^{(l)}-u) \,, \non\\
&=\begin{cases}\left(\prod_{j=1}^{m_{l}}\frac{1}{(1-\gamma_{l} (u_{j}^{(l)}-u))
	(\eta-\gamma_{l} (u_{j}^{(l)}-u))}\right)\,
t^{(l)}(\gamma_{l} u; \{\gamma_{l} u_{j}^{(l)}\}) \,, & l=1,...,k_0-3\\
\frac{1}{\eta^{m_l}}t^{(k_0-2)}(-u,\{-u_j^{(k_0-2)}\}) & l=k_0-2
\end{cases}\,,
\label{transfRRmodelC}
\end{align}
where we used \eqref{RRModelC}, and 
$t^{(k_0-2)}(u,\{u_j^{(k_0-2)}\})$ is given by
\begin{align}
t^{(k_0-2)}(u; \{u_{j}^{(k_0-2)}\}) & = 
\tr_{0}\tilde{r}^{(k_0-2)}_{01}(u-u_1^{(k_0-2)},\delta_0)\,\ldots\,\tilde{r}^{(k_0-2)}_{0 
	m_{k_0-2}}(u-u_{m_{k_0-2}}^{(k_0-2)},\delta_0) \,,
\label{transfer6vertex}
\end{align} 
which has eigenvalues\footnote{We used at this step the usual 
algebraic Bethe ansatz, since $\tilde{r}(u,\delta_0)$ is of 
six-vertex form.} 
\begin{align}
\Lambda^{(k_0-2)}(u,\{u_j^{(k_0-2)}\})=&\prod_{j=1}^{m_{k_0-2}}\left(\eta-\delta_0 (u-u_j^{(k_0-2)})\right)\prod_{j=1}^{m_{k_0-1}}\frac{\eta+\delta_0 (u-u_j^{(k_0-1)})}{-\delta_0 (u-u_j^{(k_0-1)})}\non\\
&+\prod_{j=1}^{m_{k_0-2}}\left(\delta_0 (u-u_j^{(k_0-2)})\right)\prod_{j=1}^{m_{k_0-1}}\frac{\eta-\delta_0 (u-u_j^{(k_0-1)})}{\delta_0 (u-u_j^{(k_0-1)})}.
\label{eq:eigenmodelCk0-2}
\end{align}

The auxiliary eigenvalues are therefore 
\begin{equation}
\Lambda_{\text{aux}}^{(l)}(u; \{u_{j}^{(l)}\})
=\begin{cases}\left(\prod_{j=1}^{m_{l}}\frac{1}{(1-\gamma_{l} (u_{j}^{(l)}-u))
	(\eta-\gamma_{l} (u_{j}^{(l)}-u))}\right)\,
\Lambda^{(l)}(\gamma_{l} u; \{ \gamma_{l} u_{j}^{(l)}\}) & l=1,...,k_0-3\\
\frac{1}{\eta^{m_{k_0-2}}}\Lambda^{(k_0-2)}(-u; \{- u_{j}^{(k_0-2)}\}) & l=k_0-2
\end{cases} \,.
\label{lambauxmodelC}
\end{equation}
More explicitly, for $l=1, \ldots, k_{0}-4$, the auxiliary eigenvalues are given by
\begin{align}
&\Lambda_{\text{aux}}^{(l)}(u; \{u_{j}^{(l)}\})
= \prod_{j=1}^{m_{l+1}}\frac{\eta+ u_j^{(l+1)}-\gamma_{l} u}{u_j^{(l+1)}- \gamma_{l} u}\non \\
&+ \Lambda_{\text{aux}}^{(l+1)}(\gamma_{l} u; \{u_{j}^{(l+1)}\})
\prod_{j=1}^{m_{l}}\frac{\gamma_{l}(u-u_j^{(l)})}{\eta + \gamma_{l} (u-u_j^{(l)})}
\prod_{j=1}^{m_{l+1}}\frac{-\eta+\gamma_{l+1} (u_j^{(l+1)}-\gamma_{l} u)}
{u_j^{(l+1)}-\gamma_{l} u} \,, \non \\
&\qquad\qquad l=1, \ldots, k_{0}-4\,;
\label{LamauxmodelC}
\end{align}
while for $l=k_0-3$ we have
\begin{align}
&\Lambda_{\text{aux}}^{(k_0-3)}(u; \{u_{j}^{(k_0-3)}\})
= \prod_{j=1}^{m_{k_0-2}}\frac{\eta + u_j^{(k_0-2)}-\gamma_{k_0-3} u}{u_j^{(k_0-2)}- \gamma_{k_0-3} u}\non \\
&+ \Lambda_{\text{aux}}^{(k_0-2)}(\gamma_{k_0-3}u; \{u_{j}^{(k_0-2)}\})
\prod_{j=1}^{m_{k_0-3}}\frac{\gamma_{k_0-3} (u-u_j^{(k_0-3)})}{\eta + \gamma_{k_0-3}(u-u_j^{(k_0-3)})}
\prod_{j=1}^{m_{k_0-2}}\frac{-\eta}{u_j^{(k_0-2)}- \gamma_{k_0-3}u} \,,
\label{LamauxmodelCb}
\end{align}
and finally for $l=k_0-2$
\begin{align}
\Lambda_{\text{aux}}^{(k_0-2)}(\gamma_{k_0-3}u; 
\{u_{j}^{(k_0-2)}\})=&\prod_{j=1}^{m_{k_0-2}}\frac{\eta+\delta_0 
(\gamma_{k_0-3}u-u_j^{(k_0-2)})}{\eta}\prod_{j=1}^{m_{k_0-1}}\frac{\eta-\delta_0 (\gamma_{k_0-3}u+u_j^{(k_0-1)})}{\delta_0 (\gamma_{k_0-3}u+u_j^{(k_0-1)})}\non\\
&+\prod_{j=1}^{m_{k_0-2}}\frac{-\delta_0 
(\gamma_{k_0-3}u-u_j^{(k_0-2)})}{\eta}\prod_{j=1}^{m_{k_0-1}}\frac{\eta+\delta_0 (\gamma_{k_0-3}u+u_j^{(k_0-1)})}{-\delta_0 (\gamma_{k_0-3}u+u_j^{(k_0-1)})} \,.
\label{LamauxmodelCc}
\end{align}
The equations \eqref{LamauxmodelC}-\eqref{LamauxmodelCc} were obtained using \eqref{lambauxmodelC} together with equations \eqref{eigenmodelC} and \eqref{eq:eigenmodelCk0-2}. Having obtained $\Lambda_{\text{aux}}^{(l)}(u,\{u_j^{(l)}\})$ for all values of $l$
\eqref{LamauxmodelC}-\eqref{LamauxmodelCc}, we can use them together 
with \eqref{eigenmodelC} to compute $\Lambda^{(0)}(u,\{u_j^{(0)}\})$.

\subsection{Bethe equations}\label{sec:BEmodeIII}

We now obtain the corresponding Bethe equations. The first set comes from requiring that 
$\Lambda_{\text{aux}}^{(l)}(u;\{u_j^{(l)}\})$ in 
\eqref{LamauxmodelC} have vanishing residues at the poles 
$u=\gamma_lu_i^{(l+1)}$, and then shifting $l\mapsto l-1$
\begin{align}
	\prod_{j=1}^{m_{l-1}}\frac{\eta + u_{i}^{(l)}-\gamma_{l-1}u_{j}^{(l-1)}}{u_{i}^{(l)}-\gamma_{l-1}u_{j}^{(l-1)}}
	&= 
	\prod_{j \ne i; j=1}^{m_{l}} \frac{\gamma_{l} (u_{j}^{(l)} - u_{i}^{(l)})-\eta}
	{u_{j}^{(l)} - u_{i}^{(l)} + \eta} 	\prod_{j=1}^{m_{l+1}}\frac{\eta + u_{j}^{(l+1)} - \gamma_{l} u_{i}^{(l)}}
	{u_{j}^{(l+1)} - \gamma_{l}  u_{i}^{(l)}}
	\,,  \non \\
	&\qquad i=1, 2, \ldots, m_{l}\,, 
	\qquad l= 1, 2, \ldots, k_{0} - 3\,. 
	\label{BE1modelC}
\end{align}

In order to obtain the two remaining sets of Bethe equations, we first 
substitute \eqref{LamauxmodelCc} into \eqref{LamauxmodelCb}, and then 
require the vanishing of the
residues of $\Lambda_{\text{aux}}^{(k_0-3)}(u; \{u_{j}^{(k_0-3)}\})$ at both poles $u=\gamma_{k_0-3}u_i^{(k_0-2)}$ 
and $u=\gamma_{k_0-3}u_i^{(k_0-1)}$. The result is
\begin{align}
\prod_{j=1}^{m_{k_0-3}}\frac{\eta + u_{i}^{(k_0-2)}-\gamma_{k_0-3}u_{j}^{(k_0-3)}}
{u_{i}^{(k_0-2)}-\gamma_{k_0-3}u_{j}^{(k_0-3)}}
&= 
\prod_{j \ne i; j=1}^{m_{k_0-2}} \frac{\delta_0 (u_{i}^{(k_0-2)} - 
	u_{j}^{(k_0-2)})+\eta}{u_{i}^{(k_0-2)} - u_{j}^{(k_0-2)} - \eta}\non \\
&\times 
\prod_{j=1}^{m_{k_0-1}}\frac{\eta+\delta_0 
(\gamma_{k_0-2}u_i^{(k_0-2)}-u_j^{(k_0-1)})}{-\delta_0 (\gamma_{k_0-2}u_i^{(k_0-2)}-u_j^{(k_0-1)})}\,, \non \\
\hspace{3cm} i=1,...,m_{k_0-2} \,,
\label{BE2modelC}
\end{align}
\begin{align}
1&=(-1)^{m_{k_0-2}}\prod_{j=1}^{m_{k_0-2}}\frac{u_i^{(k_0-1)}-\gamma_{k_0-1}u_j^{(k_0-2)}}{ u_i^{(k_0-1)}-\gamma_{k_0-1}u_j^{(k_0-2)}-\delta_0\eta}\prod_{j \ne i; j=1}^{m_{k_0-1}} \frac{u_{i}^{(k_0-1)} - 
	u_{j}^{(k_0-1)}-\delta_0\eta}{u_{i}^{(k_0-1)} - 
	u_{j}^{(k_0-1)} + \delta_0\eta}\,, \non\\
&\hspace{3.5cm} i=1,...,m_{k_0-1} \,.
\label{BE3modelC}
\end{align}

Following a similar procedure as for models I and II (see 
transformations \eqref{chi} and \eqref{doubletildebetheroots}), we 
find that the Bethe equations \eqref{BE1modelC}-\eqref{BE3modelC} can 
be brought to the form
\begin{equation}
	(\gamma_l)^{m_{l}-1}z_l =	\prod_{l'=1}^{k_{0}-1} \prod_{j=1}^{m_{l'}}{}^{'}
	\frac{\tilde{\tilde{u}}^{(l)}_{i} - \tilde{\tilde{u}}^{(l')}_{j} + 
		\frac{\eta}{2}c_{l, l'}}
	{\tilde{\tilde{u}}^{(l)}_{i} - \tilde{\tilde{u}}^{(l')}_{j} - 
		\frac{\eta}{2}c_{l, l'}} \,, \qquad i = 1, \ldots, m_{l}\,,
	\qquad l = 1, \ldots, k_{0}-1\,, 
	\label{simplifiedBEmodelIII}
\end{equation}where the primed product omits the $j=i$ term  
if $l'=l$, and $z_l$ is given by
\begin{equation}
	z_l=\begin{cases}
		1 & \text{ for } \quad  l=1,...,k_0-3\\
		  (-1)^{m_{l+1}}(\gamma_l\delta_0)^{m_{l}-1} & \text{ for } \quad  l=k_0-2\\
		 1 & \text{ for } \quad  l=k_0-1
	\end{cases} \,.
\end{equation}
For the first $ k_0-3 $ values of $ l $, $c_{l, l'}$ is given by the same expressions as in model I (see \eqref{Csmodel1}) for both $ \delta_0=\pm1 $. For $ l=k_0-2 $ and $ l=k_0-1 $, however, they are given by
\begin{align}
	\text{diagonal:} & &c_{l,l} &= \begin{cases}
		(\delta_0+1)\chi_l & \text{ if } \quad  l = k_0-2 \\
		-2\,\delta_0 \chi_{l} & \text{ if } \quad  l =k_0 -1
	\end{cases} \,, \non \\
\text{off-diagonal:} &  &c_{l,l'} &= \begin{cases}
	-\chi_{l} & \text{ if } \quad  l' = l - 1 \text{ and } \quad l=k_0-2 \\
	\delta_0\chi_{l} & \text{ if } \quad  l' = l - 1 \text{ and } \quad l=k_0-1 \\
	\delta_0\chi_{l^\prime} & \text{ if } \quad  l' = l +1\text{ and } \quad l=k_0-2 \\
	0        & \text{ otherwise }
	\end{cases} \,.
\label{Csdefdelta-1}
\end{align} 
In addition to the transformations performed in models I and II, for model III, for $ \delta_0=+1 $ we needed to shift the last Bethe root by $ \tilde{\tilde{u}}^{(k_0-1)}_{i}\rightarrow \tilde{\tilde{u}}^{(k_0-1)}_{i}+\chi_{k_0-1}\eta $ in order to bring the Bethe equations in the form \eqref{simplifiedBEmodelIII}. 

\begin{figure}[t]
\begin{center}
\unitlength=2.4pt
\begin{picture}(40,18)
\thicklines 
\put(22,10){\line(1,0){6}}
\put(32,10){\line(1,0){6}} 
\put(42,10){\line(1,0){6}} 
\put(20,10){\circle{4}} 
\put(30,10){\circle{4}}
\put(40,10){\circle{4}} 
\put(50,10){\circle{4}} 
\put(17.69,8.53){\large$\times$} 
\put(27.69,8.53){\large$\times$}
\end{picture}
\caption{Dynkin diagram for the model III case $\{5,4,3,2\}^{-}$
\label{fig:dynkinIII}}
\end{center}
\end{figure}
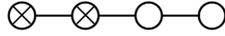

As an example, the Dynkin diagram 
for the case $\{5,4,3,2\}^{-}$ (for which $\gamma_{1} =  \gamma_{2} =  
\gamma_{3} = \gamma_{4} = -1\,, \delta_{0} = 1$), is shown in Fig. 
\ref{fig:dynkinIII}. 
As already noted, the first $k_0-3$ Bethe equations are similar to 
those of model I, but the last two are generally significantly different. 

We have checked, as for models \RN{1} and \RN{2}, the completeness of this Bethe ansatz, see Appendix \ref{app:completeness}.

\section{Discussion and outlook}

In this paper we found a new family of integrable models that we call
flag integrable models.  These models are composed of operators that
act on subspaces which have a flag structure.  Interestingly we find
that our models are rational and are characterized by a sequence of
integers corresponding to the dimensions of the subspaces.

Even though the models have a seemingly simple structure, they exhibit
interesting features.  First, we found that Model I has a symmetry
algebra of a new type.  The symmetry algebra is a generalization of
the usual graded algebra $\mathfrak{gl}(m|n)$ and we correspondingly
call it a generalized graded algebra.  The generators of this algebra
have a different type of grading element in the coproduct and
commutation relations that corresponds to the different stripes of the
flag it connects.  We found a generalization to a Yangian algebra.
Models II and III also exhibit this algebra but are not fixed by it.

Finally, we used nested algebraic Bethe ansatz to determine
the spectrum of Models \RN{1}, \RN{2} and \RN{3}, see Eqs. 
\eqref{Lamaux}-\eqref{BE2}, 
\eqref{anagenModelA}-\eqref{BEset2modelA}, and 
\eqref{eigenmodelC}-\eqref{BE3modelC}, respectively.
Although the Bethe ansatz solution for Model I appears
similar to that of the $\mathfrak{gl}(m|n-m)$ model, we argue
in Appendix \ref{app:modelAgl111}
for the case $n=3$ that these models are not equivalent.
An analysis of examples with small values of $L, d, \vec k$ (see Appendix \ref{app:completeness}) suggests that 
many of the eigenvalues are described by infinite, singular and/or 
continuous Bethe roots. 
For the XXX chain, infinite Bethe roots do not affect transfer-matrix 
eigenvalues, and describe descendant (that is, not $su(2)$ 
highest-weight) states (see, e.g. \cite{Levkovich-Maslyuk:2016kfv}). In contrast, for the models studied here, 
as for those in e.g. \cite{Frahm:2010ei, Frahm:2012eb, Frahm:2019czs, Nepomechie:2019zrx},
the infinite Bethe roots appear to be necessary to obtain the 
spectrum. The appearance of continuous Bethe roots is also unusual. 
Perhaps these features are artifacts of our choices of coordinates 
and gradings, and could be eliminated by working with different choices.

The Bethe equations of all three models can be brought to a simple 
form, similar to those of $\mathfrak{gl}(m|n-m)$,
see \eqref{BAIIcompact}, \eqref{BAIcompact}, and 
\ref{simplifiedBEmodelIII}, respectively. The first $k_0-k_{d-1}-1$
Bethe equations are the same for all the three models, but the 
remaining $k_{d-1}$ equations differ among themselves substantially. Perhaps some of 
the $(-1)^{m}$ factors appearing in these equations could be eliminated by introducing gradings or 
twists, see e.g. \cite{Beisert:2005if}.

There are some interesting further directions that can be pursued.
The physical properties (phase structure, ground state, low-lying 
excitations) of the models presented here remain to be explored.
It would be interesting to see if a universal $R$-matrix could be
formulated along the lines as was done for $\mathfrak{gl}(n)$ and
$\mathfrak{gl}(m|n)$ \cite{Khoroshkin:1994uk,Cai:1997zz,Rej:2010mu}.
It would also be interesting to clarify the remaining symmetries of
model \RN{2}, and to account for its unusual degeneracies. A proper 
treatment of repeated singular solutions in model \RN{2} is still missing.
The trigonometric models found in Appendix \ref{app:Trig}
also warrant further study.

We have restricted our attention to periodic boundary conditions 
(PBC). For the new R-matrices found here,
it would be interesting to find corresponding boundary K-matrices 
(solutions of the boundary Yang-Baxter equation) \cite{Cherednik:1984vvp, Sklyanin:1988yz, Ghoshal:1993tm}, 
to formulate the corresponding open-chain models, and to determine the 
spectrum of their transfer matrices. 
Perhaps there is a choice of 
boundary conditions for which the models have more symmetry compared 
with PBC, which could help account for degeneracies.

\paragraph{Acknowledgements } We would like to thank 
M. Bauer, H. Frahm, C. Paletta and A. Pribytok for discussions, and the 
anonymous referee for many valuable suggestions.  MdL was
supported by SFI, the Royal Society and the EPSRC for funding under
grants UF160578, RGF$\backslash$R1$\backslash$181011,
RGF$\backslash$EA$\backslash$180167 and 18/EPSRC/3590.
RIN was supported in part by a Cooper fellowship.
ALR was supported by the grant 18/EPSRC/3590 and by a UKRI Future Leaders Fellowship (grant number MR/T018909/1).
This research was supported in part by the National Science Foundation under Grant
No.  NSF PHY-1748958. We would like to thank the Kavli Institute for 
Theoretical Physics, where this work was completed, for its hospitality.

%\newpage
\appendix

\section{Comparison between 
$\mathfrak{gl}(k_0-k_1|...|k_{d-2}-k_{d-1}|k_{d-1})$\\
and $\mathfrak{gl}(m|n-m)$}\label{app:modelAgl111}

In this section, we argue that type-\RN{1} models with $d>2$ stripes, 
which have symmetry $\mathfrak{gl}(k_0-k_1|...|k_{d-2}-k_{d-1}|k_{d-1})$, are not 
equivalent to models with two stripes ($d=2$) that have symmetry $\mathfrak{gl}(m|n-m)$. 
For simplicity, we focus on the case $n=3$. We first make the 
argument in Sec. \ref{app:modelAgl111R} for the R-matrices, 
and then in Sec. \ref{app:modelAgl111BA} for the corresponding Bethe ans\"atze.

\subsection{R-matrices}\label{app:modelAgl111R}

In addition to $\mathfrak{gl}(3)$, there are three type-\RN{1} R-matrices 
with $n=3$, as given in Table \ref{table:n=3symmetry}.

\begin{table}[h!]
	\centering
	\begin{tabular}{|c|l|l|}
		\hline 
		$ d $ & $\hspace{1.7cm} \vec{k} $ & Symmetry\\ \hline
		2 & $ k_0=3, \,k_1=2 $& $ \mathfrak{gl}(1|2) $\\
		2 & $ k_0=3, \,k_1=1 $& $ \mathfrak{gl}(2|1) $\\
		3 & $ k_0=3, \,k_1=2,\,k_2=1 $& $ \mathfrak{gl}(1|1|1) $\\\hline
	\end{tabular}
\caption{Type-\RN{1} R-matrices with $n=3$.}\label{table:n=3symmetry}
\end{table} 
\noindent

We now argue that the $\mathfrak{gl}(1|1|1)$ type-\RN{1} R-matrix 
cannot be mapped to $\mathfrak{gl}(2|1)$ or $\mathfrak{gl}(1|2)$ 
R-matrices. Our argument consists of two parts:

\begin{enumerate}
	\item Showing that the eigenvalues are different, which implies 
	that the R-matrices cannot be related by similarity transformations.
	\item Showing that the R-matrices cannot be related by
	generalized (Drinfeld) twists.
\end{enumerate}

The eigenvalues and the corresponding degeneracies of the three R-matrices, 
displayed in Table \ref{table:spectrumn=3}, are evidently all
different.

\begin{table}[h!]
	\centering
	\begin{tabular}{|l|c|c|}\hline
		Model & Eigenvalues & Degeneracies\\
		\hline
		$\mathfrak{gl}(1|2)$ &  $-(1 + u)(u + \eta)$ & 1   \\ 
		&   $(1 + u) (u - \eta)$ & 2 \\
		&    $-(1 + u) (u - \eta)$ & 3 \\
		&    $ (1 + u) (u + \eta)$ & 3 \\
		\hline
		$\mathfrak{gl}(2|1)$ & $(1 + u) (u - \eta)$ & 3 \\
		&    $-(1 + u) (u  - \eta)$ & 1 \\
		&    $(1 + u) (u  + \eta)$ & 5 \\
		\hline
		$\mathfrak{gl}(1|1|1)$ & $-(1 + u) (u + \eta)$ & 1 \\
		&   $(1 + u) (u - \eta)$ &  2 \\
		&   $-(1 + u) (u - \eta)$ &  2 \\
		&   $(1 + u) (u + \eta)$ &  4 \\
	\hline
	\end{tabular}
\caption{Spectra of type-\RN{1} R-matrices with $n=3$.}\label{table:spectrumn=3}
\end{table}

Let us now check if the models can be related by a generalized twist
\begin{equation}
\tilde{R}_{12}(u)=W_{21}\, R_{12}(u)\, W_{12}^{-1} \,,
\label{eq:twist}
\end{equation}
with
\begin{equation}
\left[R_{12},W_{13}W_{23}\right]=\left[R_{23},W_{13}W_{12}\right]=0 \,.
\label{eq:twistcondition}
\end{equation}
We observe that the three R-matrices, as well as the one for 
$\mathfrak{gl}(3)$, satisfy
\begin{equation}
R(\Delta \mathfrak{h}_1)=(\Delta^{\text{op}} \mathfrak{h}_1)R \quad \text{and} \quad 
R(\Delta  \mathfrak{h}_2)=(\Delta^{\text{op}} \mathfrak{h}_2)R \,,
\end{equation}
where $\mathfrak{h}_i $ are the $\mathfrak{gl}(3)$  diagonal generators:
\begin{equation}
 \mathfrak{h}_1=e_{1,1}-e_{2,2} \quad \text{and} \quad  
 \mathfrak{h}_2=e_{2,2}-e_{3,3} \,,
\end{equation}
and 
\begin{equation}
\Delta \mathfrak{h}_i=\mathfrak{h}_i\otimes 1+1\otimes \mathfrak{h}_i 
\,.
\end{equation}
Therefore, if a twist mapping these models exists, it has to satisfy 
\begin{equation}
\left[W,\Delta\mathfrak{h}_i\right]=0\,, \quad i=1\,, 2 \,.
\label{eq:totalspinforW}
\end{equation}
Starting with a general $ 9\times 9 $ matrix $W$ and requiring that 
\eqref{eq:totalspinforW} be satisfied, we obtain that $W$ must be of ice-rule form:
\begin{equation}
W=w_{2,4}e_{2,4}+w_{4,2}e_{4,2}+w_{3,7}e_{3,7}+w_{7,3}e_{7,3}+w_{6,8}e_{6,8}+w_{8,6}e_{8,6}+\sum_{i=1}^{9}w_{i,i}e_{i,i} \,.
\label{eq:restrictedW}
\end{equation}
We readily find that the twist equation \eqref{eq:twist} cannot be satisfied with $W$ of 
the form \eqref{eq:restrictedW} and with $\tilde{R}$ and $R$ 
corresponding to $\mathfrak{gl}(1|1|1)$ and $\mathfrak{gl}(1|2)$ (or 
$\mathfrak{gl}(2|1)$). Notice that a twist is ruled out even before considering
\eqref{eq:twistcondition}.

Gauge transformations (which are particular types of similarity 
transformations) and twists (including generalized twists like
the one above) are the two types of transformations known to preserve
the quantum Yang-Baxter equation.  Since the R-matrix for 
$\mathfrak{gl}(1|1|1)$ is not related to the R-matrices for 
$\mathfrak{gl}(1|2)$ or $\mathfrak{gl}(2|1)$ by such
transformations, we believe it is new.  For more details about
this R-matrix, see section \ref{sec:SymModelA}.

% We have also studied the spectrum of the
% transfer matrices for periodic boundary conditions as well as the
% spectrum of the Hamiltonians for open spin chain with free boundary
% conditions (for a few sites).  For periodic boundary conditions, when
% compared, the spectrum of the transfer matrices for these models have
% very different spectra.  For the Hamiltonians with free boundary
% conditions, we looked into all the models for $ n=3 $ and $ n=6 $ and
% several of them have the same spectrum, although many have different
% spectrum.  We believe this is hiding some interesting property
% regarding the allowed integrable boundary conditions.  It would be
% interesting to classify all the reflection matrices for these models
% and study their corresponding transfer matrices spectrum to best
% clarify this point.

\subsection{Bethe ansatz}\label{app:modelAgl111BA}

The transfer-matrix eigenvalues for the case with symmetry
$\mathfrak{gl}(1|2)$, which corresponds to model \RN{1} with $d=2$ 
and $\vec{k}=\{3,2\}$, is given by
\begin{align}
	\Lambda^{(1|2)}(u;\theta_j)=&\prod_{j=1}^{L}(1+u-\theta_j)\left[\prod_{j=1}^{L}(u-\theta_j+\eta)\prod_{j=1}^{m_1}\frac{u-u_j^{(1)}-\eta}{u-u_j^{(1)}}\right.\nonumber\\
	& 
	\left.+(-1)^{m_1}\prod_{j=1}^{L}(u-\theta_j)\left(\prod_{j=1}^{m_1}\frac{u-u_j^{(1)}-\eta}{u-u_j^{(1)}}\prod_{j=1}^{m_2}\frac{u+u_j^{(2)}+\eta}{u+u_j^{(2)}}+\prod_{j=1}^{m_2}\frac{u+u_j^{(2)}-\eta}{u+u_j^{(2)}}\right)\right] \,,
	\label{eq:eigenvaluegl12}
\end{align}
see Eqs. \eqref{anagenModelA}, \eqref{genglkaux}, \eqref{eq:Lambdak0-kdm1v2}.
The corresponding Bethe equations are given by
\begin{align}
	\prod_{j=1}^{L}\frac{u_k^{(1)}-\theta_j+\eta}{u_k^{(1)}-\theta_j}&=(-1)^{m_1+1}\prod_{j=1}^{m_2}\frac{u_j^{(2)}+u_k^{(1)}+\eta}{u_j^{(2)}+u_k^{(1)}}\,,\quad k=1,...,m_1\,, \label{eq:BEgl12a}\\
	1&=\prod_{j=1}^{m_1}\frac{u_j^{(1)}+u_k^{(2)}}{u_j^{(1)}+u_k^{(2)}+\eta}\prod_{j\neq k,j=1}^{m_2}\frac{u_j^{(2)}-u_k^{(2)}-\eta}{u_j^{(2)}-u_k^{(2)}+\eta}\,\quad k=1,...,m_2 \,,
	\label{eq:BEgl12b}
\end{align}
see Eqs. \eqref{BE1modelA}-\eqref{BEset2modelA}. These results can be 
brought to a more symmetric form by performing the redefinitions
\begin{equation}
	u_i^{(1)} \mapsto u_i^{(1)} - \frac{\eta}{2}\,, \qquad 
	u_i^{(2)} \mapsto -u_i^{(2)} \,, 
	\label{redefinitions}
\end{equation}
leading to 
\begin{align}
	\Lambda^{(1|2)}(u;\theta_j)=&\prod_{j=1}^{L}(1+u-\theta_j)\Bigg\{
	(-1)^{m_1}\prod_{j=1}^{L}(u-\theta_j)\prod_{j=1}^{m_2}\frac{u-u_j^{(2)}-\eta}{u-u_j^{(2)}}
	\nonumber\\
	&+ \prod_{j=1}^{m_1}\frac{u-u_j^{(1)}-\frac{\eta}{2}}{u-u_j^{(1)}+\frac{\eta}{2}}\Big[
	\prod_{j=1}^{L}(u-\theta_j+\eta) + (-1)^{m_1}\prod_{j=1}^{L}(u-\theta_j)
	\prod_{j=1}^{m_2}\frac{u-u_j^{(2)}+\eta}{u-u_j^{(2)}}\Big]\Bigg\} \,,
	\label{eq:eigenvaluegl12p}
\end{align}
and
\begin{align}
	\prod_{j=1}^{L}\frac{u_k^{(1)}-\theta_j+\frac{\eta}{2}}{u_k^{(1)}-\theta_j-\frac{\eta}{2}}
	&=(-1)^{m_1+1}\prod_{j=1}^{m_2}\frac{u_k^{(1)}-u_j^{(2)}+\frac{\eta}{2}}{u_k^{(1)}-u_j^{(2)}-\frac{\eta}{2}}\,,
	\quad k=1,...,m_1\,, \label{eq:BEgl12ap}\\
	1&=\prod_{j=1}^{m_1}\frac{u_k^{(2)}-u_j^{(1)}+\frac{\eta}{2}}{u_k^{(2)}-u_j^{(1)}-\frac{\eta}{2}}
	\prod_{j\neq k,j=1}^{m_2}\frac{u_k^{(2)}-u_j^{(2)}-\eta}{u_k^{(2)}-u_j^{(2)}+\eta}\,, \quad k=1,...,m_2 \,,
	\label{eq:BEgl12bp}
\end{align}
respectively.

The transfer-matrix eigenvalues for the case with symmetry 
$\mathfrak{gl}(1|1|1)$, which corresponds to model \RN{1} with 
$d=3$ and $\vec{k}=\{3,2,1\}$, is given by\footnote{This case is not 
included in section \ref{sec:BAmodelA}, since there we restrict for 
simplicity to the cases with $k_{d-1}>1$.}
\begin{align}
	\Lambda^{(1|1|1)}(u;\theta_j)=&\prod_{j=1}^{L}(1+u-\theta_j)\left[\prod_{j=1}^{L}(u-\theta_j+\eta)\prod_{j=1}^{m_1}\frac{u-u_j^{(1)}-\eta}{u-u_j^{(1)}}\right.\nonumber\\
	& 
	\left.+(-1)^{m_1}\prod_{j=1}^{L}(u-\theta_j)\left(\prod_{j=1}^{m_1}\frac{u-u_j^{(1)}-\eta}{u-u_j^{(1)}}+(-1)^{m_2}\right)\prod_{j=1}^{m_2}\frac{u-u_j^{(2)}+\eta}{u-u_j^{(2)}}\right] \,,
	\label{eq:eigenvaluegl111}
\end{align}
with corresponding Bethe equations
\begin{align}
	 \prod_{j=1}^{L}\frac{u_k^{(1)}-\theta_j+\eta}{u_k^{(1)}-\theta_j}&=(-1)^{m_1+1}\prod_{j=1}^{m_2}\frac{u_j^{(2)}-u_k^{(1)}-\eta}{u_j^{(2)}-u_k^{(1)}}\,,\quad k=1,...,m_1\,, \label{eq:BEgl111a}\\
	 1&=(-1)^{m_2+1}\prod_{j=1}^{m_1}\frac{u_j^{(1)}-u_k^{(2)}}{u_j^{(1)}-u_k^{(2)}+\eta}\,,\quad k=1,...,m_2\,. \label{eq:BEgl111b}
\end{align}
Using the redefinition
\begin{equation}
	u_i^{(1)} \mapsto u_i^{(1)} - \frac{\eta}{2}\,, 
	\label{redefinitions2}
\end{equation}
these results become
\begin{align}
	\Lambda^{(1|1|1)}(u;\theta_j)=&\prod_{j=1}^{L}(1+u-\theta_j)\Bigg\{  	
	(-1)^{m_1+m_{2}}\prod_{j=1}^{L}(u-\theta_j)\prod_{j=1}^{m_2}\frac{u-u_j^{(2)}+\eta}{u-u_j^{(2)}}+\nonumber\\
	&
	+\prod_{j=1}^{m_1}\frac{u-u_j^{(1)}-\frac{\eta}{2}}{u-u_j^{(1)}+\frac{\eta}{2}}\Big[
	\prod_{j=1}^{L}(u-\theta_j+\eta) + (-1)^{m_1}\prod_{j=1}^{L}(u-\theta_j)
	\prod_{j=1}^{m_2}\frac{u-u_j^{(2)}+\eta}{u-u_j^{(2)}}\Big]	\Bigg\} \,,
	\label{eq:eigenvaluegl111p}
\end{align}
and
\begin{align}
	\prod_{j=1}^{L}\frac{u_k^{(1)}-\theta_j+\frac{\eta}{2}}{u_k^{(1)}-\theta_j-\frac{\eta}{2}}
	&=(-1)^{m_1+1}\prod_{j=1}^{m_2}\frac{u_k^{(1)}-u_j^{(2)}+\frac{\eta}{2}}{u_k^{(1)}-u_j^{(2)}-\frac{\eta}{2}}\,,
	\quad k=1,...,m_1\,, \label{eq:BEgl111ap}\\
	1&=(-1)^{m_2+1}\prod_{j=1}^{m_1}\frac{u_k^{(2)}-u_j^{(1)}-\frac{\eta}{2}}{u_k^{(2)}-u_j^{(1)}+\frac{\eta}{2}}
	\,, \quad k=1,...,m_2 \,,
	\label{eq:BEgl111bp}
\end{align}
respectively.

% From the expressions above, one can see that there is no shift or 
% redefinition of the Bethe roots that would map the results from 
% $\mathfrak{gl}(1|2)$ to the ones for $\mathfrak{gl}(1|1|1)$.

\subsubsection{Remark about gradings}

As already remarked, all the R-matrices in this paper satisfy the non-graded 
(ordinary) Yang-Baxter equation.

Let us compare our non-graded $\mathfrak{gl}(1|2)$ results with corresponding results
obtained using a graded R-matrix in the FFB grading with reference 
state $(0,0,1)^{\otimes L}$ \cite{Lai:1974, Essler:1992nk}. Setting in 
\eqref{eq:eigenvaluegl12p} $\eta=i$ and all inhomogeneities $\theta_j$ to zero, we obtain
% \begin{align}
% \Lambda(u)=&(1+u)^L\left[(u+i)^L
% \prod_{j=1}^{m_1}\frac{u-u_j^{(1)}-\frac{i}{2}}{u-u_j^{(1)}+\frac{i}{2}}\right.\nonumber\\
% & \left.+(-1)^{m_1} u^L \prod_{j=1}^{m_1}\frac{u-u_j^{(1)}-\frac{i}{2}}{u-u_j^{(1)}+\frac{i}{2}}
% \prod_{j=1}^{m_2}\frac{u-u_j^{(2)}+i}{u-u_j^{(2)}}
% +(-1)^{m_1} u^L 
% \prod_{j=1}^{m_2}\frac{u-u_j^{(2)}-i}{u-u_j^{(2)}}\right] \,.
% \end{align}
\begin{align}
	\Lambda^{(1|2)}(u;\theta_j=0)=&(1+u)^{L}\Bigg\{
	\prod_{j=1}^{m_1}\frac{u-u_j^{(1)}-\frac{i}{2}}{u-u_j^{(1)}+\frac{i}{2}}\Big[
	(u+i)^{L} + (-1)^{m_1}u^{L}
	\prod_{j=1}^{m_2}\frac{u-u_j^{(2)}+i}{u-u_j^{(2)}}\Big]\nonumber\\
	& + 	
	(-1)^{m_1}u^{L}\prod_{j=1}^{m_2}\frac{u-u_j^{(2)}-i}{u-u_j^{(2)}}\Bigg\} \,.
\end{align}
Apart from differences in conventions and notations, this result is 
the same as (3.50) in \cite{Essler:1992nk}, {\em except} that the latter 
has $-1$ in place of our factors $(-1)^{m_1}$, which can be 
attributed to the fact that our R-matrix is not graded. Indeed, a 
similar phenomenon can be seen in the $OSp(1|2)$ model 
\cite{Kulish:1985, Martins:1994nq, Nepomechie:2019zrx}, 
compare in \cite{Nepomechie:2019zrx} the non-graded result (2.16) 
that has $(-1)^{m}$ factors  vs. the corresponding graded-result (3.12) 
that does not have such factors.

\subsubsection{Fermionic duality transformation}

It is interesting to investigate whether the above Bethe ansatz results for 
$\mathfrak{gl}(1|2)$ and $\mathfrak{gl}(1|1|1)$ can be related by a 
fermionic duality transformation. Following the approach in 
\cite{Gohmann:2003} (see also \cite{Beisert:2005di} and references 
therein), we define the polynomial $P(u)$
\begin{equation}
	P(u) = \prod_{k=1}^{L}(u - \theta_{k} - \frac{\eta}{2}) 
	\prod_{k=1}^{m_{2}}(u - u^{(2)}_{k} + \frac{\eta}{2}) - 
	(-1)^{m_{1}+1} \prod_{k=1}^{L}(u - \theta_{k} + \frac{\eta}{2}) 
	\prod_{k=1}^{m_{2}}(u - u^{(2)}_{k} - \frac{\eta}{2}) \,,
	\label{Pdef}
\end{equation}
in terms of which the first $\mathfrak{gl}(1|2)$ Bethe equation
\eqref{eq:BEgl12ap} becomes
\begin{equation}
	P(u^{(1)}_{k}) = 0 \,, \qquad k =1, \ldots, m_{1} \,.
\end{equation}
Since $P(u)$ has (for $m_{1}$ even) degree $L+m_{2}$, it has 
$m' = L+m_{2}-m_{1}$ additional zeros
\begin{equation}
	P(u'_{k}) = 0 \,, \qquad k =1, \ldots, m' \,.
	\label{uprime}
\end{equation}
If we identify $u' \leftrightarrow u^{(1)}$, then \eqref{uprime}
is the same as the first $\mathfrak{gl}(1|1|1)$ Bethe equation 
\eqref{eq:BEgl111ap}, except for $(-1)^{m}$ factors.
We see that $P(u)$ has the factorized form
\begin{equation}
	P(u) \propto \prod_{k=1}^{m_{1}}(u - u^{(1)}_{k}) 
	\prod_{k=1}^{m'}(u - u'_{k})\,.
	\label{Pfac}
\end{equation}
We observe that
\begin{align}
	\frac{P(u^{(2)}_{j} + \frac{\eta}{2})}{P(u^{(2)}_{j} - 
	\frac{\eta}{2})} &= (-1)^{m_{1}+1} \prod_{k\ne j}^{m_{2}}
	\frac{u^{(2)}_{j} - u^{(2)}_{k} + \eta}{u^{(2)}_{j} - u^{(2)}_{k} - \eta} \non \\
	&= \prod_{k=1}^{m_{1}}\frac{u^{(2)}_{j} - u^{(1)}_{k}+ 
	\frac{\eta}{2}}{u^{(2)}_{j} - u^{(1)}_{k} - \frac{\eta}{2}} 
	\prod_{k=1}^{m'}\frac{u^{(2)}_{j} - u'_{k}+ 
	\frac{\eta}{2}}{u^{(2)}_{j} - u'_{k} - \frac{\eta}{2}} \,, \qquad 
	j = 1, \ldots, m_{2} \,,
	\label{dualid1}
\end{align}
where the first equality follows from \eqref{Pdef}, and the second 
equality follows from \eqref{Pfac}. That is, we have the identity
\begin{equation}
\prod_{k=1}^{m_{1}}\frac{u^{(2)}_{j} - u^{(1)}_{k} + 
\frac{\eta}{2}}{u^{(2)}_{j} - u^{(1)}_{k} - \frac{\eta}{2}} 
\prod_{k\ne j}^{m_{2}}\frac{u^{(2)}_{j} - u^{(2)}_{k} - 
\eta}{u^{(2)}_{j} - u^{(2)}_{k} + \eta} = (-1)^{m_{1}+1} 
\prod_{k=1}^{m'}\frac{u^{(2)}_{j} - u'_{k} - \frac{\eta}{2}}
{u^{(2)}_{j} - u'_{k} + \frac{\eta}{2}}\,, \quad j = 1, \ldots, m_{2} \,.
\label{dualid2}
\end{equation}
The LHS of \eqref{dualid2} coincides with the RHS of the second $\mathfrak{gl}(1|2)$ Bethe equation
\eqref{eq:BEgl12bp}; while the RHS of \eqref{dualid2} is the same as the RHS of the second $\mathfrak{gl}(1|1|1)$ Bethe equation
\eqref{eq:BEgl111bp} if we again identify $u' \leftrightarrow 
u^{(1)}$, except for $(-1)^{m}$ factors. In summary, up to
$(-1)^{m}$ factors, the $\mathfrak{gl}(1|2)$ and 
$\mathfrak{gl}(1|1|1)$ Bethe equations are related by a fermionic 
duality transformation.\footnote{We thank the referee for bringing  
this fact to our attention.}

However, the eigenvalue expressions for $\mathfrak{gl}(1|2)$ \eqref{eq:eigenvaluegl12p} 
and $\mathfrak{gl}(1|1|1)$ \eqref{eq:eigenvaluegl111p}
are {\em not} related in such a way. Indeed, we now observe that
\begin{align}
	\frac{P(u - \frac{\eta}{2})}{P(u + \frac{\eta}{2})} 
	&= \prod_{k=1}^{m_{1}}\frac{u - u^{(1)}_{k} - \frac{\eta}{2}}{u - u^{(1)}_{k} + \frac{\eta}{2}} 
	\prod_{k=1}^{m'}\frac{u - u'_{k} - \frac{\eta}{2}}{u - u'_{k} + \frac{\eta}{2}} \non \\
	&= \frac{\prod_{k=1}^{L}(u - \theta_{k} - \eta) + (-1)^{m_{1}} \prod_{k=1}^{L}(u - \theta_{k})
	\prod_{k=1}^{m_{2}} \frac{u - u^{(2)}_{k} - \eta}{u - u^{(2)}_{k}}}
	{\prod_{k=1}^{L}(u - \theta_{k})
	\prod_{k=1}^{m_{2}} \frac{u - u^{(2)}_{k} + \eta}{u - u^{(2)}_{k}}
	+ (-1)^{m_{1}}\prod_{k=1}^{L}(u - \theta_{k} + \eta)}\,,
	\label{dualid3}
\end{align}
where the first equality follows from \eqref{Pfac}, and the second 
equality follows from \eqref{Pdef}. That is, we have the identity
\begin{align}
& \prod_{k=1}^{m_{1}}\frac{u - u^{(1)}_{k} - \frac{\eta}{2}}{u - u^{(1)}_{k} + \frac{\eta}{2}} 
\left[\prod_{k=1}^{L}(u - \theta_{k} + \eta) + (-1)^{m_{1}}\prod_{k=1}^{L}(u - \theta_{k})
	\prod_{k=1}^{m_{2}} \frac{u - u^{(2)}_{k} + \eta}{u - u^{(2)}_{k}}
	 \right]\non \\
&\qquad = \prod_{k=1}^{m'}\frac{u - u'_{k} + \frac{\eta}{2}}{u - 
u'_{k} - \frac{\eta}{2}} \left[
(-1)^{m_{1}}\prod_{k=1}^{L}(u - \theta_{k} - \eta) +  \prod_{k=1}^{L}(u - \theta_{k})
	\prod_{k=1}^{m_{2}} \frac{u - u^{(2)}_{k} - \eta}{u - 
	u^{(2)}_{k}} \right] \,.
	\label{dualid4}
\end{align}
The LHS of \eqref{dualid4} coincides with part of the expression 
for the $\mathfrak{gl}(1|2)$ eigenvalue \eqref{eq:eigenvaluegl12p}. 
However, after the identification  $u' \leftrightarrow u^{(1)}$,
the RHS of \eqref{dualid4} does not appear to be related to the 
$\mathfrak{gl}(1|1|1)$ eigenvalue \eqref{eq:eigenvaluegl111p}, which 
has a similar factor but with $\eta \mapsto -\eta$. We 
conclude that the $\mathfrak{gl}(1|2)$ and 
$\mathfrak{gl}(1|1|1)$ models are {\em not} related by a 
fermionic duality transformation.

\section{Trigonometric solution}\label{app:Trig}

We can generalize our analysis to contain trigonometric
models.  Similar to \cite{Maassarani:1998} these models contain
several constants.  In order to achieve this, we split our matrices
into upper/lower triangular and diagonal parts:
\begin{align}
&\mathbb{P}_+^{(k,n)} = \sum_{i<j}^{k} e_{ij}\otimes e_{ji} \,,
&&\mathbb{P}_-^{(k,n)} = \sum_{i>j}^{k} e_{ij}\otimes e_{ji} \,,
&&\mathbb{P}_0^{(k,n)} = \sum_{i=j}^{k} e_{ij}\otimes e_{ji} \,,\\
&\mathbb{K}_+^{(k,n)} = \sum_{i<j}^{k} e_{ij}\otimes e_{ij} \,,
&&\mathbb{K}_-^{(k,n)} = \sum_{i>j}^{k} e_{ij}\otimes e_{ij} \,,
&&\mathbb{K}_0^{(k,n)} = \sum_{i=j}^{k} e_{ij}\otimes e_{ij} \,,
\end{align}
where $e_{ij}$ are $n \times n$ matrices as before 
\eqref{eq:P}-\eqref{eq:K}.
Notice that $\mathbb{K}_0^{(k,n)} =\mathbb{P}_0^{(k,n)}$, so in our
Ansatz for the Hamiltonian we only need to consider one of them.
Hence, we now consider a Hamiltonian of the form
\begin{align}
\mathcal{H}^{\vec{k}} = \sum_{i=0}^{d-1}\left( a_i \, 
\mathbb{I}^{(k_i,n)} + b^{\pm}_i\,  \mathbb{P}_\pm^{(k_i,n)} + 
b^{0}_i\,  \mathbb{P}_0^{(k_i,n)} + c^\pm_i\, 
\mathbb{K}_\pm^{(k_i,n)} \right) \,.
\end{align}
Obviously, we can recover our previous Ansatz \eqref{HAnsatz} from
this one by putting $b_{i}^{+} = b_{i}^{-} = b_{i}^{0} \equiv b_{i}$, 
and similarly for the $c$'s. Let us now again solve 
this system recursively.

At the highest level, we recover the rational $SO(n)$ type models 
from the previous section. However, when $c^\pm_0=0$, we find the solution
\begin{align}
&b^0_0 = 1 \,, && b^-_0 = x_0 \,, && b^+ =\frac{1}{x_0} \,,
\end{align}
for $x_0$ a constant. In the next step, there are more possibilities that generalize the rational cases. First there is the recurrence step
\begin{align}
&a_1=0\,, && b_1^0 = -2,0\,, && b^-_1 = x_1\,, && b^+_1 = 
-\frac{1}{x_0} + \frac{1}{x_0+x_1}\,, &&c^\pm_1=0 \,.
\end{align}
The second solution is the termination step
\begin{align}
& a_1= \pm 1\,, && b_1^0 = -1\,,  && b^-_1 = -x_0\,, && b^+_1 = 
-\frac{1}{x_0}\,, &&c^\pm_1=0 \,.
\end{align}
For $x_1=1$ we recover the rational solutions. Hence, we are lead to 
the trigonometric generalizations of models \RN{1} and \RN{2}
\begin{align}
\tilde{H}^{\RN{1},\vec{k},\vec{y}}=a_0\,\mathbb{I}^{(n,n)} + 
\sum_{j=0}^{d-1} (-1)^j(1+y_j) \,\mathbb{P}^{(k_j,n)}_0
+
\sum_{j=0}^{d-1}x_j\, \mathbb{P}_-^{(k_j,n)} 
+
\sum_{j=0}^{d-1}\Big[\frac{1}{\sum_{i=0}^{j} 
x_i}-\frac{1}{\sum_{i=0}^{j-1} x_i} \Big]\mathbb{P}_+^{(k_j,n)} \,,
\label{eq:HmodelAtrig}
\end{align}
where the vector $\vec{y} = \{0,\pm1,\pm1,\ldots\}$, where each of 
the signs can be different. We recover the rational model \RN{1}  by setting $x_i = (-1)^i$. Similarly, we find
\begin{align}
\tilde{H}^{\RN{2},\vec{k},\vec{y}}=&\,a_0\,\mathbb{I}^{(n,n)} + 
\sum_{j=0}^{d-2} (-1)^j(1+y_j) \,\mathbb{P}^{(k_j,n)}_0
+
\sum_{j=0}^{d-2}x_j\, \mathbb{P}_-^{(k_j,n)} 
+
\sum_{j=0}^{d-2}\Big[\frac{1}{\sum_{i=0}^{j} x_i}-\frac{1}{\sum_{i=0}^{j-1} x_i} \Big]\mathbb{P}_+^{(k_j,n)}  \nonumber\\
& - (-1)^{d}\mathbb{P}_0^{(k_{d-1},n)} - \big(\sum_{i=0}^{d-2} x_i 
\big) \mathbb{P}^{(k_{d-1},n)}_- - \big(\sum_{i=0}^{d-2} 
\frac{1}{x_i} \big)  \mathbb{P}^{(k_{d-1},n)}_- \pm 
\mathbb{I}^{(k_{d-1},n)} \,.
\label{eq:HmodelBtrig}
\end{align}
The original model \RN{2} can again be easily recovered from this solution.

\section{Completeness checks}\label{app:completeness}

We present here Bethe roots $\{ u^{(l)}_{k} \}$ corresponding to each 
of the eigenvalues of 
the homogeneous transfer matrices (all $\theta_{j} =0$) with small 
values $L, d, \vec k$ for 
model \RN{2} (Tables \ref{table:B+d2k42}, \ref{table:B+d3k432}, 
\ref{table:B-d3k432}),  model \RN{1} (Tables \ref{table:modelA42}, 
\ref{table:modelA43}, \ref{table:modelA432}) and model \RN{3} (Tables 
\ref{table:modelC42}, \ref{table:modelC432}), which serve as 
completeness checks of the Bethe ansatz.
The columns in the tables labeled ``deg'' display the degeneracy 
(multiplicity) of an eigenvalue.
We emphasize the presence of numerous eigenvalues described by infinite, 
singular and/or continuous (arbitrary) Bethe roots.  For model \RN{2}, we 
find instances with repeated singular Bethe roots
(such as the last line of Table \ref{table:B+d3k432}), where
the roots indeed give the eigenvalue through the TQ equation
\eqref{TQ0}, but the Bethe equations are not all satisfied (at 
least naively). For models \RN{1} and \RN{3}, we do not find such Bethe root 
configurations, so their Bethe ans\"atze appear to be complete.

%\subsection{Model B}

\begin{table}[h!]
\centering
\begin{tabular}{|c|c|c|c|c|c|c|}
\hline
$L$ & $m_{1}$ & $m_{2}$ & $p$ & deg & $\{ u^{(1)}_{k} \}$ & $\{ u^{(2)}_{k} \}$\\   
\hline
1 & 0  & 0 & - & 2  & - & - \\
1 & 1  & 1 & 0 & 2  & $\infty$ & $\infty$  \\
\hline
2 & 0  & 0 & - & 3  & - & - \\
2 & 1  & 0 & - & 1  & $-\frac{\eta}{2}$ & -  \\
2 & 1  & 1 & 0 & 4  & $-\frac{\eta}{2}$ & $\infty$  \\
2 & 1  & 1 & 0 & 7  & $\infty$  & $\infty$  \\
2 & 2  & 2 & 1 & 1  & \textcolor{blue}{0, $-\eta$}  & $(-1\pm\frac{1}{\sqrt{2}})\eta$\\
\hline
\end{tabular}
\caption{\small Model \RN{2}$^{+}$ with $d=2$, $\vec k = 
\{4,2\}$. Bethe roots in blue are singular solutions.}
\label{table:B+d2k42}
\end{table}

\begin{table}[h!]
\centering
\begin{tabular}{|c|c|c|c|c|c|c|}
\hline
$L$ & $m_{1}$ & $m_{2}$ & $p$ & deg & $\{ u^{(1)}_{k} \}$ & $\{ u^{(2)}_{k} \}$\\   
\hline
1 & 0  & 0 & - & 1 & - & - \\
1 & 1  & 0 & - & 1 & $\infty$ & -  \\
1 & 1  & 1 & 0 & 2 & $\infty$ & $\infty$  \\
\hline
2 & 0  & 0 & - & 1 & - & - \\
2 & 1  & 0 & - & 1 & $-\frac{\eta}{2}$ & -  \\
2 & 1  & 0 & - & 1 & $\infty$ & -  \\
2 & 1  & 1 & 0 & 2 & $-\frac{\eta}{2}$ & $\infty$  \\
2 & 1  & 1 & 0 & 2 & $\infty$  & $\infty$  \\
2 & 2  & 0 & - & 1 & $(-1 \pm i)\frac{\eta}{2}$ & - \\
2 & 2  & 1 & 0 & 3 & \textcolor{blue}{0, $-\eta$}   & \textcolor{blue}{0} \\
2 & 2  & 1 & 0 & 2 & $(-1 \pm i)\frac{\eta}{2}$   & $\infty$ \\
2 & 2  & 2 & 0 & 3 & \textcolor{blue}{$-\eta$, 
$-\eta$} & \textcolor{blue}{0, 0 }\\ 
\hline
\end{tabular}
\caption{\small Model \RN{2}$^{+}$ with $d=3$, $\vec k = \{4,3,2\}$. Bethe roots in blue are singular solutions.}
\label{table:B+d3k432}
\end{table}

\begin{table}[h!]
\centering
\begin{tabular}{|c|c|c|c|c|c|c|}
\hline
$L$ & $m_{1}$ & $m_{2}$ & $p$ & deg & $\{ u^{(1)}_{k} \}$ & $\{ u^{(2)}_{k} \}$\\   
\hline
1 & 0  & 0 &  - & 1 & - & - \\
1 & 1  & 0 &  - & 1 & $\infty$ & -  \\
1 & 1  & 1 &  0 & 2 & $\infty$ & $\infty$  \\
\hline
2 & 0  & 0 &  - & 1 & - & - \\
2 & 1  & 0 &  - & 1 & $-\frac{\eta}{2}$ & -  \\
2 & 1  & 0 &  - & 1 & $\infty$ & -  \\
2 & 1  & 1 &  0 & 2 & $-\frac{\eta}{2}$ & $\infty$  \\
2 & 1  & 1 &  0 & 2 & $\infty$  & $\infty$  \\
2 & 2  & 0 &  - & 1 & $(-1 \pm i)\frac{\eta}{2}$ & - \\
2 & 2  & 1 &  0 & 2 & \textcolor{blue}{0, $-\eta$}   & 
\textcolor{blue}{0}  \\
2 & 2  & 1 &  0 & 5 & $(-1 \pm i)\frac{\eta}{2}$   & $\infty$ \\
2 & 2  & 2 &  1 & 1 & \textcolor{blue}{0, $-\eta$}   & \textcolor{blue}{0, $-\eta$} \\
\hline
\end{tabular}
\caption{\small Model \RN{2}$^{-}$ with $d=3$, $\vec k = \{4,3,2\}$. Bethe roots in blue are singular solutions.}
\label{table:B-d3k432}
\end{table}

%\subsection{Model A}

%From the checks we performed we believe that model A is complete. We will present all the models with $ n=4 $, with $ L=1,2 $. These are just to show which type of Bethe roots appear, for example, but several other checks were made, for both higher $ L $ and higher $ n $. 

\begin{table}[h!]
	\centering
	\begin{tabular}{|c|c|c|c|c|c|c|c|}
		\hline
		$L$ & $m_{1}$ & $m_{2}$& $m_{3}$ & deg & $\{ u^{(1)}_{k} \}$ & $\{ u^{(2)}_{k} \}$ & $\{ u^{(3)}_{k} \}$\\   
		\hline
		1 & 0  & 0 & 0 &2 & - & - & - \\
		1 & 1  & 0 & 0 &2 & $\infty$ & -& -   \\
		\hline
		2 & 0  & 0 & 0 &3 & - & - & - \\
		2 & 1  & 0 & 0 &1 & $-\frac{\eta}{2}$ & - & -  \\ 
		2 & 1  & 1 & 0 &4 & $-\frac{\eta}{2}$ & $ \infty $ & -  \\
		2 & 1  & 1 & 1 &4 & $\infty$ & \textcolor{red}{$u^{(2)}_{1} $} & $ \infty $   \\
		2 & 2  & 2 & 0 &3 & $-\frac{(1\pm i)\eta}{2}$ & $-\left(1\pm\frac{i}{\sqrt{2}}\right)\eta$& -   \\
		2 & 2  & 2 & 1 &1 & \textcolor{blue}{$\{-\eta,\,0 \} $}  & $-\left(1\pm\frac{i}{\sqrt{2}}\right)\eta$ & $ \frac{\eta}{2} $  \\
		\hline
	\end{tabular}
	\caption{\small Model \RN{1} with $d=2$, $\vec k = \{4,2\}$. Bethe 
	roots in blue are singular Bethe solutions; Bethe roots in red 
	are arbitrary.} \label{table:modelA42}
\end{table}

\begin{table}[h!]
	\centering
	\begin{tabular}{|c|c|c|c|c|c|c|c|}
		\hline
		$L$ & $m_{1}$ & $m_{2}$& $m_{3}$ & deg & $\{ u^{(1)}_{k} \}$ & $\{ u^{(2)}_{k} \}$ & $\{ u^{(3)}_{k} \}$\\   
		\hline
		1 & 0  & 0 & 0 &1 & - & - & - \\
		1 & 1  & 0 & 0 &3 & $\infty$ & -& -   \\
		\hline
		2 & 0  & 0 & 0 &1 & - & - & - \\
		2 & 1  & 1 & 0 &3 & $-\frac{\eta}{2}$ & $ \infty $ & -  \\ 
		2 & 1  & 1 & 1 &3 & $ \infty $ & \textcolor{red}{$u^{(2)}_{1} $} & $ \infty $  \\
		2 & 2  & 0 & 0 &6 & $-\frac{(1\pm i)\eta}{2}$ & - & -   \\
		2 & 2  & 2 & 0 &3 &\textcolor{blue}{$\{-\eta,\,0 \} $}  & \textcolor{blue}{0} & - \\
		\hline
	\end{tabular}
	\caption{\small Model \RN{1} with $d=2$, $\vec k = \{4,3\}$. Bethe 
	roots in blue are singular solutions; Bethe roots in red are 
	arbitrary.} \label{table:modelA43}
\end{table}

\begin{table}[h!]
	\centering
	\begin{tabular}{|c|c|c|c|c|c|c|c|}
		\hline
		$L$ & $m_{1}$ & $m_{2}$& $m_{3}$ & deg & $\{ u^{(1)}_{k} \}$ & $\{ u^{(2)}_{k} \}$ & $\{ u^{(3)}_{k} \}$\\   
		\hline
		1 & 0  & 0 & 0 &1 & - & - & - \\
		1 & 1  & 0 & 0 &1 & $\infty$ & -& -   \\
		1 & 1  & 1 & 0 &2 & $\infty$ & $ \infty $& -   \\
		\hline
		2 & 0  & 0 & 0 &4 & - & - & - \\
		2 & 1  & 0 & 0 &1 & $\infty$ & - & -  \\ 
		2 & 1  & 0 & 0 &1 & $ -\frac{\eta}{2} $ & - & -  \\
		2 & 1  & 1 & 0 &2 &$ -\frac{\eta}{2} $ & $ \infty $ & -   \\
		2 & 1  & 1 & 1 &2 &$ \infty $ & \textcolor{red}{$ u_1^{(2)} $} & $ \infty $ \\
		2 & 2  & 0 & 0 &1 & $-\frac{(1\pm i)\eta}{2}  $ & - & - \\
		2 & 2  & 1 & 0 &2 &  $-\frac{(1\pm i)\eta}{2} $ & $ \infty $ & - \\
		2 & 2  & 1 & 1 &1 & \textcolor{blue}{$ \left\{-\eta,0\right\} $} & \textcolor{blue}{0}& $ \infty $ \\		
		2 & 2  & 2 & 0 &2 &$ \left\{ 
		\textcolor{red}{u_1^{(1)}},-\eta\left( 
		\frac{\eta+\textcolor{red}{u_1^{(1)}}}{\eta+2
		\textcolor{red}{u_1^{(1)}}} \right)  \right\} $ & $ 
		\left\{0,-2 \textcolor{red}{u_1^{(1)}}\left( \frac{\eta+ 
		\textcolor{red}{u_1^{(1)}}}{\eta+2 \textcolor{red}{u_1^{(1)}}} \right)\right\} $ & 0 \\		
		\hline
	\end{tabular}
	\caption{\small Model \RN{1} with $d=3$, $\vec k = \{4,3,2\}$. Bethe 
	roots in blue are singular solutions; Bethe roots in red are 
	arbitrary. } \label{table:modelA432}
\end{table}

\begin{table}[h!]
	\centering
	\begin{tabular}{|c|c|c|c|c|c|c|c|}
		\hline
		$L$ & $m_{1}$ & $m_{2}$& $m_{3}$ & deg & $\{ u^{(1)}_{k} \}$ & $\{ u^{(2)}_{k} \}$ & $\{ u^{(3)}_{k} \}$\\   
		\hline
		1 & 0  & 0 & 0 &2 & - & - & - \\
		1 & 1  & 0 & 0 &2 & $\infty$ & -& -   \\
		\hline
		2 & 0  & 0 & 0 &3 & - & - & - \\
		2 & 1  & 0 & 0 &1 & $ -\frac{\eta}{2} $ & - & -  \\
		2 & 1  & 1 & 0 &4 & $\infty$ & \textcolor{red}{$ u_1^{(2)} $} & -  \\ 
		2 & 1  & 1 & 0 &4 &$ -\frac{\eta}{2} $ & $ \infty $ & -   \\
		2 & 2  & 2 & 0 &2 &$ -(1\pm i)\frac{\eta}{2} $ & $ i(2i\pm \sqrt{2})\frac{\eta}{2} $ & - \\
		2 & 2  & 0 & 0 &1 & $\left\{\infty,\infty \right\}  $ & $\left\{\textcolor{red}{ u_1^{(2)}} ,\textcolor{red}{ u_2^{(2)}} \right\} $& $ \infty $ \\
		2 & 2  & 2 & 1 &1 &  $ \left\{ 
		\textcolor{red}{u_1^{(1)}},-\eta\left( 
		\frac{\eta+ \textcolor{red}{u_1^{(1)}}}{\eta+2
			\textcolor{red}{u_1^{(1)}}} \right)  \right\} $ & $\left\{\infty,\infty \right\}  $ &$  \infty $\\
		\hline
	\end{tabular}
	\caption{\small Model \RN{3} with $d=2$, $\vec k = \{4,2\}$ and $ \mu_0=+1 $.  Bethe roots in red are 
		arbitrary.} \label{table:modelC42}
\end{table}

\begin{table}[h!]
	\centering
	\begin{tabular}{|c|c|c|c|c|c|c|c|}
		\hline
		$L$ & $m_{1}$ & $m_{2}$& $m_{3}$ & deg & $\{ u^{(1)}_{k} \}$ & $\{ u^{(2)}_{k} \}$ & $\{ u^{(3)}_{k} \}$\\   
		\hline
		1 & 0  & 0 & 0 &1 & - & - & - \\
		1 & 1  & 0 & 0 &1 & $\infty$ & -& -   \\
		1 & 1  & 1 & 0 &2 & $\infty$ & $ \textcolor{red}{u_1^{(2)}} $& -   \\
		\hline
		2 & 0  & 0 & 0 &3 & - & - & - \\
		2 & 1  & 0 & 0 &1 & $\infty$ & - & -  \\ 
		2 & 1  & 0 & 0 &1 & $ -\frac{\eta}{2} $ & - & -  \\
		2 & 1  & 1 & 0 &2 &$ -\frac{\eta}{2} $ & $ \infty $ & -   \\
		2 & 1  & 1 & 0 &2 &$ \infty $ & \textcolor{red}{$ u_1^{(2)} $} & - \\
		2 & 2  & 0 & 0 &1 & $-\frac{(1\pm i)\eta}{2}  $ & - & - \\
		2 & 2  & 1 & 0 &2 &  $-\frac{(1\pm i)\eta}{2} $ & $ \infty $ & - \\
		2 & 2  & 1 & 0 &2 & \textcolor{blue}{$ \left\{0,-\eta\right\} $} & \textcolor{blue}{0}& - \\	
		2 & 2  & 2 & 1 &1 & $-\frac{(1\pm i)\eta}{2} $ & $\left\{\infty,\infty \right\}  $& - \\	
		2 & 2  & 2 & 1 &1 & $-\frac{(1\pm i)\eta}{2} $ & $\left\{\infty,0 \right\}  $& $ \infty $ \\
		\hline
	\end{tabular}
	\caption{\small Model \RN{3} with $d=3$, $\vec k = \{4,3,2\}$ and $ \mu_0=-1 $. Bethe 
		roots in blue are singular solutions; Bethe roots in red are 
		arbitrary.} \label{table:modelC432}
\end{table}

\clearpage
\bibliographystyle{nb}
\bibliography{References}

\end{document}